\documentclass[%
  reprint,
 superscriptaddress,
 amsmath,amssymb,
 aps,
 showpacs,
prd,
]{revtex4-1}

\usepackage{graphicx}
\usepackage{dcolumn}
\usepackage{bm}
\usepackage{enumerate}
\usepackage{multirow}

\def\peppo{\(p \rightarrow e^+ \pi^0\)}

\def\pmppo{\(p \rightarrow \mu^+ \pi^0\)}

\def\pepom{\(p \rightarrow e^+ \omega\)}

\def\eqepi0{\(p \rightarrow e^+ \pi^0\)}
\def\eqmupi0{\(p \rightarrow \mu^+ \pi^0\)}
\def\eqeomega{\(p \rightarrow e^+ \omega\)}
\def\eqmuomega{\(p \rightarrow \mu^+ \omega\)}
\def\eqeeta{\(p \rightarrow e^+ \eta\)}
\def\eqmueta{\(p \rightarrow \mu^+ \eta\)}
\def\eqerho{\(p \rightarrow e^+ \rho^0\)}
\def\eqmurho{\(p \rightarrow \mu^+ \rho^0\)}
\def\eqepim{\(n \rightarrow e^+ \pi^-\)}
\def\eqmupim{\(n \rightarrow \mu^+ \pi^-\)}
\def\eqerhom{\(n \rightarrow e^+ \rho^-\)}
\def\eqmurhom{\(n \rightarrow \mu^+ \rho^-\)}
\newcommand{\etal}{{\it et\ al}.}

\begin{document}

\title{Search for Nucleon Decay into Charged Anti-lepton plus Meson in Super-Kamiokande I and II}

\newcommand{\AFFicrr}{\affiliation{Kamioka Observatory, Institute for Cosmic Ray Research, University of Tokyo, Kamioka, Gifu 506-1205, Japan}}
\newcommand{\AFFkashiwa}{\affiliation{Research Center for Cosmic Neutrinos, Institute for Cosmic Ray Research, University of Tokyo, Kashiwa, Chiba 277-8582, Japan}}
\newcommand{\AFFipmu}{\affiliation{Institute for the Physics and
Mathematics of the Universe, Todai Institutes for Advanced Study,
University of Tokyo, Kashiwa, Chiba 277-8582, Japan }}
\newcommand{\AFFbu}{\affiliation{Department of Physics, Boston University, Boston, MA 02215, USA}}
\newcommand{\AFFbnl}{\affiliation{Physics Department, Brookhaven National Laboratory, Upton, NY 11973, USA}}
\newcommand{\AFFuci}{\affiliation{Department of Physics and Astronomy, University of California, Irvine, Irvine, CA 92697-4575, USA }}
\newcommand{\AFFcsu}{\affiliation{Department of Physics, California State University, Dominguez Hills, Carson, CA 90747, USA}}
\newcommand{\AFFcnm}{\affiliation{Department of Physics, Chonnam National University, Kwangju 500-757, Korea}}
\newcommand{\AFFduke}{\affiliation{Department of Physics, Duke University, Durham NC 27708, USA}}
\newcommand{\AFFfukuoka}{\affiliation{Junior College, Fukuoka Institute of Technology, Fukuoka, Fukuoka 811-0295, Japan}}
\newcommand{\AFFgmu}{\affiliation{Department of Physics, George Mason University, Fairfax, VA 22030, USA }}
\newcommand{\AFFgifu}{\affiliation{Department of Physics, Gifu University, Gifu, Gifu 501-1193, Japan}}
\newcommand{\AFFuh}{\affiliation{Department of Physics and Astronomy, University of Hawaii, Honolulu, HI 96822, USA}}
\newcommand{\AFFkanagawa}{\affiliation{Physics Division, Department of Engineering, Kanagawa University, Kanagawa, Yokohama 221-8686, Japan}}
\newcommand{\AFFkek}{\affiliation{High Energy Accelerator Research Organization (KEK), Tsukuba, Ibaraki 305-0801, Japan }}
\newcommand{\AFFkobe}{\affiliation{Department of Physics, Kobe University, Kobe, Hyogo 657-8501, Japan}}
\newcommand{\AFFkyoto}{\affiliation{Department of Physics, Kyoto University, Kyoto, Kyoto 606-8502, Japan}}
\newcommand{\AFFumd}{\affiliation{Department of Physics, University of Maryland, College Park, MD 20742, USA }}
\newcommand{\AFFmit}{\affiliation{Department of Physics, Massachusetts Institute of Technology, Cambridge, MA 02139, USA}}
\newcommand{\AFFmiyagi}{\affiliation{Department of Physics, Miyagi University of Education, Sendai, Miyagi 980-0845, Japan}}
\newcommand{\AFFnagoya}{\affiliation{Solar Terrestrial Environment Laboratory, Nagoya University, Nagoya, Aichi 464-8602, Japan}}
\newcommand{\AFFsuny}{\affiliation{Department of Physics and Astronomy, State University of New York, Stony Brook, NY 11794-3800, USA}}
\newcommand{\AFFniigata}{\affiliation{Department of Physics, Niigata University, Niigata, Niigata 950-2181, Japan }}
\newcommand{\AFFokayama}{\affiliation{Department of Physics, Okayama University, Okayama, Okayama 700-8530, Japan }}
\newcommand{\AFFosaka}{\affiliation{Department of Physics, Osaka University, Toyonaka, Osaka 560-0043, Japan}}
\newcommand{\AFFseoul}{\affiliation{Department of Physics, Seoul National University, Seoul 151-742, Korea}}
\newcommand{\AFFshizuokasc}{\affiliation{Department of Informatics in
Social Welfare, Shizuoka University of Welfare, Yaizu, Shizuoka, 425-8611, Japan}}
\newcommand{\AFFskk}{\affiliation{Department of Physics, Sungkyunkwan University, Suwon 440-746, Korea}}
\newcommand{\AFFtohoku}{\affiliation{Research Center for Neutrino Science, Tohoku University, Sendai, Miyagi 980-8578, Japan}}
\newcommand{\AFFtokyo}{\affiliation{The University of Tokyo, Bunkyo, Tokyo 113-0033, Japan }}
\newcommand{\AFFtokai}{\affiliation{Department of Physics, Tokai University, Hiratsuka, Kanagawa 259-1292, Japan}}
\newcommand{\AFFtit}{\affiliation{Department of Physics, Tokyo Institute
for Technology, Meguro, Tokyo 152-8551, Japan }}
\newcommand{\AFFtsinghua}{\affiliation{Department of Engineering Physics, Tsinghua University, Beijing, 100084, China}}
\newcommand{\AFFwarsaw}{\affiliation{Institute of Experimental Physics, Warsaw University, 00-681 Warsaw, Poland }}
\newcommand{\AFFuw}{\affiliation{Department of Physics, University of Washington, Seattle, WA 98195-1560, USA}}

\AFFicrr
\AFFkashiwa
\AFFbu
\AFFbnl
\AFFuci
\AFFcsu
\AFFcnm
\AFFduke
\AFFfukuoka
\AFFgifu
\AFFuh
\AFFkanagawa
\AFFkek
\AFFkobe
\AFFkyoto
\AFFmiyagi
\AFFnagoya
\AFFsuny
\AFFniigata
\AFFokayama
\AFFosaka
\AFFseoul
\AFFshizuokasc
\AFFskk
\AFFtokai
\AFFtokyo
\AFFipmu
\AFFtsinghua
\AFFwarsaw
\AFFuw

\author{H.~Nishino} 
\altaffiliation{Present address: University of California, Berkeley, CA 94720, USA}
\AFFkashiwa

\author{K.~Abe}
\AFFicrr
\author{Y.~Hayato}
\AFFicrr
\AFFipmu
\author{T.~Iida}
\author{M.~Ikeda}
\author{J.~Kameda}
\author{Y.~Koshio}
\author{M.~Miura} 
\AFFicrr
\author{S.~Moriyama} 
\author{M.~Nakahata} 
\AFFicrr
\AFFipmu
% ...

\collaboration{The Super-Kamiokande Collaboration}
\noaffiliation

\date{\today}

\begin{abstract}

Searches for a nucleon decay into a charged anti-lepton
($e^+$ or $\mu^+$) plus a light meson ($\pi^0$, $\pi^-$, $\eta$,
$\rho^0$, $\rho^-$, $\omega$) were performed using the Super-Kamiokande
I and II data. Twelve nucleon decay modes
were searched for. The total exposure is 140.9 kiloton$\cdot$years, which
includes a 91.7 kiloton$\cdot$year exposure (1489.2 live days) of
Super-Kamiokande-I and a 49.2 kiloton$\cdot$year exposure (798.6 live days) of
Super-Kamiokande-II.
The number of candidate events in the data was consistent with the
atmospheric neutrino background expectation.
No significant evidence for a nucleon decay was observed in the data.
Thus, lower limits on the nucleon partial lifetime at 90\% confidence
level were obtained.
The limits range from $3.6\times 10^{31}$ to $8.2\times 10^{33}$ years,
depending on the decay modes.

\end{abstract}


\maketitle

\section{Introduction}

The standard model of particle physics has been successful in explaining 
most experimental results.
However, it contains many
empirical parameters, such as masses and generations of fermions,
coupling constants, mixing angles, and so on.
Grand Unified Theories (GUTs) have been proposed to account for these parameters.
The basic idea of GUTs is that the SU(2)$\times$U(1)
symmetry of electroweak interactions and the SU(3) color symmetry of
strong interactions 
are
incorporated into 
a larger symmetry group broken at an ultra-high energy scale.
GUTs predict new interactions in which leptons and quarks can
transform one into the other by exchanging a super-heavy gauge
boson. This type of interaction can lead to baryon
number violating nucleon decays.
The simplest GUT, minimal SU(5)~\cite{Georgi:1974sy},
predicts the partial lifetime of a proton via the mode \eqepi0 to be
$\sim 10^{31\pm1}$ years~\cite{Langacker:1980js, Langacker:1994vf},
which was excluded by the IMB and KAMIOKANDE
experiments~\cite{McGrew:1999nd, Hirata:1989kn}.

There are various viable GUTs (see, e.g., \cite{Pati:1974yy,
Lee:1994vp, Shaban:1992vv, Pati:2003qia, Kim:2002im, Buchmuller:2004eg, Ellis:2002vk}),
such as models incorporating supersymmetry (SUSY),
models with a symmetry group like SO(10), flipped SU(5) models, and models
in extra dimensions.
Predictions for the lifetime of the nucleon strongly depend on the models 
and also have large uncertainties.
Some models predict the partial lifetime of the proton to be in the
accessible range of the Super-Kamiokande experiment.

The Super-Kamiokande collaboration 
previously
published the results of \eqepi0\ and
\eqmupi0\ searches~\cite{Nishino:2009gd}.
Updated results 
have also been presented since that time
\cite{epi0_sk1234}.
No candidate events for proton
decay were found.
Although the \eqepi0\ mode is considered to be the dominant decay mode
in many GUT models, branching ratios for the other modes are not
necessarily negligible.
Many possibilities for nucleon decays are theoretically proposed.
In some SU(5) and SO(10) GUT models
(\cite{Machacek:1979tx, Gavela:1980at, Donoghue:1979pr,Buccella:1989za}),
branching ratios for \eqeeta, \eqerho\ or
\eqeomega\ can be as high as 10 $\sim$ 20\% and the proton and bound neutron
lifetimes are expected to be comparable to each other.
Also, $N\rightarrow \mu^+ + meson$ modes can be induced by mixing
effects between lepton families. 
Though typical SU(5) GUT models predict very small branching ratios for
the muon modes (see, e.g., \cite{Berezinsky:1981qb, Machacek:1979tx}), the flipped SU(5) GUT~\cite{Ellis:2002vk} shows that
the \eqmupi0\ mode can have a comparable branching ratio with the
\eqepi0\ mode.
Therefore, nucleon decay modes mediated by the exchange of super-heavy
gauge bosons are important for most of the GUT models.
In spite of the importance of investigating all possible decay modes, so far 
only the \eqepi0\ and \eqmupi0\ searches, and the searches for several modes favored by SUSY
GUTs~\cite{Kobayashi:2005pe} have been published by the
Super-Kamiokande experiment.
In this paper, we search for 
nucleon decays into a charged anti-lepton ($e^+$ and
$\mu^+$) plus an
un-flavored light meson ($\pi$, $\eta$, $\rho$ and $\omega$) 
using the Super-Kamiokande I and II data.
Eight modes were studied for proton decay, and four for neutron decay. 
The results of the \eqepi0\ and \eqmupi0 mode
are identical to those reported
in an earlier Letter~\cite{Nishino:2009gd}, however
more details of the analysis are described in this paper. In addition,
we briefly report the results of the search in these two channels based on
a longer exposure including Super-Kamiokande III and IV data. 

\section{Super-Kamiokande Detector}

Super-Kamiokande is a 50-kiloton water Cherenkov detector located in the
Kamioka Observatory in Japan, under $\sim$1000~m of rock overburden.
The Super-Kamiokande detector is made of a cylindrical stainless steel
tank, 39.3~m in diameter and 41.4~m in height.
The detector is optically separated into two regions: an inner detector
(ID), and an outer detector (OD).
On the surface of the ID, 20-inch photomultiplier tubes (PMTs)
are uniformly attached to detect Cherenkov light radiated by
relativistic charged particles.
The OD, which surrounds the ID with a 2~m thickness of water, is used to
reject cosmic ray muon events and to tag exiting charged particles with
1,885 outward-facing 8-inch PMTs.
The OD region also serves as a shield from radioactivity from materials
outside the detector wall.

The Super-Kamiokande experiment started data-taking in April, 1996, and
had continued observation for five years until the detector maintenance
in July 2001 (SK-I). On November 2001, while
filling the tank with water after the maintenance, an accident occurred
which destroyed more than half of the PMTs.
The detector was partially rebuilt with half the density of photo-sensor
coverage in the ID.
Observation began again in October 2002 and stopped in October 2005
(SK-II) for a full detector reconstruction.
The next phase, SK-III, started October 2006 and was switched to SK-IV
with newly developed electronics and online systems in September 2008.
This paper 
reports nucleon decay search results from the first two
periods: SK-I and SK-II.
The ID photo-sensor coverage in the SK-I period was 40\% with 11,146
20-inch PMTs.
In the SK-II period, the coverage reduced to 19\% with 5,182 20-inch
PMTs. All of the ID PMTs have been equipped with acrylic covers and
fiber-reinforced plastic cases since the start of SK-II to avoid a
cascade of
implosions of PMTs.
Details of the detector configuration 
and performance for SK-I have been published~\cite{Fukuda:2002uc}.

 \section{Simulation}
  \subsection{Nucleon Decay Simulation}\label{sec:sim_ndk}
  In order to estimate the detection efficiencies of nucleon decays,
  a nucleon decay Monte Carlo (MC) simulation was developed.
  The number of background events was estimated by simulations of the
  atmospheric neutrino flux and neutrino interactions.
  
  In an H$_2$O molecule, there are two free protons in the hydrogen
  nuclei, 
  and eight bound protons and eight bound neutrons in the oxygen nucleus.
  The decay probabilities of free protons and bound
  protons 
  were assumed
  to be equal in the simulation.
  All of the decay modes studied in this analysis are two-body decays,
  meaning that the charged lepton and meson are back-to-back with a monochromatic
  momentum in the nucleon rest frame.

  For the decay of bound nucleons in oxygen,
  the simulation takes into account the effect of 
  Fermi motion of the nucleons, correlation with another nucleon, 
  and meson-nucleon interactions in the nucleus (nuclear effects).
  Nucleon momentum and nuclear binding energy in the nucleus were
  calculated as described in \cite{Nakamura:1976mb}.
  Good agreement of the calculations with an electron scattering
  experiment on $^{12}$C is also shown in \cite{Nakamura:1976mb}.
  
  Mesons generated in an oxygen nucleus interact with nucleons
  until they escape from the nucleus.
  The position of the nucleon decay in a nucleus was determined by the
  Woods-Saxon nuclear density distribution in the simulation.
  From this position, 
  $\pi$, $\eta$ and $\omega$ mesons were tracked in an oxygen nucleus.
  The lifetime of the $\rho$ meson is so short ($\beta\gamma\tau\simeq$ 0.3~fm) 
  that it decays
  immediately inside the nucleus into 2 $\pi$ mesons; therefore, the nuclear effects of the $\rho$ meson
  itself were not considered in the simulation.
  The mass of the $\rho$ meson was assumed to have a Breit-Wigner type
  distribution with a width of $\Gamma = 149$~MeV.
  For these reasons, particles generated via the modes including a $\rho$ meson
  do not have monochromatic momentum, even in the case of free proton decay.
  
  Ten percent of nucleons in oxygen were assumed to decay
  correlating with another nucleon~\cite{Yamazaki:1999gz}.
  In such correlated decay events, 
  the invariant mass of $e^+$ and $\pi^0$ is smaller than
  the mass of the nucleon because of the invisible momentum of another
  correlated nucleon.

  Simulated particles of nucleon decays and atmospheric neutrino
  interactions were passed through a
  GEANT-3~\cite{geant} custom detector simulation.
  Hadronic interactions were treated by CALOR~\cite{Gabriel:1989ri}
  for nucleons and charged pions of $p_{\pi} > 500~$ MeV/$c$,
  and by a custom simulation program~\cite{Nakahata:1986zp} for charged
  pions of $p_{\pi} \leq 500~$ MeV/$c$.

   \subsubsection{$\pi$ Meson Nuclear Effects}\label{subsec:pinuc}
   In considering pion nuclear effects, inelastic scattering, absorption, charge
   exchange and pion production were taken into account in the
   simulation.
   Pion production hardly occurs in nucleon decay events since the
   cross-section for low momentum pions ($p_\pi<500$~MeV/$c$) is
   negligibly small.
   Cross-sections for these processes were calculated based on the model of
   Oset \etal~\cite{Salcedo:1987md}.
   The angular and momentum distributions of pions were determined from the
   results of $\pi$-$N$ scattering experiments~\cite{Rowe:1978fb}.
   Because of the Pauli exclusion principle,
   the momentum of the scattered nucleon is required to be greater than the
   Fermi surface momentum, given by
   $p_F(r)=\left(\frac{3}{2}\pi^2\rho(r)\right)^{\frac{1}{3}}$,
   where $\rho(r)$ is the nuclear density distribution and $r$ is the
   distance from the center of the nucleus.
   
   The $\pi^0$ momentum from the decay mode of \peppo\ is 459~MeV/$c$ in
   the rest frame of the nucleon.
   At this momentum, 37\% of $\pi^0$s were simulated to be absorbed or
   charged-exchanged in the nucleus.
   This is a major reason for the inefficiency of the \peppo\ and \pmppo\
   modes.
   The probability  for a $\pi^0$ to escape from a nucleus without any
   scattering was estimated to be 44\%.
   
   \subsubsection{$\eta$ Meson Nuclear Effect}\label{subsec:etanuc}
\begin{figure}[htbp]
    \begin{center}
     \includegraphics[width=.9\linewidth]{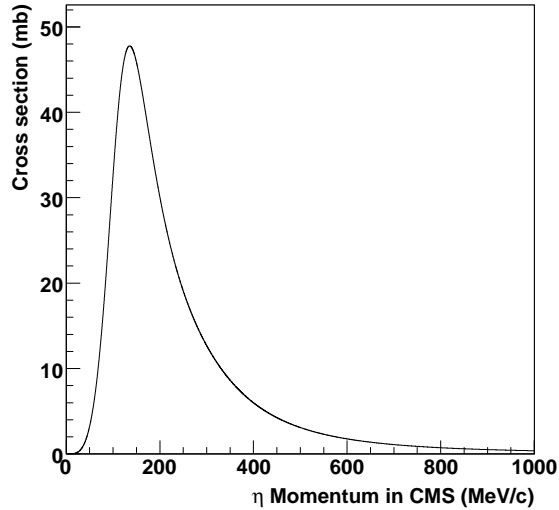}
     \caption{Calculated cross-section of $\eta$-nucleon interaction
     considering $S_{11}$(1535) resonance in the custom simulation code.
     }
     \label{fig:etasigma}
    \end{center}
   \end{figure}

   The interactions of $\eta$s and nucleons in the nucleus
   were
   considered through a
   baryon resonance of $S_{11}(1535)$.
   \begin{eqnarray}
    \eta + N &\rightarrow& S_{11}(1535) \nonumber\\
    S_{11}(1535) &\rightarrow& N' + meson\ (\pi, \eta, \pi\pi)
   \end{eqnarray}
   The cross-section for $\eta$-nucleon interactions 
   shown in Fig.~\ref{fig:etasigma} 
   was calculated by the Breit-Wigner formula as follows:
   \begin{equation}
    \sigma = \frac{\pi}{k^2}\frac{\Gamma_{\eta N}\left(\Gamma_{\textrm{total}}-\Gamma_{\eta N}\right)}{\left(E_{\textrm{CMS}}-M_{\textrm{res}}\right)^2+\Gamma_{\textrm{total}}^2/4},
   \end{equation}
   where $E_{\textrm{CMS}}$ is the 
   center-of-mass energy 
   of $\eta$-$N$,
   $M_{\textrm{res}}$ is the mass of the resonance, 
   $\Gamma_\textrm{total}$ is the total width of $S(1535)$ resonance,
   $\Gamma_{\eta N}$ is the partial width of $S(1535)\rightarrow \eta N$,
   and $k$ is the wave number~($p/\hbar$).
   The net nuclear effect of $\eta$-meson in an oxygen nucleus is
   estimated by using this cross-section as well as considering the
   nuclear density distribution, nucleon momentum  distribution and the
   Pauli exclusion principle effect in an nucleus.
   Since pions can be generated by the decay of the resonance,
   nuclear effects for $\pi$ mesons were also considered according to
   Section~\ref{subsec:pinuc}.
   Approximately 56\% of $\eta$ mesons generated by $p\rightarrow
   e^+\eta$ escaped from the nucleus without any scattering,
   while 38\% of $\eta$ mesons were absorbed or decayed into other particles
   from the resonance.
   
   The $\eta$ meson nuclear effects simulation 
   were
   checked by 
   a comparison with the experimental cross-section of 
   $\eta$ photoproduction on a $^{12}$C target measured at
   MAINZ~\cite{RoebigLandau:1996xa}.
   Since photons are insensitive to strong interactions and able to
   probe the inside of nuclei,
   meson photoproduction data were used for the study of the nuclear
   effects.
   The peak momentum of the $\eta$ meson in the photoproduction
   experiment is approximately 300~MeV/c, which
   is equivalent to the $\eta$ momentum generated by
   $p\rightarrow e^+(\textrm{or}\ \mu^+)\ \eta$.
   In order to simulate the $\eta$ photoproduction on a $^{12}$C target,
   $\eta$ mesons were generated following the cross-section
   data of a proton target from the SAID
   calculation~\cite{sim:etaphotopro-nucleon},
   and tracked in nuclei while suffering the $\eta$-$N$ interaction
   described above.
   The cross-section on a neutron 
   was assumed to be 2/3 of that of a proton target,
   since this assumption reproduces the experimental cross-sections of
   a deuteron target~\cite{Krusche:1995zx}.
   
\begin{figure}[htbp]
    \begin{center}
     \includegraphics[width=.9\linewidth]{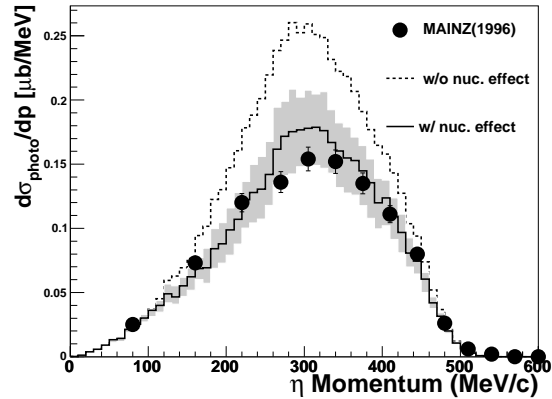}
     \caption{The differential cross-sections of photoproduction of
     an $\eta$ meson on
     a $^{12}$C target with $\gamma$ energy of 735~$\sim$~765~MeV.
     The experimental data at MAINZ~\cite{RoebigLandau:1996xa} are shown as
     circles. The solid line shows the simulated
     production cross-section, and its shaded region corresponds to
     a factor 2 difference in the cross-section of $\eta$ nuclear effects.
     The dashed line shows the production cross-section without nuclear
     effects.
     }
     \label{fig:etanucl}
    \end{center}
   \end{figure}

   Figure~\ref{fig:etanucl} shows a comparison of the experimental $\eta$
   photoproduction cross-section for a $^{12}$C target
   and the cross-section simulated by our nuclear effects simulation.
   The simulation predicted well the reduction of the photoproduction
   cross-section by the nuclear effects.
   The difference between the experiment and the simulation was taken
   as the uncertainty of the $\eta$ nuclear effects.
   The uncertainty of the cross-section of the $\eta$ nuclear effects was
   estimated to be a factor of two.
   
  \subsubsection{$\omega$ Meson Nuclear Effect}\label{subsec:omeganuc}
  The width of the $\omega$ meson resonance is $\Gamma = 8.49$~MeV and its
  lifetime is $\tau = \hbar/\Gamma = 0.77\times10^{-22}$~sec.
  In the proton decay of \pepom, the $\omega$ meson momentum is $\sim$140 MeV/$c$
  so that the mean free path ($\beta\gamma\tau$) of the $\omega$ meson is $\sim$4~fm.
  The decay of $\omega$'s was taken into account in the $\omega$ meson tracking
  in the nucleus because
  the mean free path is comparable to the size of the radius of a nucleus.

  The $\omega$ meson interactions with a nucleon in an oxygen nucleus were
  calculated with a boson exchange model by Lykasov \etal~\cite{Lykasov:1998ma}.
  The coupling constants and the form factor of this model
  were fixed by the experimental data.
  The simulation takes into account reactions such as $\omega N \rightarrow \omega N$,
  $\omega N \rightarrow \pi N'$, 
  $\omega N \rightarrow \rho N'$, $\omega N \rightarrow \rho \pi N'$,
  $\omega N \rightarrow \pi \pi N'$ and $\omega N \rightarrow \sigma
  N$.
  Approximately 53\% of $\omega$ mesons generated in 
  $p\rightarrow e^+\omega$ decayed in the 
  nucleus.
  Therefore,
  the nuclear effects of secondary pions was also considered.

  The CBELSA/TAPS collaboration measured the $\omega$ photoproduction cross
  section on the various nucleus targets of C, Ca, Nb and
  Pb~\cite{Kotulla:2008xy}.
  They extracted the $\omega N$ cross section from the photoproduction
  cross-sections and compared it with the calculated cross-section by
  Lykasov \etal.
  The cross-section obtained by the experiment is 
  approximately 3 times larger at most than the calculated cross-section.
  We took this difference as an uncertainty in the cross-section of
  $\omega$ meson nuclear effects.
  
  \subsection{Atmospheric Neutrino Simulation}\label{sim:atm}
  
  Atmospheric neutrino interactions are backgrounds for nucleon
  decay searches in Super-Kamiokande.
  Charged current single (multi) pion production can be the dominant
  background source for the search modes.
  A charged current quasi-elastic scattering can also contribute the
  background because a nucleon produced by the neutrino interaction can
  produce pions by hadronic interactions in water.
  Atmospheric neutrino events were simulated
  using the NEUT~\cite{Hayato:2002sd} neutrino interaction MC with
  an atmospheric neutrino flux
  calculated by Honda \etal~\cite{Honda:2006qj}.
  Complete details of the simulation are described in \cite{Ashie:2005ik}.
  Atmospheric neutrino simulation MC equivalent to a 500-year
  observation (11.25~megaton$\cdot$year exposure) were generated to
  estimate the nucleon decay search background for each SK-I and SK-II.
  Pure $\nu_\mu$ to $\nu_\tau$ oscillation with 
  $\Delta m^2 = 2.5\times 10^{-3}$~eV$^2$ 
  and $\sin^22\theta = 1.0$ was assumed for the estimation.

\section{Data Set and Data Reduction}

We used data from a 91.7 kiloton$\cdot$year exposure of 1489.2 live days
during SK-I and a 49.2 kiloton$\cdot$year exposure of 798.6 live days of
during SK-II.
The data acquisition trigger threshold for this analysis corresponds to
a 5.7~MeV electron in SK-I and 8 MeV in SK-II.
The trigger rate was about 11~Hz,
resulting in approximately $10^6$ events every day.
Most of those are events caused by a cosmic ray muon, or a low energy
background
from the radioactivity of materials around the detector wall,
or flashing PMTs.
Several stages of data reduction were applied to these events before
proceeding to further detailed event reconstruction processes that are described in 
Section~\ref{sec:reconstruction}. The details of the data reduction can be
found in \cite{Ashie:2005ik}.

For nucleon decay searches, fully contained (FC) events, which have
an activity only in the inner detector, were selected 
by requiring a vertex to be inside the 22.5 kiloton fiducial volume (2m
away from the ID detector wall),
visible energy to be greater than 30~MeV
and no hit-PMT clusters in the outer detector.
The rate of FC events was 
$8.18\pm0.07$ (stat.) events per day for SK-I and $8.26\pm 0.10$(stat.) 
events per day for SK-II.
In total, we obtained 12232 and 6584 FC events in the SK-I
and SK-II data, respectively.
The background contamination other than atmospheric neutrinos was estimated
to be less than 1\%.

The same reduction criteria were also applied to the nucleon decay MC.
For nucleon decays via the $N\rightarrow e^+$ meson modes, the
reduction survival efficiencies were estimated to be greater than 97\%.
On the other hand, for the $N\rightarrow \mu^+ (\rho\ \textrm{or}\ \omega)$
modes, the survival efficiencies are relatively lower because the muon
can be
invisible due to the large meson mass, and because the meson sometimes cannot
escape from the nucleus or it can be immediately absorbed by nucleons in
water. 
That results in no detectable event signal in a water Cherenkov detector.
The reduction survival efficiencies for all of the search modes are shown
in Table~\ref{tab:reduction_ndkmc}.

 \begin{table}[htbp]
  \begin{center}
   \begin{tabular}{l|rr||l|rr}
    \hline \hline
    Mode & SK-I & SK-II & Mode & SK-I & SK-II \\
    \hline
    \eqepi0     & $>$99\% & $>$99\%
    & \eqeomega & $>$99\% & $>$99\%\\
    \eqmupi0    & $>$99\% & $>$99\%
    & \eqmuomega& 93\% & 94\% \\
    \eqeeta     & $>$99\% & $>$99\%
    &  \eqepim  & $>$99\% & $>$99\% \\
    \eqmueta    & $>$99\% & $>$99\%
    & \eqmupim  & $>$99\% & $>$99\% \\
    \eqerho     & 98\% & 98\%
    & \eqerhom  & 97\% & 97\%\\
    \eqmurho    & 81\% & 82\%
    & \eqmurhom & 83\% & 84\%\\
    \hline \hline
   \end{tabular}
   \caption{The survival efficiency by the data reduction
   criteria for the nucleon decay MC events in the fiducial volume for
   SK-I and SK-II. The fiducial volume cut is not included.}
    \label{tab:reduction_ndkmc}
  \end{center}
 \end{table}

\section{Event Reconstruction}\label{sec:reconstruction}

\subsection{Event Reconstruction for Nucleon Decay Search}\label{sec:rec_rec}
Event reconstruction processes were applied to the fully contained events
which passed the data reduction.
The same reconstruction algorithms were applied both for the observed
data and the MC simulation.
Most of the processes are common with atmospheric neutrino analyses
in Super-Kamiokande.
The details of their algorithms are described in \cite{Ashie:2005ik}.

A vertex position was determined by the time-of-flight subtracted
timing distribution.
Vertex resolutions for the free proton decay events of \eqepi0\
are 18.1 (20.1)~cm in SK-I (SK-II).
A particle identification (PID) algorithm classified found Cherenkov
rings into shower-type ($e$-like) or non shower-type
($\mu$-like). This classification basically exploits the ring pattern
difference between a fuzzy ring pattern of electromagnetic showers by
an electron or a gamma-ray and a sharper Cherenkov ring edge produced by a
muon or a charged pion. 
The misidentification probabilities for the free proton decay MC
samples were estimated to be 3.3\% (3.4\%) and 4.8\% (5.4\%) for
\eqepi0\ and \eqmupi0\, respectively, in SK-I (SK-II).
Additionally, for nucleon decay searches expecting a low momentum muon
and/or a charged pion, a reconstructed Cherenkov opening angle was also
used to improve the PID efficiency.

\begin{figure}[htbp] 
 \begin{center}
  \includegraphics[width=.9\linewidth]{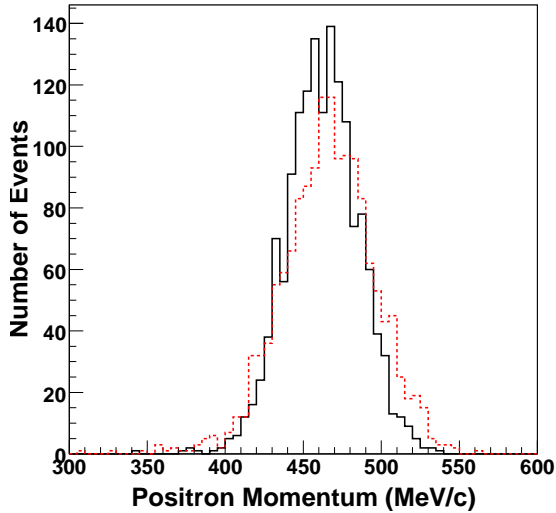}
  \caption{The reconstructed momentum distributions for positrons in
  the free proton decay of \eqepi0\ in SK-I (solid) and SK-II (dashed).
  The momentum resolution for the positron, as shown by the width of the
  distribution, is 4.9\% in SK-I and 6.0\% in SK-II.
  }
  \label{fig:epi0_separation}
 \end{center}
\end{figure}

\begin{figure}[htbp] 
 \begin{center}
  \includegraphics[width=.8\linewidth,angle=-90,clip]{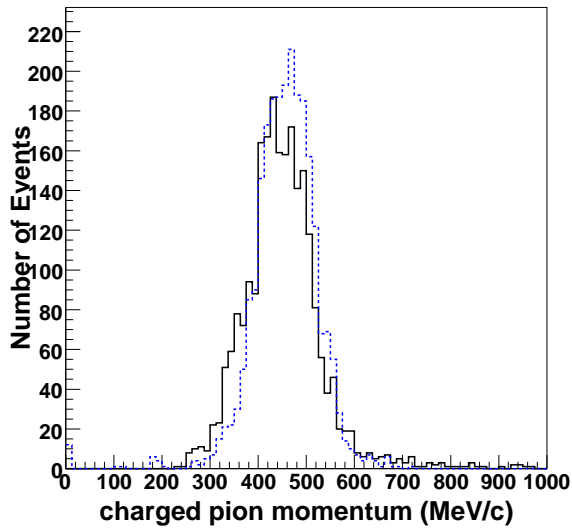}
  \caption{The charged pion momentum distribution for the \eqepim\
  MC in SK-I.
  The solid line and the dashed line show the
  reconstructed and true pion momentum distributions, respectively.
  The charged pion momentum can be well reconstructed with
  a resolution of 9.2\% (9.7\%) in SK-I (SK-II).
  }
  \label{fig:ppi_reconst}
 \end{center}
\end{figure}

The momentum of each ring was estimated by the charge detected inside each
Cherenkov ring cone.
In multi-ring events such as nucleon decay events, 
Cherenkov photons radiated from different particles pile up in each PMT.
The detected charge at each PMT was separated into the contribution from
each particle using an expected charge distribution which takes into
account light scattering in water, reflection on PMT surfaces and the
vertex shift due to the $\gamma$'s conversion length.
The momentum resolutions for the proton decay MC were estimated by
the monochromatic momentum positron of \eqepi0\ to be
4.9\% in SK-I and 6.0\% in SK-II, as shown in Fig.~\ref{fig:epi0_separation}.
Likewise, the muon momentum resolution was found from (\eqmupi0) to be
2.8\% in SK-I and 4.1\% in SK-II.
The momentum of charged pions is more difficult to
reconstruct using
only the total detected charge because charged pions suffer
from interactions with nucleons in the water.
The momentum resolution for charged pions was improved by using the
Cherenkov opening angle as well as the total charge.
The momentum resolution for charged pions in the \eqepim\ MC
was estimated to be 9.2\% (9.7\%) in SK-I (SK-II).
Figure~\ref{fig:ppi_reconst} shows the charged pion momentum distribution in
\eqepim, including the smearing of the true pion momentum due to the Fermi
motion of a decaying nucleon.

\subsection{Calibration}\label{sec:calibration}

The characterization of the Super-Kamiokande detector has been performed in
a variety of calibrations, which are described in detail in~\cite{Fukuda:2002uc, Ashie:2005ik}.
One of the most important calibrations for nucleon decay searches is
the determination of the energy scale because we distinguish nucleon decay
events from atmospheric neutrino events using their total invariant mass
and momentum.

The absolute 
momentum 
scale was checked by the Michel electron momentum
spectrum from stopping muons, the invariant mass of $\pi^0$s, 
and the Cherenkov opening angles and track lengths of stopping muons.
The calibrated 
momentum 
ranges from a few tens of MeV/$c$ to about 10 GeV/$c$.
The uncertainty of the absolute
momentum 
scale was estimated to be less
than 0.74\% (1.60\%) for SK-I (SK-II).
The time variation of the
momentum 
scale was monitored by stopping
muon and Michel electron events and estimated to be 0.83\% (0.53\%)
in RMS for SK-I (SK-II).

The detector non-uniformity of the energy scale
is also important for the systematic error
of the total momentum reconstruction.
It was checked by the Michel electron momentum to be uniform within
$\pm$0.6\% for both SK-I and SK-II.

\section{Nucleon Decay Analysis}

  \subsection{Event Selection}
  
  \subsubsection{Summary of Selection Criteria}\label{sec:summary_criteria}
   The data used in this nucleon decay search are the 
   FC data from the SK-I and SK-II periods.
   We have to extract only nucleon decay signals from the FC data which
   are dominated by atmospheric neutrino events.
   Using the nucleon decay simulation and the atmospheric neutrino
   simulation, optimal criteria were determined.
   The atmospheric neutrino MC was normalized
   with the observed data using the number of single-ring $e$-like
   events, which are assumed to have a negligible neutrino oscillation
   effect.
   In order to study many nucleon decay modes systematically, event
   selection criteria were chosen to be as simple and common as
   possible. The primary for determining the criteria was
   to obtain a low enough background level.

   The following reconstructed information
   was used in the event selection criteria:
   \begin{enumerate}
    \item{the number of Cherenkov rings,}
    \item{the particle type of each ring,}
    \item{the meson invariant mass (if it is possible to be reconstructed),}
    \item{the number of Michel electrons,}
    \item{the total invariant mass and the total momentum (if it is
	 possible to be reconstructed).}
   \end{enumerate}
   The detailed numbers of event selection criteria for all the searches
   are summarized in Table~\ref{tab:summary_criteria}.

\begin{figure}[htbp]
    \begin{center}
     \includegraphics[width=\linewidth]{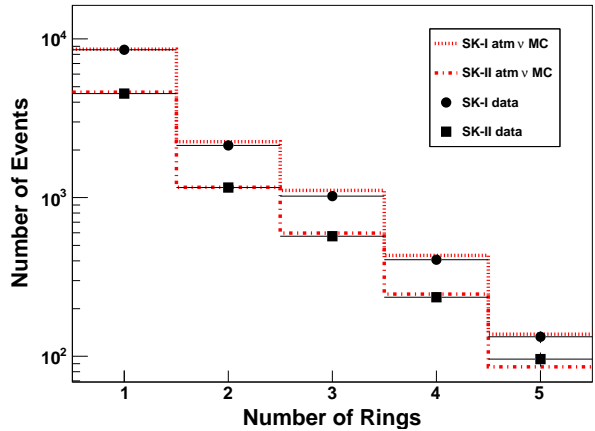}
     \caption{
     Number of rings distribution for atmospheric neutrino MC
     and data. 
     The number of MC events are normalized to
     the observed data by the
     number of single ring $e$-like events (also in other figures
     unless otherwise noted).
     }
     \label{fig:nring_mcdata}
    \end{center}
   \end{figure}

   All of the nucleon decay modes in this study are expected to have
   multiple Cherenkov rings.
   The criterion and efficiency of the number of rings cut for each mode are
   shown in Table~\ref{tab:nring}.
   More than half of the atmospheric neutrino MC events are single-ring events, as shown
   in Fig.~\ref{fig:nring_mcdata}, and they were rejected by the cut on number of rings.

   For the particle identification, two types of
   algorithm were used in this analysis, as described in Section~\ref{sec:rec_rec}.
   The PID using both a Cherenkov ring pattern and an opening angle was
   used for the modes in which we search for low momentum muons ($p_\mu<\sim 300$ MeV/$c$)
   or charged pions.

   The number of Michel electrons was required to be consistent with a
   nucleon decay signal. With this requirement, the background events can be effectively
   reduced, while the loss of the signal detection efficiency is negligible
   for the modes in which no Michel electrons are expected.
   
\begin{figure*}[htbp]
    \begin{center}
     \includegraphics[width=\linewidth]{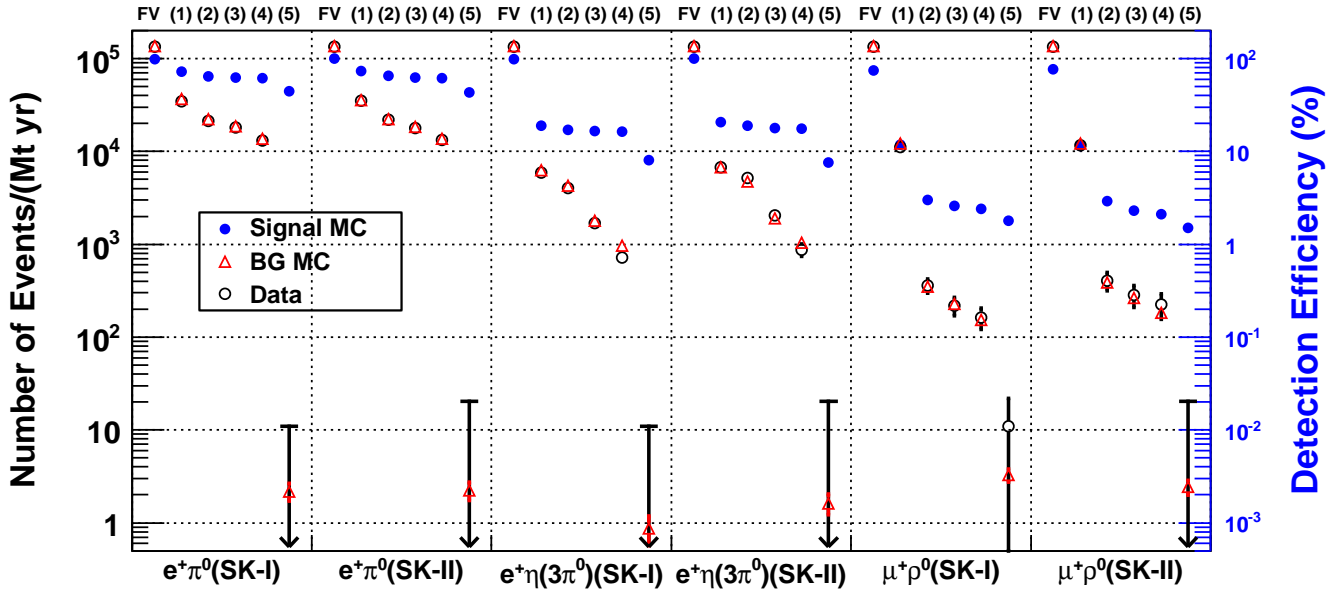}
     \caption{Number of events and detection efficiency at each
     event selection step: (FV) fiducial volume, (1) number of
     Cherenkov rings, (2) PID, (3) meson invariant mass, (4) number of
     Michel electrons and (5) total invariant mass and total momentum.
     The plot shows three nucleon decay mode searches:
     the mode with highest efficiency (\eqepi0), the mode with the greatest number of rings
     (\eqeeta\ ($\eta \rightarrow 3\pi^0$)) and the mode using PID
     with both Cherenkov ring pattern and opening angle (\eqmurho). 
     The number of observed events and
     the estimated background rates agree with each other.
     Also, the searches in SK-I and SK-II are compared.
     }
     \label{fig:nevent_step}
    \end{center}
   \end{figure*}

   The total momentum $P_\textrm{tot}$, the total energy $E_\textrm{tot}$,
   and the total invariant mass
   $M_\textrm{tot}$ are defined as:
   \begin{eqnarray}
    P_\textrm{tot} &=& \left|\sum_i^{\textrm{all}} \vec{p_i}\right|,\\
    E_\textrm{tot} &=& \sum_i^{\textrm{all}}
     \sqrt{\left|\vec{p_i}\right|^2+m_i^2},\\
    M_\textrm{tot} &=& \sqrt{E_\textrm{tot}^2- P_\textrm{tot}^2},
   \end{eqnarray}
   where $\vec{p_i}$ is the momentum of each Cherenkov ring, and $m_i$ is
   the mass of a particle ($\gamma$, $e^\pm$, $\mu^\pm$, $\pi^\pm$).
   The meson mass is reconstructed in a similar way by summing up the
   momenta and energies of secondary particles from the meson decay.
   Although our particle identification algorithm can only classify a
   Cherenkov
   ring into a shower-type ring ($e^\pm$ or $\gamma$) or a non shower-type ring
   ($\mu^\pm$ or $\pi^\pm$),
   the invariant mass reconstruction in some modes requires the ability to distinguish
   $e^\pm$ from $\gamma$ or $\mu^\pm$ from $\pi^\pm$.
   For example, for the $\pi^0$ invariant mass reconstruction in the
   \eqepi0\ mode search, we should identify each Cherenkov ring as originating from a positron or a
   $\gamma$-ray.
   However, that cannot be done in a large water Cherenkov detector.
   In these cases, the invariant mass was calculated for all
   possible combinations of particle type assignment.
   Then, the best combination in which the reconstructed mass is the
   closest to the expected nucleon (or meson) mass, is selected.
   All of the studied nucleon decay modes are two-body decays with
   back-to-back kinematics, and have isotropic event signatures,
   which is the most significant difference from typical atmospheric
   neutrino events.
   Therefore, the event selection by total momentum and total
   invariant mass is a powerful tool to eliminate the atmospheric
   neutrino background as can be found in Fig.~\ref{fig:nevent_step}.
   The total momentum cut threshold is set to be 250~MeV/$c$ 
   for most searches considering the Fermi motion of
   bound nucleons.
   To keep the background rates low enough (below 0.5 events in the total
   exposure),
   a tighter total momentum cut with a threshold of 150~(or
   200)~MeV/$c$
   is applied in some relatively high-background mode searches.  
   
   \begin{table*}[htbp]
    \begin{center}
     \begin{tabular}{l|ccccccc}
      \hline \hline
      & (1)    & (2)   & (3)                  & (4)
      & \multicolumn{2}{c}{(5)} &\\
      Modes & Ring & PID & $M_\textrm{meson}$ & decay-$e$ 
		      &$M_\textrm{tot}$ &
			      $P_\textrm{tot}$ & Note\\
      \hline
      \eqepi0 &2, 3&  SS(S)& 85$\sim$185 ($\pi^0$)& 0 
		      & 800$\sim$1050 & $<$250\\
      \eqmupi0 &2, 3& NS(S)& 85$\sim$185 ($\pi^0$)& 1 
		      & 800$\sim$1050 & $<$250\\
      \eqeeta\ ($2\gamma$) &3&  SSS& 480$\sim$620 ($\eta$)&0
		      & 800$\sim$1050 & $<$250\\
      \eqmueta\ ($2\gamma$) &3&  NSS*& 480$\sim$620 ($\eta$)&1
		      & 800$\sim$1050 & $<$250\\
      \eqeeta\ ($3\pi^0$) &4, 5&  SSSS(S)& 400$\sim$700 ($\eta$)&0
		      & 800$\sim$1050 & $<$150\\
      \eqmueta\ ($3\pi^0$) &4, 5& NSSS(S)*& 400$\sim$700 ($\eta$)&1
		      & 800$\sim$1050 & $<$250\\
      \eqerho &3&  SNN*& 600$\sim$900 ($\rho^0$)& 0, 1
		      & 800$\sim$1050 & $<$150\\
      \eqmurho &3&  NNN*& 600$\sim$900 ($\rho^0$)& 1, 2
		      & 800$\sim$1050 & $<$250\\
      \eqeomega\ ($\pi^0\gamma$) &3, 4&  SSS(S)& 650$\sim$900 ($\omega$)&0
		      & 800$\sim$1050 & $<$150\\
      \eqmuomega\ ($\pi^0\gamma$) &2, 3&  SS(S)& 650$\sim$900 ($\omega$)&1
		      & - & $<$200\\
      \eqeomega\ ($\pi^+\pi^-\pi^0$) &4&  SSSN*& 85$\sim$185 ($\pi^0$)&0, 1
		      & 600$\sim$800 & $<$200 & $P_{e^+}$\\
      \eqmuomega\ ($\pi^+\pi^-\pi^0$) &3&  SSN*& 85$\sim$185 ($\pi^0$)&2
		      & 450$\sim$700 & $<$200\\
      \eqepim &2&  SN& - & 0 
		      & 800$\sim$1050 & $<$250\\
      \eqmupim &2&  NN& - & 1
		      & 800$\sim$1050 & $<$250\\
      \eqerhom &4&  SSSN*& 600$\sim$900 ($\rho^-$)& 0
		      & 800$\sim$1050 & $<$250 & $M_{\pi^0}$\\
      \eqmurhom &3&  SSN*& 600$\sim$900 ($\rho^-$)& 1
		      & - & $<$150 & $M_{\pi^0}$\\
      \hline \hline
     \end{tabular}
     \caption{Summary of the selection criteria.
     The numbers in parentheses correspond to the numbers of selection
     criteria described in Section~\ref{sec:summary_criteria}.
     For example, for the \eqmueta\ ($3\pi^0$) mode, 
     this table means that the selection criteria requires that (1) the
     number of Cherenkov rings is 4 or 5, (2) one
     ring is non shower-type and all other rings are shower-type,
     (3) reconstructed $\eta$ meson invariant mass is in between 400 and 700
     MeV/$c^2$, (4) the number of Michel electron is 1, and (5) reconstructed total
     invariant mass is in between 800 and 1050 MeV/$c^2$ and
     reconstructed total momentum is less than 250 MeV/$c$.
     S and N in the PID column stand for shower-type rings and non shower-type
     rings, respectively.
     Asterisks in PID indicate that a PID using both Cherenkov ring pattern and
     opening angle was used for identifying low momentum muons
     ($p_\mu<\sim 300$ MeV/$c$) or
     charged pions.
     All the invariant mass and momenta are written in units of MeV/$c^2$ and
     MeV/$c$, respectively.
     For the \eqeomega\ (3$\pi$) mode, a positron momentum cut is
     applied
     ($100<P_{e^+}<200$).
     For the $n\rightarrow l^+ \rho^-$ mode,
     a $\pi^0$ invariant mass cut is also applied.
     }
     \label{tab:summary_criteria}
    \end{center}
   \end{table*}
   
   \begin{table}[htbp]
    \begin{center}
     \begin{tabular}{l||c|rr|rr}
      \hline \hline
      Mode & $N_\textrm{ring}$ & \multicolumn{4}{c}{Efficiency \%}\\
      & & \multicolumn{2}{c|}{SK-I} & \multicolumn{2}{c}{SK-II}\\
      \hline
      \eqepi0   & 2 or 3& 73 & (98) & 74 & (98) \\
      \eqmupi0  & 2 or 3& 74 & (98) & 74 & (98) \\
      \eqeeta\ ($2\gamma$) & 3 & 44 &(92) & 45 & (89) \\
      \eqmueta\ ($2\gamma$) & 3 & 44 &(91) & 44 & (89) \\
      \eqeeta\ ($3\pi^0$) & 4 or 5 & 19 & (76) & 21 & (76) \\
      \eqmueta\ ($3\pi^0$) & 4 or 5 & 22 & (82) & 21 & (81) \\
      \eqerho & 3 & 26 & (50) & 26 & (47) \\
      \eqmurho & 3 & 12 & (26) & 11 & (25) \\
      \eqeomega\ ($\pi^0\gamma$) & 3 or 4 & 58 & (93) & 59 & (89)\\
      \eqmuomega\ ($\pi^0\gamma$) & 2 or 3 & 61 & (97) & 61 & (96)\\
      \eqeomega\ (3$\pi$) & 4 & 19 & (27) & 19 & (26)\\
      \eqmuomega\ (3$\pi$) & 3 & 27 & (41) & 26 & (39)\\
      \eqepim & 2 & 49 && 49 &   \\
      \eqmupim & 2 & 49 && 49 & \\
      \eqerhom & 4 & 9 & & 9 &   \\
      \eqmurhom & 3 & 17 && 17 & \\
      \hline \hline
     \end{tabular}
     \caption{Efficiency of cut on number of rings for nucleon decay MC for
     SK-I and SK-II. 
     Numbers in parentheses are efficiencies for free proton decay events.
     These efficiencies include the efficiency for the
     FC data reduction and the fiducial volume cut.
     }
     \label{tab:nring}
    \end{center}
   \end{table}

   \subsubsection{$p\rightarrow l^+ \pi^0$ Mode Search}

\begin{figure}[htbp]
    \begin{center}
     \includegraphics[width=.9\linewidth]{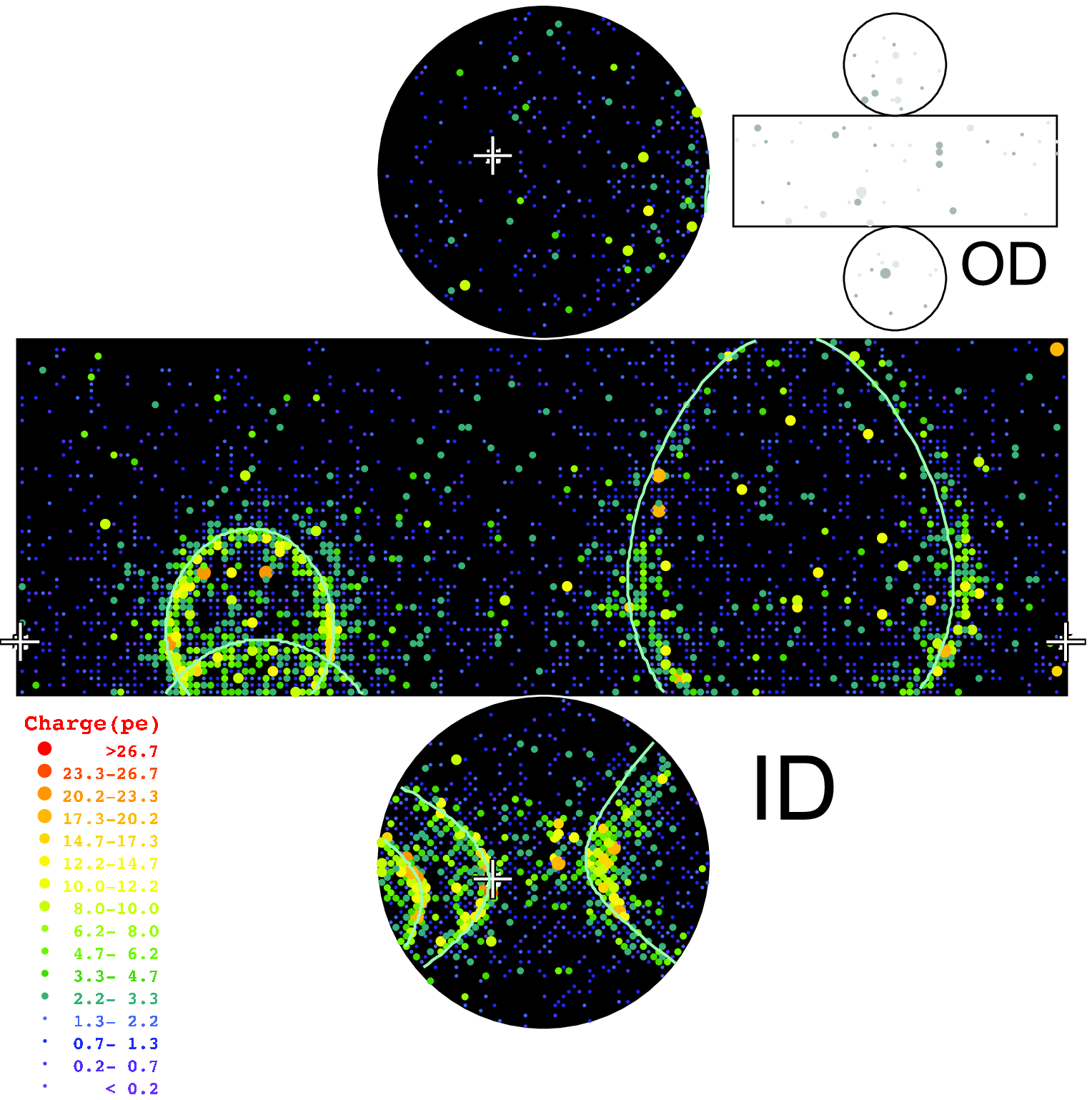}
     \caption{
     A typical proton decay MC event of \eqepi0 in SK-I.
     One Cherenkov ring from a positron and two Cherenkov rings from the two
     $\gamma$-rays from $\pi^0$ decay can be observed on the right and
     the left of the figure, respectively.
     The size of circles indicate the amount of detected charge.
     The crosses on the plot show a reconstructed vertex position
     horizontally and vertically projected on the detector wall.
     Solid lines show the reconstructed rings. The reconstruction algorithms
     correctly find all rings and identify their particle for this event.
     }
     \label{fig:display_epi0}
    \end{center}
   \end{figure}

   In the proton decay of \eqepi0 (\eqmupi0), a neutral pion and a
   positron (muon) are back-to-back and have the same momentum of 459
   (453)~MeV/$c$ in the proton rest frame.
   The neutral pion immediately decays into two $\gamma$-rays.
   Figure~\ref{fig:display_epi0} shows the event signature of a typical
   proton decay MC event for \eqepi0. One shower-type ring from the
   positron and two shower-type rings from the $\pi^0$ are clearly seen
   in this figure.
   
   The two $\gamma$-rays from the $\pi^0$ decay are back-to-back in the
   $\pi^0$ rest frame.
   They are Lorentz-boosted by the $\pi^0$ momentum of $\sim$ 450~MeV/$c$, so
   the momentum of one of the two $\gamma$-rays in the laboratory frame can be very
   low depending on the
   direction of the Lorentz boost.
   In such a case, 
   it sometimes happens that only one ring from $\pi^0$ decay is identified.
   For free proton decays of \eqepi0, 
   the fraction of two-ring and three-ring events identified was 39\% and 60\%,
   respectively.
   
\begin{figure}[htbp]
   \begin{center}
    \includegraphics[width=\linewidth]{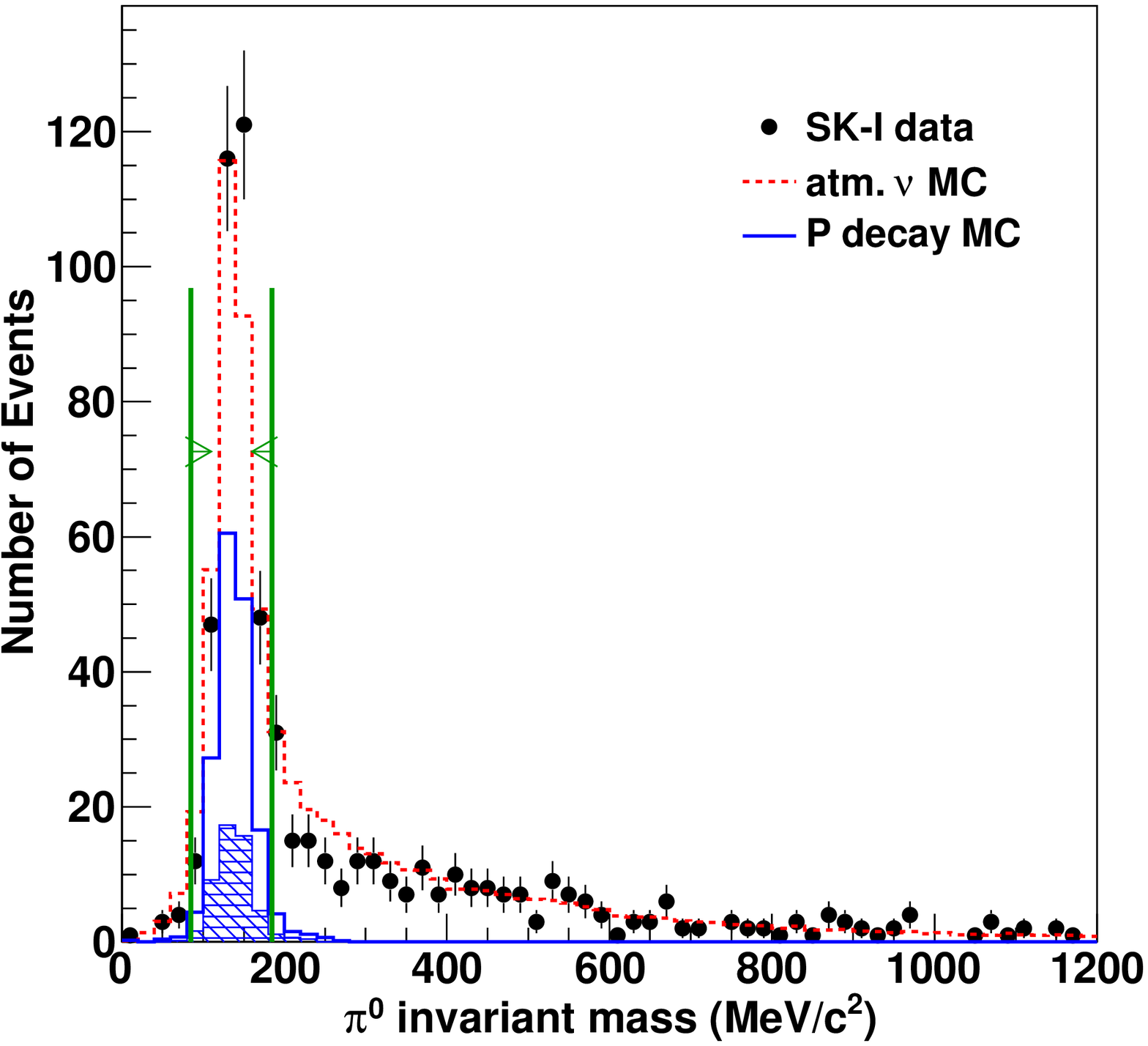}
    \caption{The invariant mass of $\pi^0$ distributions for SK-I proton decay
    MC (\eqepi0), atmospheric neutrino MC, and SK-I data.
    The shaded histogram is the distribution of free
    proton decay events.
    The bars and arrows indicate the $\pi^0$ invariant mass cut threshold.
    }
    \label{fig:epi0_pi0mass}
   \end{center}
   \end{figure}

   If all three rings were found,
   the $\pi^0$ invariant mass, which is reconstructed by two of the three
   rings, was required to be between 85 and 185~MeV/$c$.
   The reconstructed $\pi^0$ invariant mass distribution for the proton
   decay MC is shown in
   Fig.~\ref{fig:epi0_pi0mass}.
   For free proton decay events of \eqepi0,
   the mean of the reconstructed $\pi^0$ invariant mass is
   137~MeV/$c^2$, and its resolution is 20~MeV/$c^2$.

\begin{figure}[htbp]
    \begin{center}
     \includegraphics[width=\linewidth]{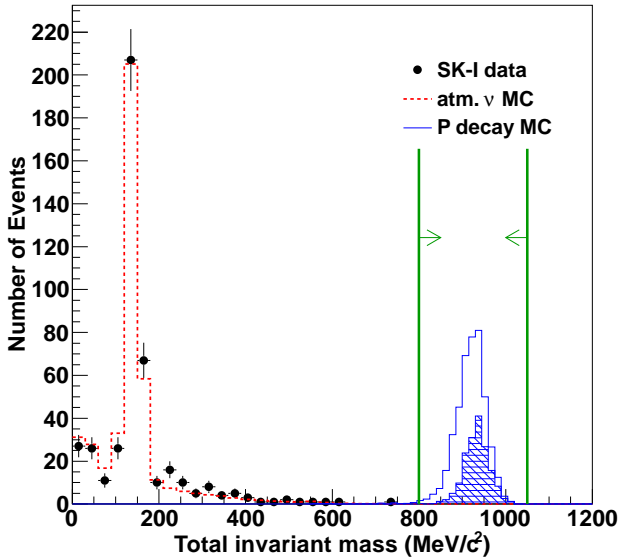}
     \caption{The total invariant mass distributions for proton
     decay MC (\eqepi0), atmospheric neutrino MC, and observed data in SK-I.
     The shaded histogram is the distribution of free
     proton decay events.
     The bars and arrows indicate the total invariant mass cut threshold.
     Only events which satisfy all other selection criteria except this
     are plotted.
     The distribution of atmospheric neutrino MC has the $\pi^0$
     invariant mass peak, which is consistent with the observed data.
     }
     \label{fig:epi0_mtot}
    \end{center}
   \end{figure}

\begin{figure*}[htbp]
    \begin{center}
     \includegraphics[width=\linewidth]{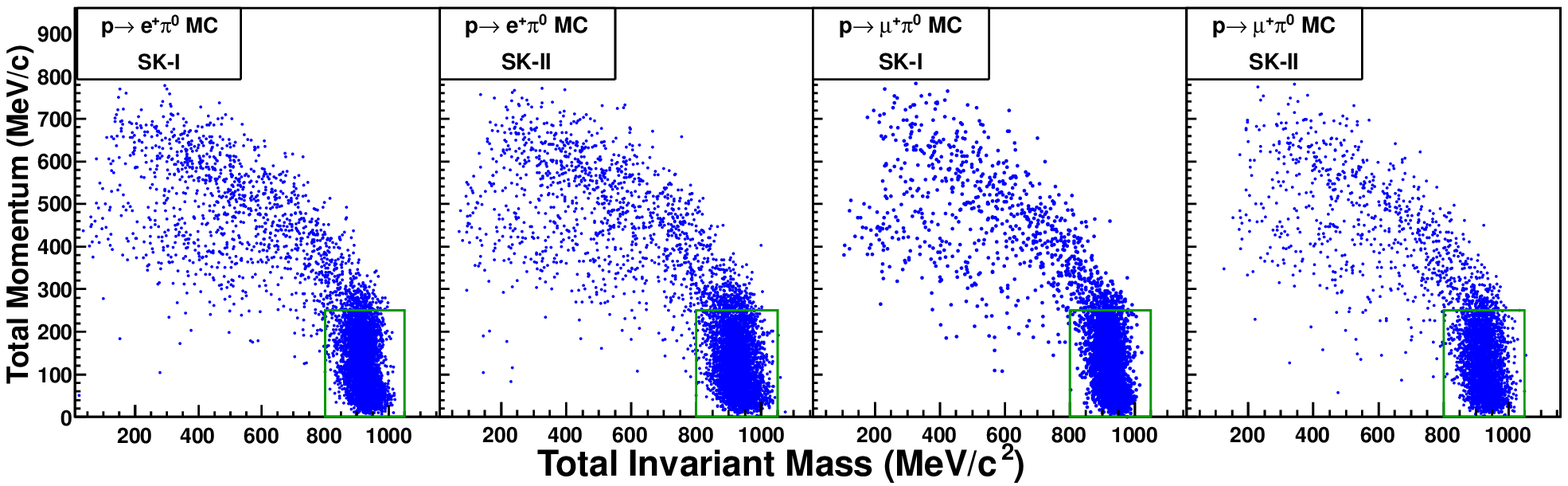}
     \includegraphics[width=\linewidth]{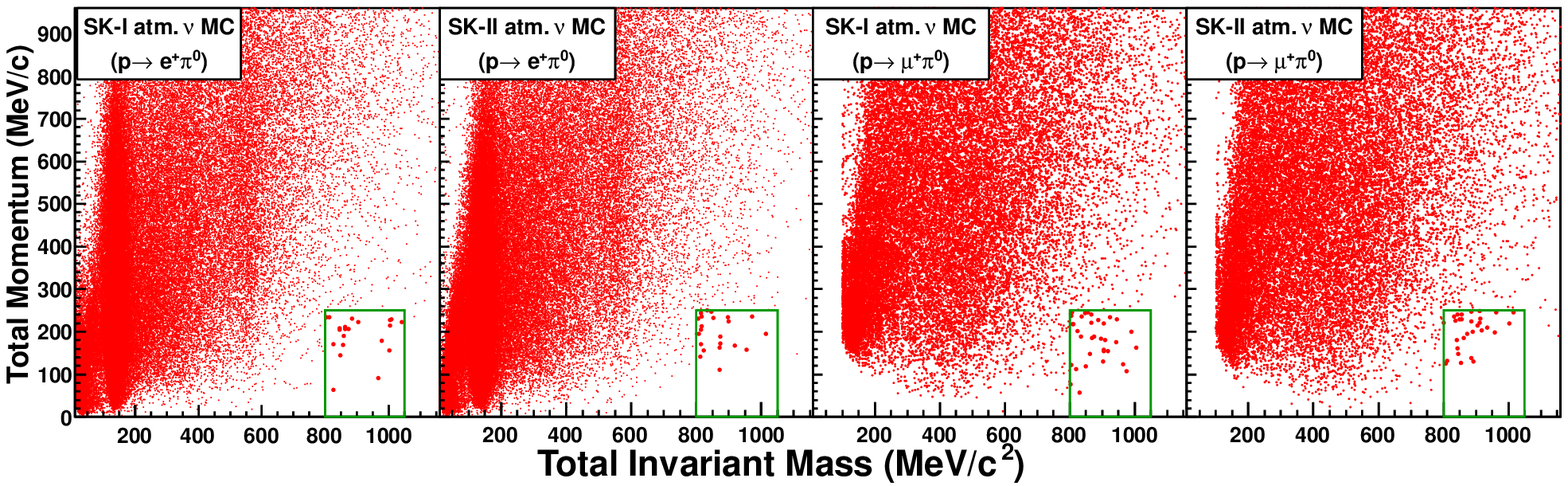}
     \includegraphics[width=\linewidth]{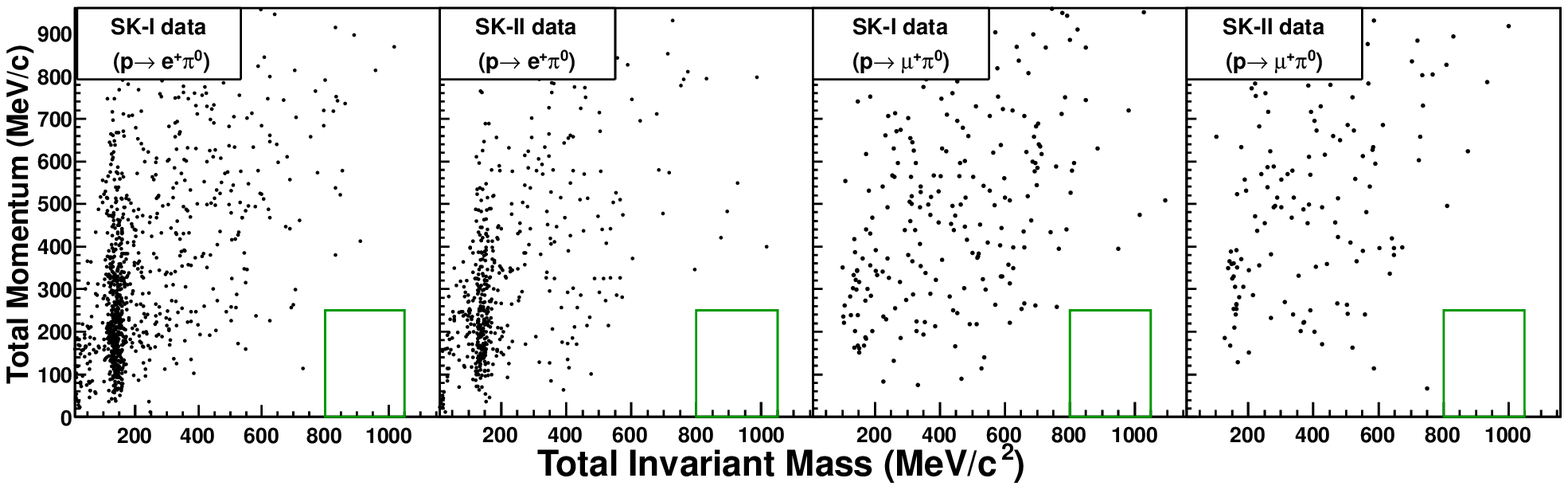}
     \caption{Total momentum versus total invariant mass distributions,
     from left to right: SK-I \eqepi0; SK-II \eqepi0; SK-I \eqmupi0; and
     SK-II \eqmupi0,
     from top to bottom: the proton decay MC; the atmospheric neutrino MC
     (11.25 megaton$\cdot$years for each SK-I and SK-II); and the observed data (91.7
     kiloton$\cdot$years in SK-I and 49.2 kiloton$\cdot$years in SK-II).
     These events satisfy the event selection criteria
     except for the selection by the total momentum and total mass.
     The boxes in figures indicate the total momentum and mass
     criteria. The points in the boxes of the atmospheric neutrino MC
     are shown in a larger size to show the distributions in the signal box.
     No candidates were found for either \eqepi0\ or \eqmupi0\ in the data.
     }
     \label{fig:epi0_mptot}
    \end{center}
   \end{figure*}

   The final criterion of total invariant mass and momentum selects the
   events which are consistent with
   the momentum and mass of a parent proton.
   Figure~\ref{fig:epi0_mtot} shows the
   distribution of the total invariant mass.
   For free proton decay events,
   the total invariant mass was well reconstructed.
   The resolution of the total invariant mass distribution
   is 28.2 (36.2)~MeV/$c^2$ in SK-I (SK-II).
   The resolution of the total momentum is 29.8 (32.5)~MeV/$c$.
   The total momentum and invariant mass cut was wide enough for free
   proton decay events.
   The total momentum and total invariant mass distributions of the
   proton decay MC and the atmospheric neutrino MC are clearly different
   from each other, as shown in  Fig.~\ref{fig:epi0_mptot}.
   This criterion reduces the atmospheric neutrino background by
   more than three orders of magnitude, as shown in Fig.~\ref{fig:nevent_step}.

   The detection efficiency for \eqepi0\ was estimated to be 44.6
   (43.5)\% in SK-I (SK-II). 
   The inefficiency is mainly due to nuclear interaction effects of
   pions in $^{16}$O.
   For free proton decay events,
   high efficiencies of 87\% (86\%) were achieved for \eqepi0.
   As for the \eqmupi0 mode search, differences in the selection
   criteria from the \eqepi0\ mode 
   are the requirement of one non shower-type ring and 
   one Michel electron.
   The detection efficiency for a Michel electron from the decay of $\mu$
   was approximately 80\%.
   This is the reason for the detection efficiency difference between
   \eqepi0\ and \eqmupi0.
   The detection efficiency for the \eqmupi0\ mode was estimated to be
   35.5\% (34.7\%) in
   SK-I (SK-II).

   The background events for \eqepi0\ and \eqmupi0\ were estimated
   to be 0.31 and 0.34 events in total for SK-I and SK-II, respectively.
   The background rates for \eqepi0\ in SK-I and SK-II are 2.1 and 2.2
   events/(megaton$\cdot$years), respectively.
   Therefore, low-background observations with high efficiency
   were achieved for these modes in the Super-Kamiokande
   detector,
   both in the SK-I and SK-II periods.
   The background for \eqepi0\ was also
   estimated from
   the experimental data of the K2K $\nu$ beam and 1-kiloton water Cherenkov
   detector~\cite{Mine:2008rt}.   
   The number of expected background events
   estimated from the K2K data is
   $0.23^{+0.06}_{-0.05}(\textrm{stat.})^{+0.06}_{-0.07}(\textrm{sys.})$
   events for the exposure of SK-I plus SK-II, which is consistent with the
   estimate from our MC.
   The background estimate was also compared with a different neutrino
   interaction MC, NUANCE~\cite{Casper:2002sd}.
   The number of background events estimated by NUANCE was $0.27 \pm
   0.10$ events (in SK-I+SK-II), which is
   also consistent with our primary estimate.
   These results are summarized in Table~\ref{tab:summaryall} with
   the results of all the other studied modes.

  \subsubsection{$p\rightarrow l^+ \eta$ Mode Search}
  The $\eta$ meson has three dominant decay modes;
  we search for two of the three modes:
  $\eta\rightarrow 2\gamma$ ($Br$=39\%) and $\eta\rightarrow 3\pi^0$ ($Br$=33\%).

  The mass of $\eta$ meson is 548~MeV/$c^2$.
  Because its mass is larger than the $\pi^0$ mass,
  the generated charged lepton and $\eta$ meson have smaller momentum,
  $\sim$300~MeV$/c$ in the proton rest frame, 
  compared with the momentum
  of generated particles in the $p\rightarrow l^+ \pi^0$ modes.
  For the \eqmueta\ mode, 
  PID with Cherenkov ring pattern and opening angle
  was used to improve the PID for a low momentum
  $\mu$ ring.
  This is common for
  both meson decay modes of \eqmueta.

\begin{figure*}[htbp]
    \begin{center}
     \includegraphics[width=\linewidth]{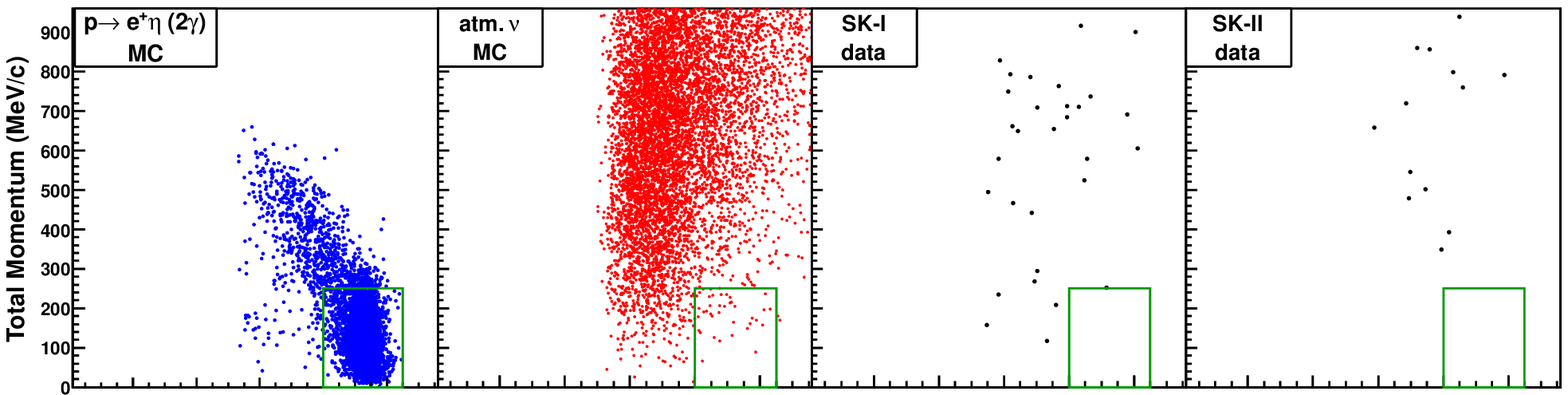}
     \includegraphics[width=\linewidth]{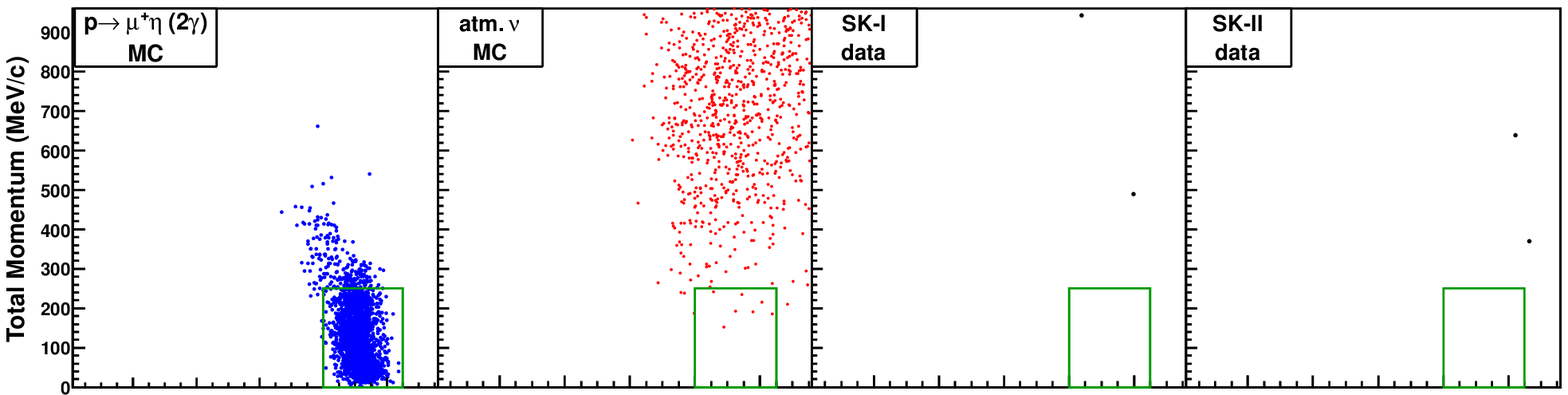}
     \includegraphics[width=\linewidth]{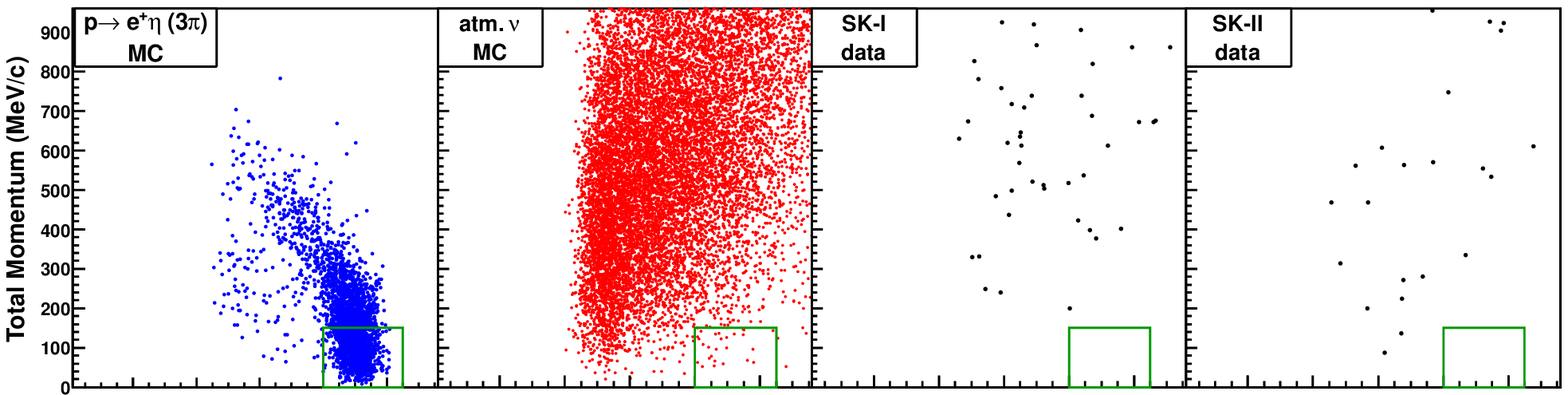}
     \includegraphics[width=\linewidth]{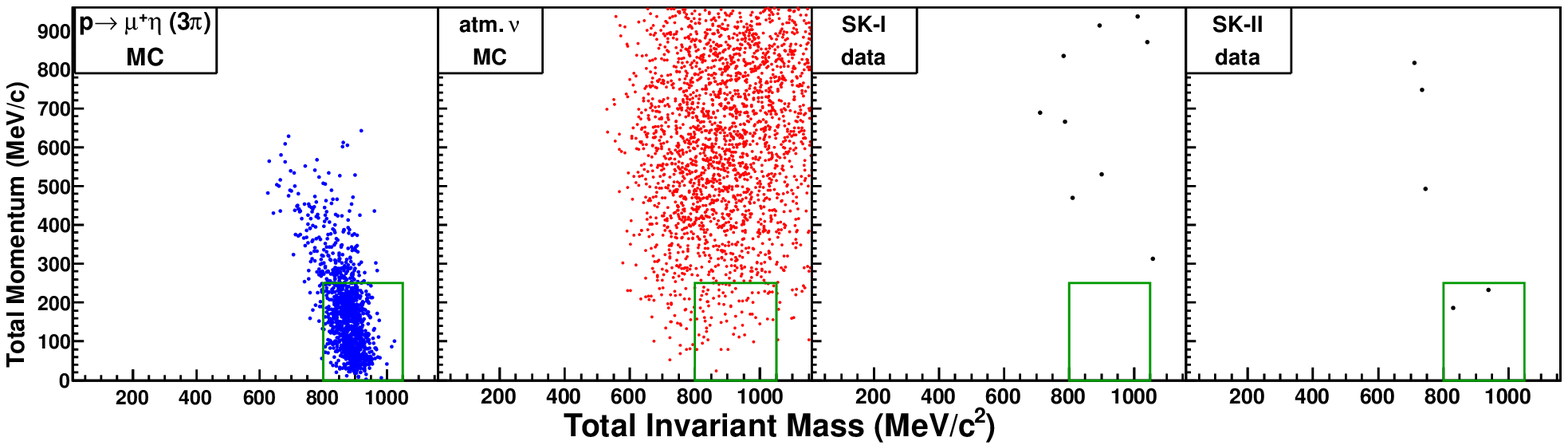}
     \caption{Total momentum versus total invariant mass distributions
     of proton decay MC (SK-I+II combined), the atmospheric neutrino MC
     (SK-I+II combined; 11.25 megaton$\cdot$years for each SK-I and
     SK-II) and SK-I (91.7 kiloton$\cdot$years) and SK-II
     (49.2 kiloton$\cdot$years) data, 
     from top to bottom: \eqeeta\ ($2\gamma$), \eqmueta\
     ($2\gamma$), \eqeeta\ ($3\pi^0$) and \eqmueta\ ($3\pi^0$).
     The boxes in figures indicate the total momentum and mass criteria.
     These events satisfy the event selection criteria for each mode
     except for the selection on the total momentum and total mass.
     Two candidates for \eqmueta\ ($3\pi^0$) were found in the SK-II data.
     }
     \label{fig:eta_mptot_fcdst}
    \end{center}
   \end{figure*}

   Figure~\ref{fig:eta_mptot_fcdst} shows the total momentum and total
   invariant mass distributions of the $p\rightarrow l^+\eta$ modes.
   Compared with the $p\rightarrow l^+ \pi^0$ modes
   many fewer events in both the atmospheric neutrino MC and observed
   data survive the selection criteria except for the total momentum and
   invariant mass cut.
   That is because 
   a significant number of atmospheric neutrino events
   with $\pi^0$ production
   can survive the selection criteria (except for the total momentum and
   invariant mass cut) for the $p\rightarrow l^+\pi^0$ modes
   while
   the number of atmospheric neutrino events
   with $\eta$ production are negligible.

   \paragraph{$p\rightarrow l^+ \eta \ (\eta \rightarrow 2\gamma)$ Mode}

   These modes have very similar event signatures with \eqepi0\ and \eqmupi0.
   However, the momentum of the two $\gamma$-rays in the $\eta$ meson
   rest frame
   is 274~MeV/$c$, which is much larger than that in \eqepi0\ and \eqmupi0.
   The opening angle between the two $\gamma$-rays from the
   $\eta$ meson decay is about 132 degrees in the laboratory frame.
   Therefore, 
   three Cherenkov rings are clearly visible and can easily be separated
   so that the number of Cherenkov rings is required to be three.
   The fraction of three-ring events in the free proton decay event
   is greater than 90\%.

\begin{figure*}[htbp]
    \begin{center}
     \begin{minipage}{.32\linewidth}
      \begin{center}
       \includegraphics[width=1.05\linewidth]{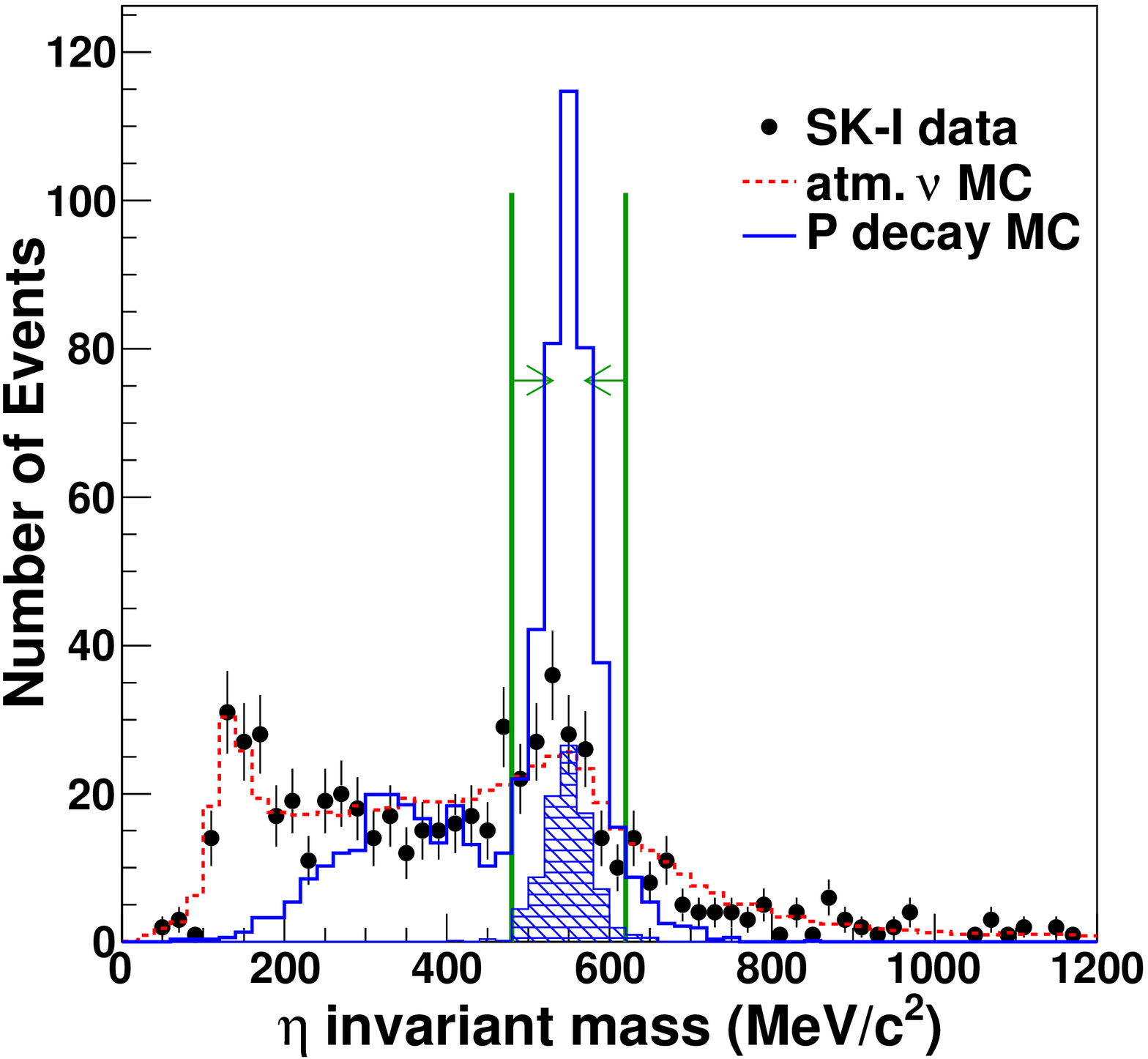}
      \end{center}
     \end{minipage}
     \begin{minipage}{.32\linewidth}
      \begin{center}
       \includegraphics[width=1.05\linewidth]{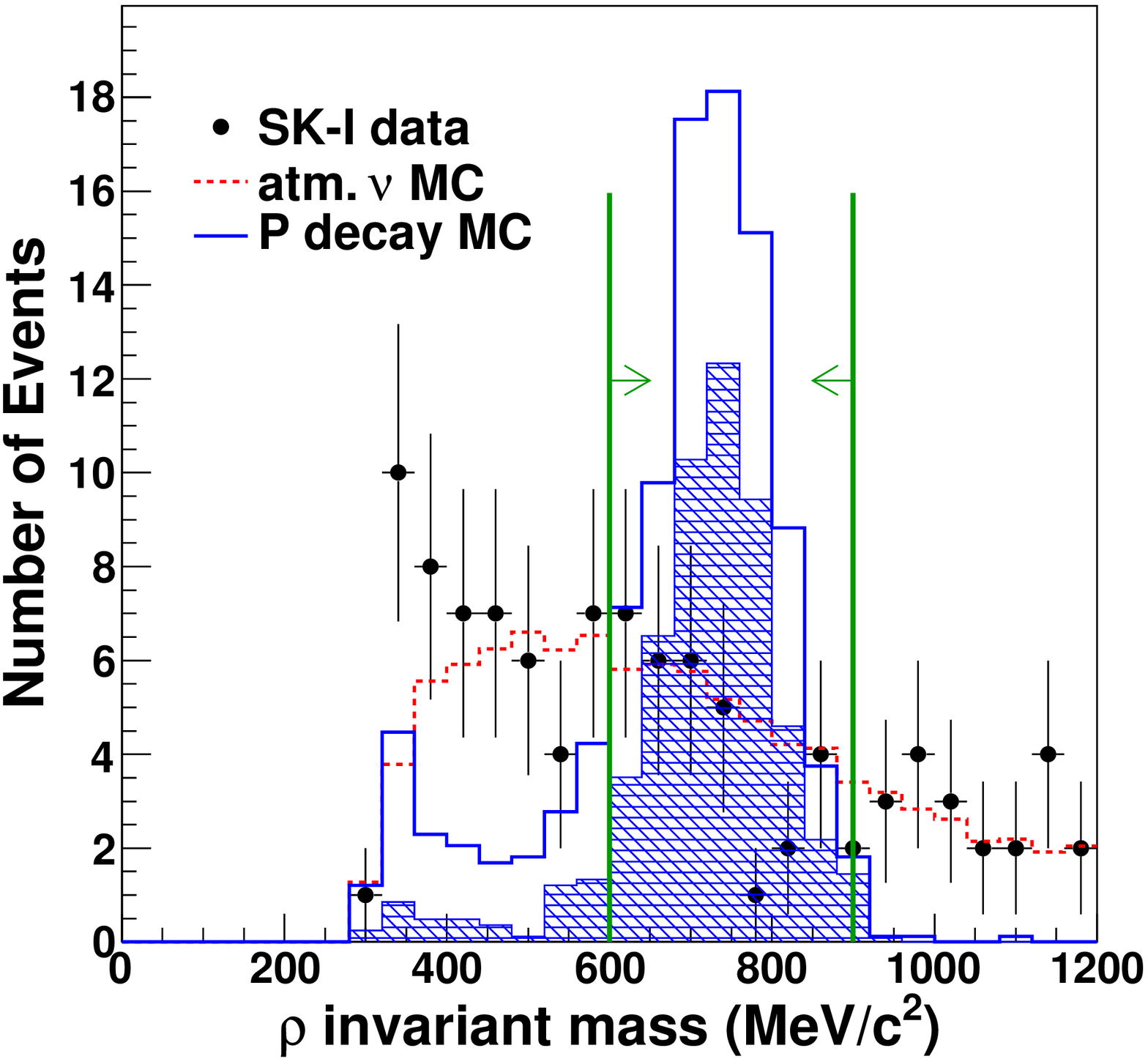}
      \end{center}
     \end{minipage}
     \begin{minipage}{.32\linewidth}
      \begin{center}
       \includegraphics[width=1.05\linewidth]{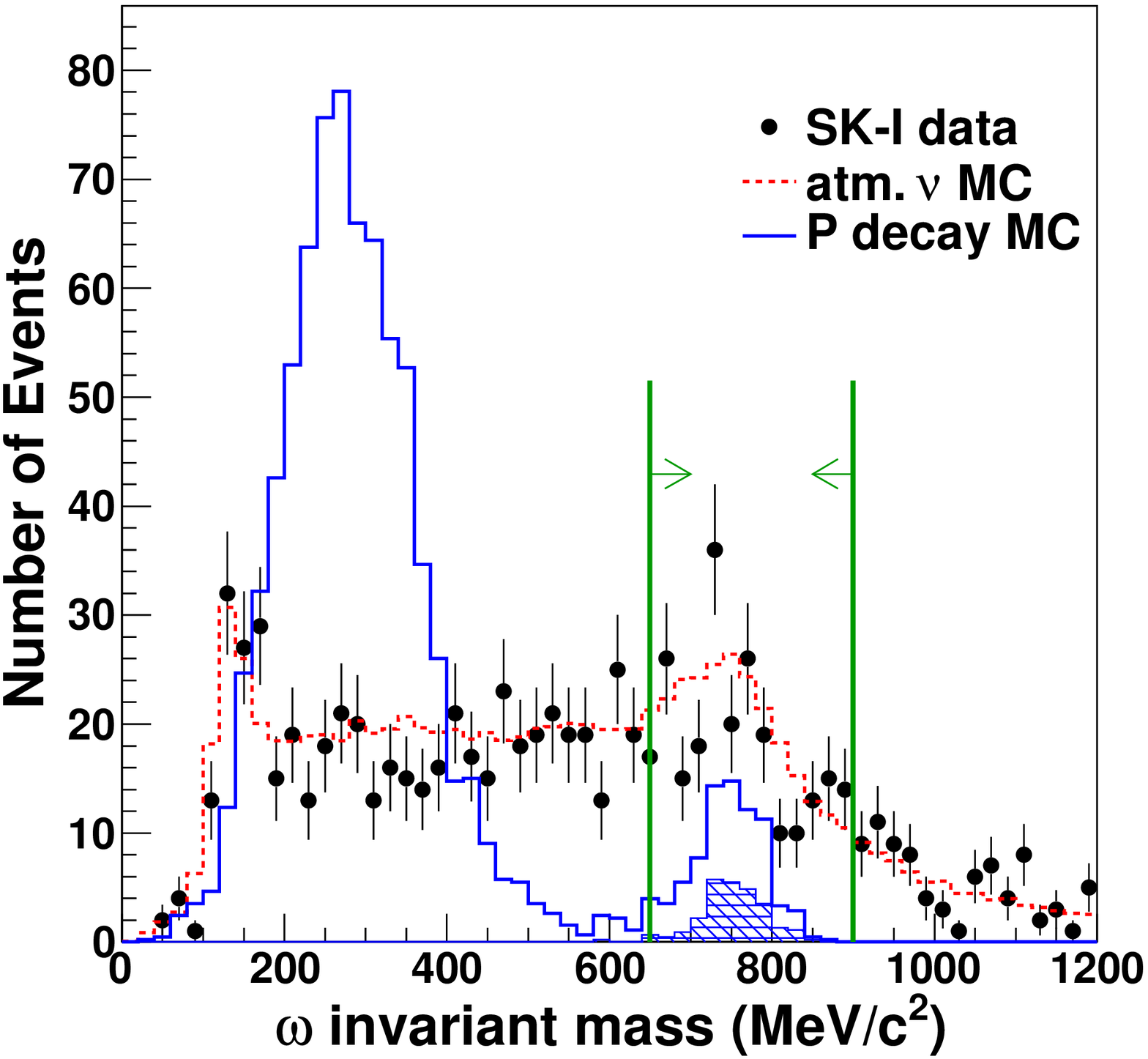}
       \end{center}
     \end{minipage}

     \begin{minipage}{.32\linewidth}
      \begin{center}
       \includegraphics[width=1.05\linewidth]{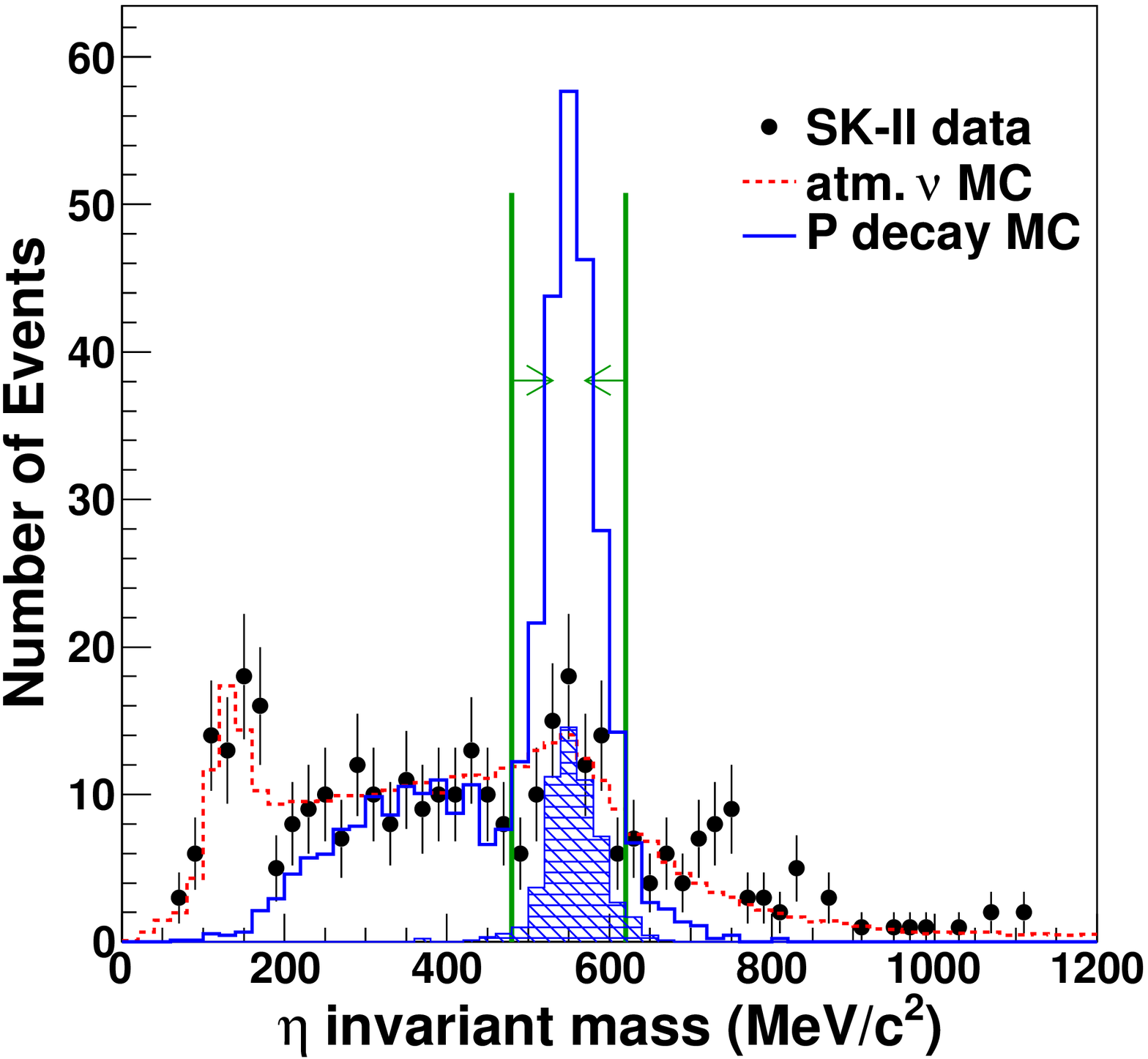}
      \end{center}
     \end{minipage}
     \begin{minipage}{.32\linewidth}
      \begin{center}
       \includegraphics[width=1.05\linewidth]{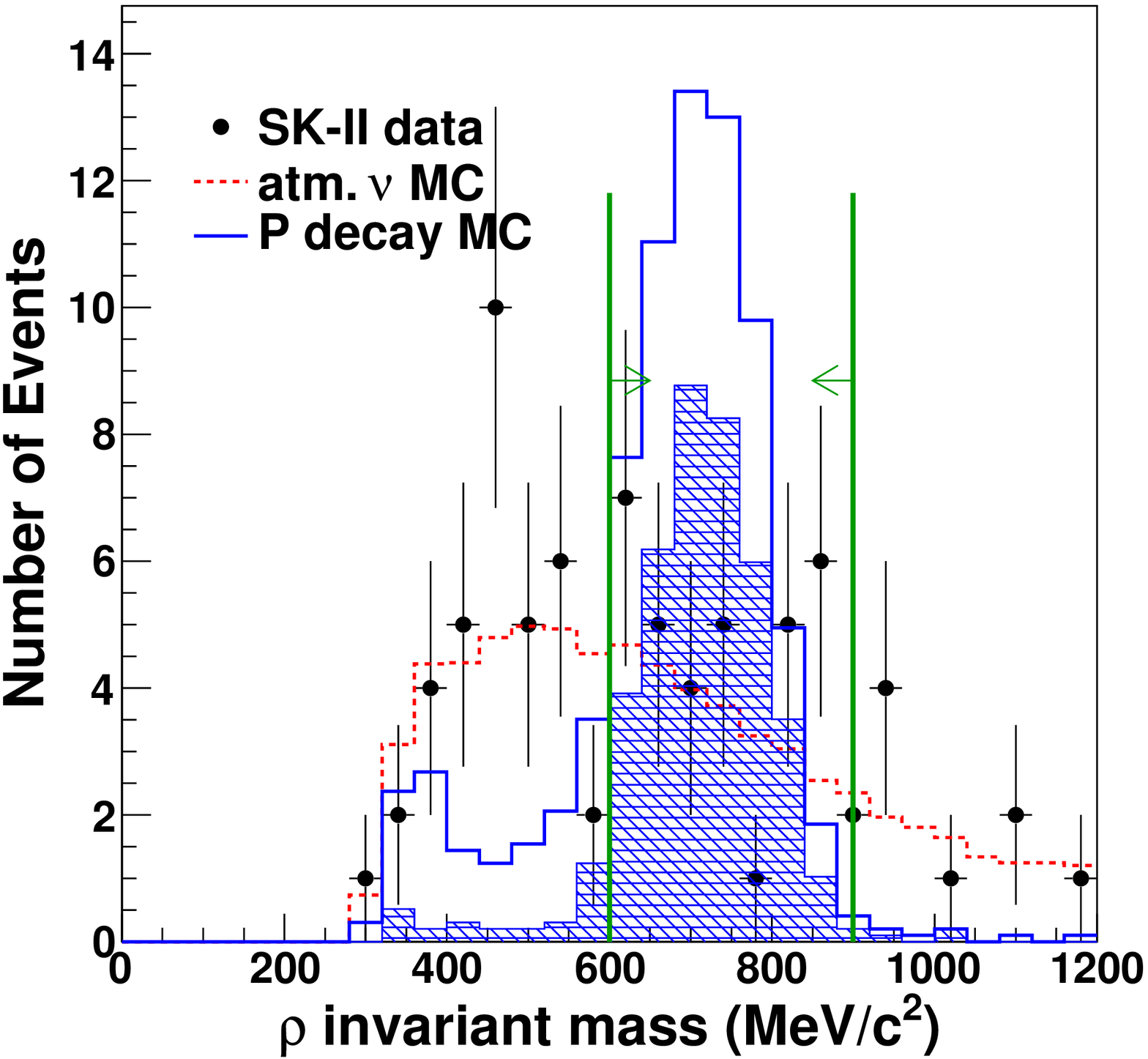}
      \end{center}
     \end{minipage}
     \begin{minipage}{.32\linewidth}
      \begin{center}
       \includegraphics[width=1.05\linewidth]{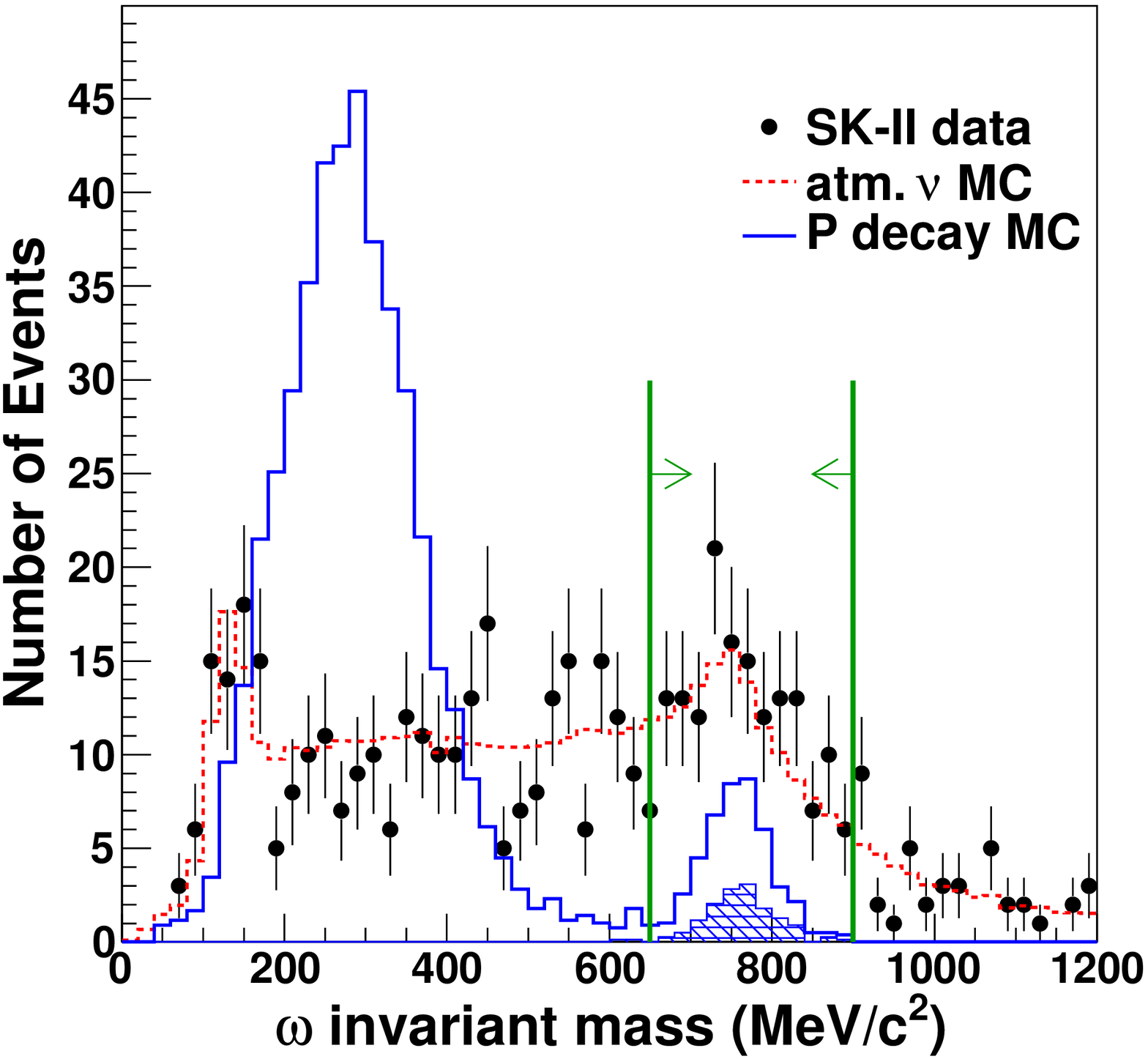}
       \end{center}
     \end{minipage}

     \caption{The meson invariant mass distributions
     for the proton decay MC (solid lines), the atmospheric neutrino
     MC (dashed lines) and the
     observed data (points) in SK-I (top) and SK-II (bottom), from left
     to right; \eqeeta\ mode (2$\gamma$),
     \eqerho\ mode and \eqeomega\ mode ($\pi^0\gamma$).
     Free proton decay events are indicated by shaded histograms.
     All decay branches of the mesons are filled in the histograms
     for the proton decay MCs. The other peaks outside of the selection
     window are events from the other decay branches.
     }
     \label{fig:mesonmass}
    \end{center}
   \end{figure*}

   Because only the three-ring events survive the number of rings cut,
   an $\eta$ invariant mass cut was applied to all of the surviving events.
   Figure~\ref{fig:mesonmass} (left panels) shows the reconstructed invariant mass
   of the $\eta$ meson for the proton decay MC in the \eqeeta\ mode.
   The $\eta$ invariant mass was well reconstructed.

   \paragraph{$p\rightarrow l^+ \eta \ (\eta \rightarrow 3\pi^0)$ Mode}
   In these proton decay modes,
   the $\eta$ meson decays into three $\pi^0$s,
   and the three $\pi^0$s immediately decay into six $\gamma$-rays.
   Therefore, one ring from a charged lepton and six rings from
   the $\gamma$-rays can be observed in principle.
   However, our Cherenkov ring counting algorithm is only capable of
   finding up to five rings.
   Consequently, the criterion of the number of rings was applied to
   select events with four or five
   rings.
   In SK-I free proton decay events 
   of the $\eta\rightarrow3\pi^0$ mode of \eqeeta\ (\eqmueta),
   77\% (84\%) of the events satisfy the criterion.

   The invariant mass of the $\eta$ meson was reconstructed using only three or
   four shower-type
   rings, though there should be 6 rings from the $3\pi^0$s.
   This resulted in a worse invariant mass resolution for the $\eta$
   meson.
   Therefore,
   the event selection window of the $\eta$ invariant mass was larger than
   that in the search via the
   $\eta\rightarrow2\gamma$ mode as shown in
   Table~\ref{tab:summary_criteria}.
   
   As shown in Fig.~\ref{fig:eta_mptot_fcdst},
   the background rates for \eqeeta\ ($3\pi$) were not low enough if the
   standard event selection criteria of $P_\textrm{tot}<250$ MeV/$c$ was applied.
   In order to further reduce the background,
   the tighter total momentum cut of $P_\textrm{tot}<150$ MeV/$c$ was
   applied for the \eqeeta\ mode.

   \subsubsection{$p\rightarrow l^+ \omega$ Mode Search}

  Two of the $\omega$ meson decay modes were searched for in this study.
  One is the $\omega\rightarrow \pi^+\pi^-\pi^0$ mode ($Br$=89\%), and the
  other is the $\omega\rightarrow \pi^0\gamma$ mode ($Br$=9\%).
  The momentum of a generated charged lepton and an $\omega$ meson is 143
  (105)~MeV/$c^2$ in \eqeomega\ (\eqmuomega).
  For the \eqmuomega\ mode, the muon momentum is lower than the Cherenkov
  threshold, and the muon ring cannot be observed.
  Therefore, the existence of the muon is indicated only by detection of the Michel
  electron from the muon decay.
  The $\omega$ meson suffers from nuclear effects in the case of a
  decay in an $^{16}$O nucleus.
  Only $\sim$~20\% of the $\omega$ mesons could escape from the nucleus.
  This causes the inefficiency for these two search modes.

   \paragraph{$p\rightarrow l^+ \omega \ (\omega \rightarrow
    \pi^0\gamma)$ Mode}

   In the $\omega\rightarrow\pi^0\gamma$ decay mode,
   the $\pi^0$ decays into two $\gamma$-rays, and three shower-type rings
   can be observed from the decays of the $\omega$ meson.
   The $\pi^0$ momentum from the $\omega$ decay is approximately 380~MeV/$c$
   in the $\omega$ meson rest frame.
   For same reason as in \eqepi0\ or \eqmupi0, one of the two rings
   from the $\pi^0$ decay has a certain probability of not being 
   identified.
   Therefore,
   two or three shower-type rings were required from the $\omega$ meson
   decay,
   and one more shower-type rings from $e^+$ was required only in the case of
   the \eqeomega\ mode.
   The fraction of events with 3 or 4 rings (2 or 3 rings) was 95\% (97\%)
   for the SK-I free
   proton decay of the $\omega\rightarrow\pi^0\gamma$ mode for
   \eqeomega\ (\eqmuomega).

   The $\omega$ invariant mass was reconstructed using all detected rings
   for the \eqmuomega\ mode.
   For the \eqeomega\ mode,
   the invariant mass was reconstructed by all but one of the detected rings,
   which was assumed to be $e^+$ ring.
   The $\omega$ invariant mass distribution for the proton decay MC is
   shown in the right panels of
   Fig.~\ref{fig:mesonmass}.
   The lower invariant mass peak in the proton decay MC was due to
   the another $\omega$ meson decay mode of $\omega \rightarrow \pi^+\pi^-\pi^0$.

\begin{figure*}[htbp]
    \begin{center}
     \includegraphics[width=\linewidth]{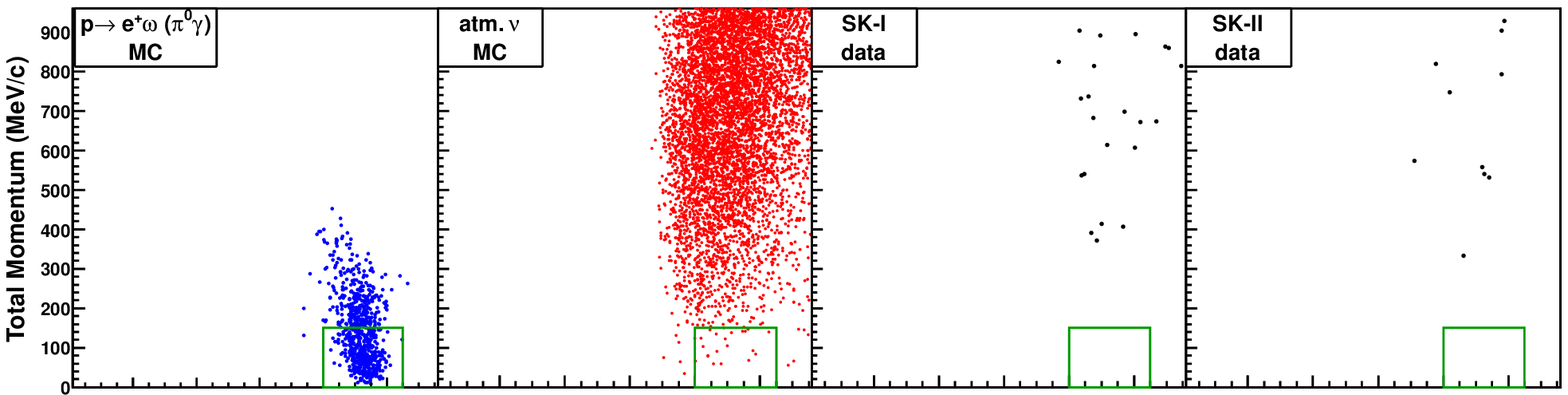}
     \includegraphics[width=\linewidth]{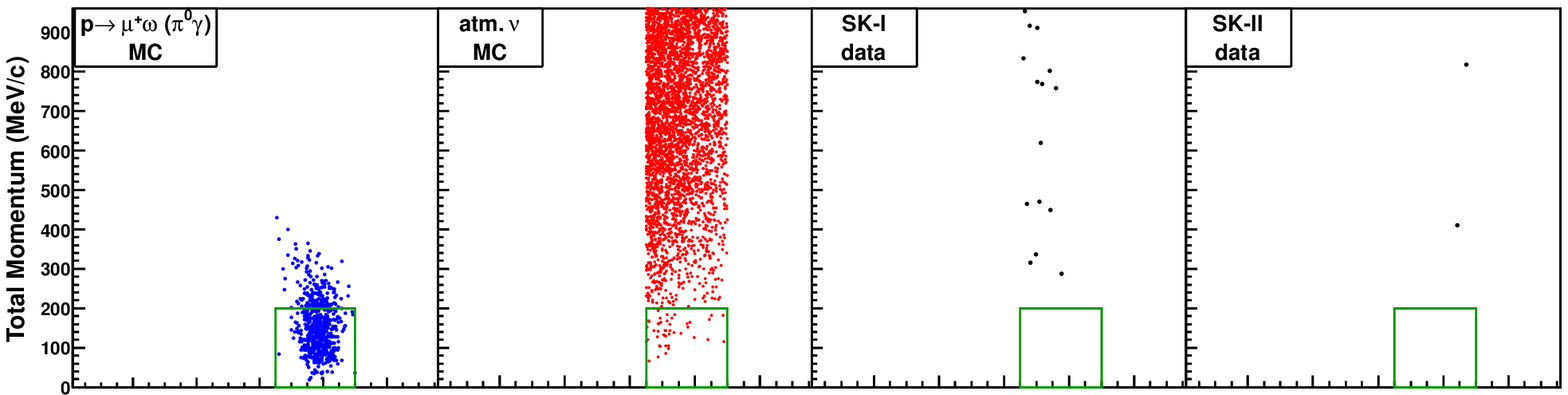}
     \includegraphics[width=\linewidth]{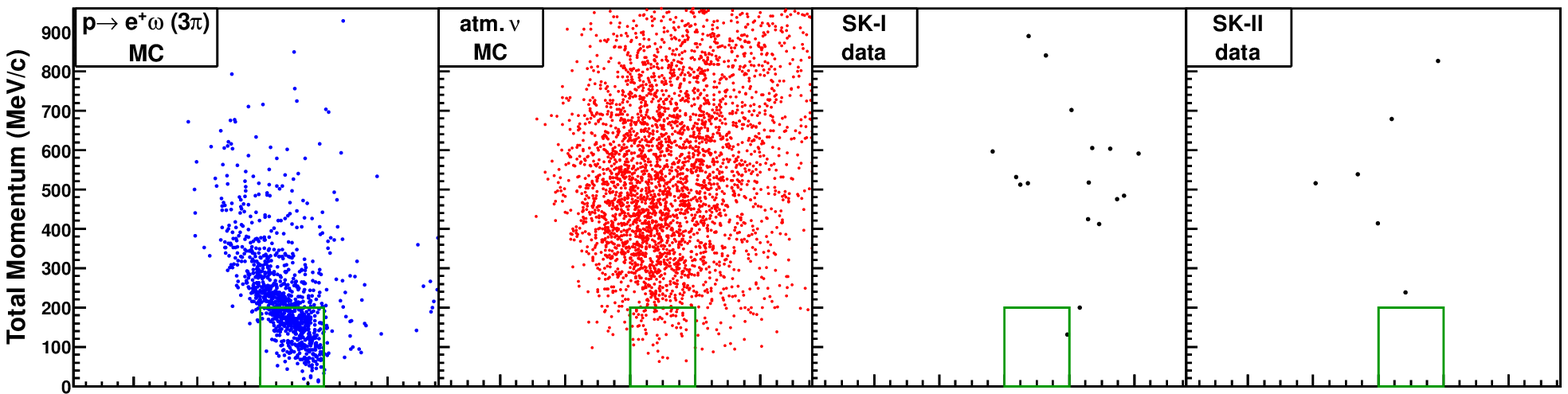}
     \includegraphics[width=\linewidth]{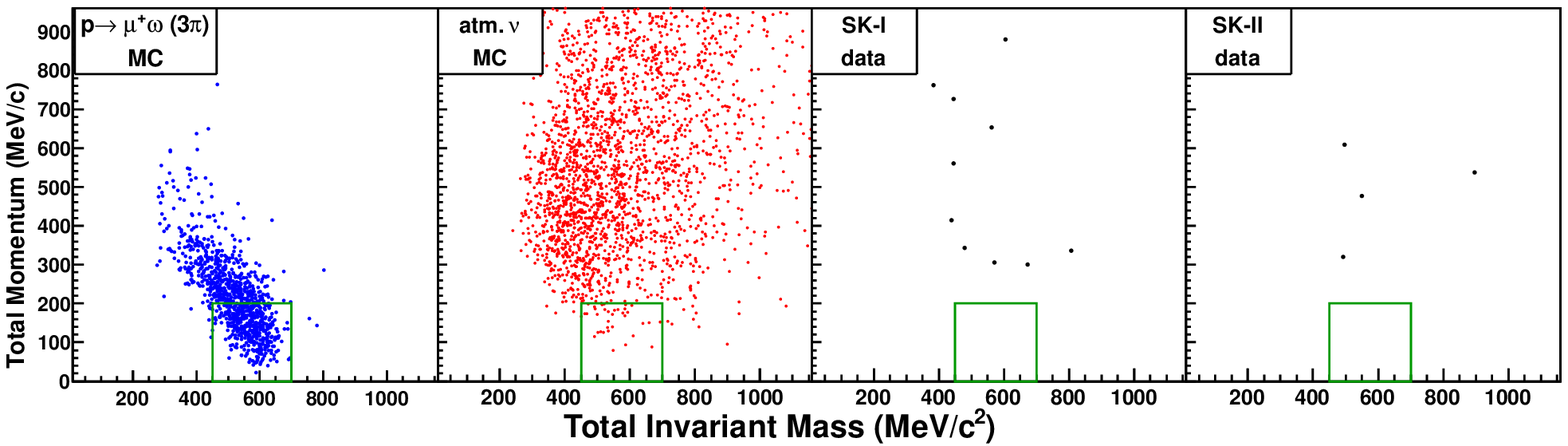}
     \caption{Total momentum versus total invariant mass distributions
     of proton decay MC (SK-I+II combined), the atmospheric neutrino MC
     (SK-I+II combined; 11.25 megaton$\cdot$years for each SK-I and
     SK-II) and SK-I (91.7 kiloton$\cdot$years) and SK-II
     (49.2 kiloton$\cdot$years) data, 
     from top to bottom: \eqeomega\ ($\pi^0\gamma$), \eqmuomega
     ($\pi^0\gamma$), \eqeomega\ ($3\pi$) and \eqmuomega\ ($3\pi$).
     These events satisfy the event selection criteria for each mode
     except for the selections on the total momentum and total mass.
     The boxes in the figures indicate the total momentum and mass
     criteria.
     For the \eqmuomega\ ($\pi^0\gamma$) mode, 
     the total invariant mass is equivalent to the meson invariant mass
     because a muon is invisible. The sharp cutoffs in
     the total invariant mass distribution correspond to the $\omega$
     invariant mass cut. For the \eqmuomega\ ($3\pi$) mode, there are no
     such sharp cutoffs because the $\omega$ invariant mass cut is not applied
     to this mode.
     A candidate for \eqeomega\ ($3\pi$) was found in the SK-I
     data. 
     }
     \label{fig:omega_mptot_fcdst}
    \end{center}
   \end{figure*}

   Since the muon is invisible for the \eqmuomega\ mode,
   the total momentum corresponds not to the proton momentum, but
   to the $\omega$ meson momentum.
   Thus,
   the reconstructed momentum for the free proton decay in SK-I
   peaked 
   around 100~MeV/$c$,
   which can be seen in Fig.~\ref{fig:omega_mptot_fcdst}.
   In order to eliminate the background sufficiently,
   tighter total momentum cuts were applied for both modes:
   $P_\textrm{tot}<150$~MeV/$c$ for \eqeomega\, and
   $P_\textrm{tot}<200$~MeV/$c$ for \eqmuomega.
   The sharp cutoffs in the total invariant mass distributions for
   \eqmuomega\ ($\pi^0\gamma$) in Fig.~\ref{fig:omega_mptot_fcdst}
   correspond to the $\omega$
   invariant mass cut. There is no total invariant mass cut applied to 
   this mode.

   \paragraph{$p\rightarrow l^+ \omega \ (\omega \rightarrow \pi^+\pi^-\pi^0)$ Mode}
   
   In this $\omega$ decay mode, 
   the $\omega$ decays into 
   two low momentum charged pions and a neutral pion.
   Their momenta are about 220~MeV/$c$.
   In order to find the low momentum non shower-type ring,
   PID with both a Cherenkov ring pattern and an
   opening angle was used for these modes.
   Due to the strong interaction of charged pions in water,
   finding both of the charged pion rings is difficult.
   Therefore, we required finding only one of the two charged pion rings.
   For the \eqeomega\ (\eqmuomega) mode, 
   4 (3) rings were required to be found in total.
   The efficiency of the number of rings cut was
   27\% (42\%) for SK-I free proton decay MC 
   of the \eqeomega\ (\eqmuomega), $\omega\rightarrow\pi^+\pi^-\pi^0$
   mode.
   
   Since one of the two charged pions was assumed to be invisible in
   the selection criteria,
   the invariant mass of the $\omega$ and the proton cannot be
   reconstructed.
   Instead of the $\omega$ mass reconstruction,
   the reconstructed $\pi^0$ invariant mass was required to be
   consistent with $\pi^0$ mass.
   As for the total invariant mass,
   the event selection criteria for the total mass were set to be lower
   than the normal event selection criteria, as shown in
   Fig.~\ref{fig:omega_mptot_fcdst} and Table~\ref{tab:summary_criteria}.
   The selection windows were determined to reduce the atmospheric
   neutrino MC, though they were not wide enough to also allow in the free proton
   decay events.

   For the \eqmuomega\ mode,
   both Michel electrons from the decay of
   $\mu^+$ and $\pi^+$ were required to be found.
   Although the detection efficiency is decreased by a factor of 2, 
   this cut reduced the background by an order of magnitude.

   In order to reduce the background further,
   the 
   reconstructed 
   positron momentum was also required to be consistent
   with the positron momentum
   for the \eqeomega\ mode.
   This criterion reduced the background by a factor of 2,
   while the detection efficiency was decreased by about 20\%.

   \subsubsection{$p\rightarrow l^+ \rho^0$ Mode Search}
   In the proton decay of \eqerho\ and \eqmurho,
   the momentum of the charged lepton and $\rho$ meson,
   which depends on the $\rho$ meson mass with a width of $\Gamma = $149 MeV,
   is about 170~MeV/$c$.
   For \eqmurho, the muon momentum can be lower than the Cherenkov threshold.
   The $\rho$ meson decays into $\pi^+\pi^-$ with a branching ratio of
   $\sim$100\%.
   The two pions suffer from strong interactions with nucleons in
   water, and also in the nucleus in the case of a proton decay in an
   $^{16}$O nucleus.
   Accordingly, a probability for finding all three Cherenkov rings
   from the charged
   lepton and the two charged pions is intrinsically very low.
   However, 
   in order to reduce the background in the selection by total
   invariant mass and total momentum,
   all three rings were required to be found in order to
   reconstruct the mass and the momentum of the proton.
   The fraction of three-ring events was 50\% and 27\% for 
   the free proton decay of \eqerho\ and \eqmurho, respectively.
   The lower efficiency for \eqmurho\ was due to proton decay events with
   an invisible muon.   
  
   The charged pions from the decay of the $\rho$ meson
   have a low momentum of about 300~MeV/$c$.
   Therefore, 
   PID with both Cherenkov ring pattern and opening angle was used for
   both \eqerho\ and \eqmurho.

   The invariant mass of the $\rho$ meson was reconstructed by two
   non shower-type rings
   and required to be between 600 and 900~MeV/$c^2$,
   as shown in the center panels of Fig.~\ref{fig:mesonmass}.
   Most of the events shown in $\rho$ meson distributions were from 
   free proton decay events.
   This means that the events from proton decays in an $^{16}$O nucleus
   rarely survived the selection criteria on the number of rings and the PID.

   For these mode searches,
   one extra Michel electron is expected from the $\pi^+$ decay, in
   addition to the one Michel electron from the $\mu$ decay.
   The number of Michel electrons were required to be consistent with
   this expectation.

\begin{figure*}[htbp]
    \begin{center}
     \includegraphics[width=\linewidth]{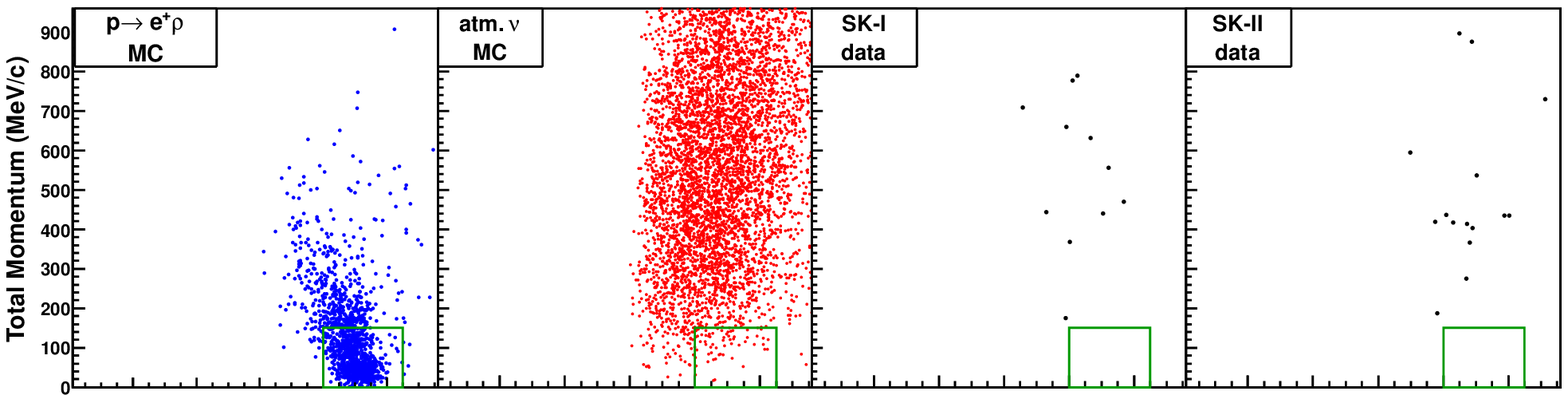}
     \includegraphics[width=\linewidth]{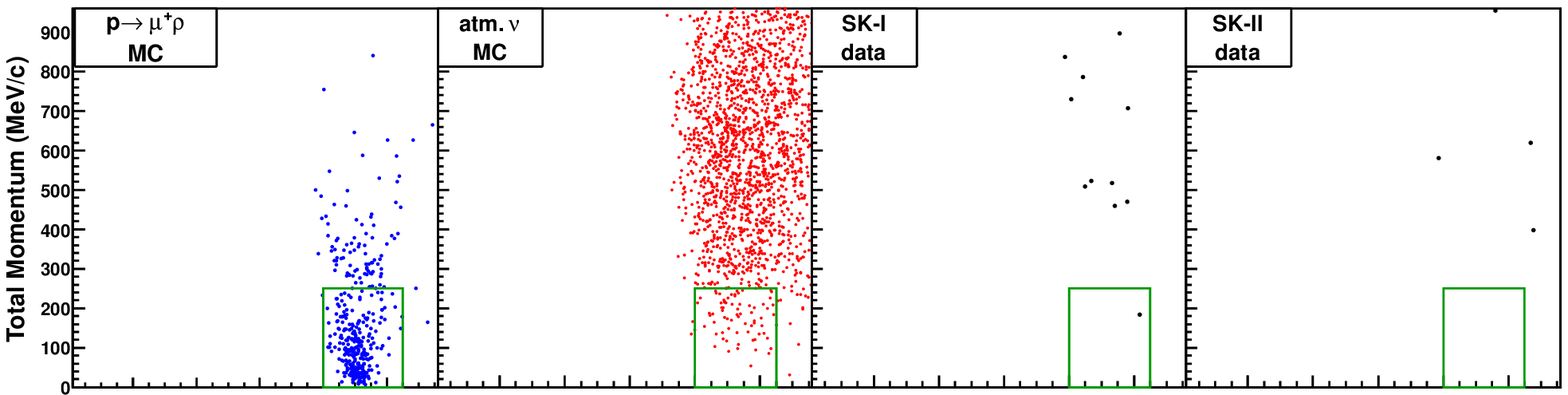}
     \includegraphics[width=\linewidth]{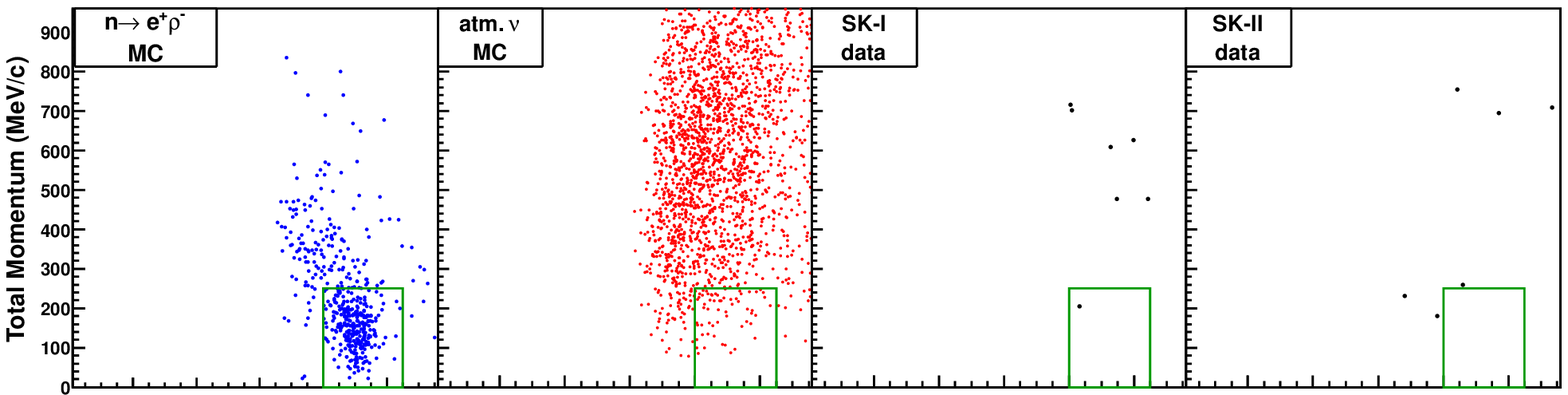}
     \includegraphics[width=\linewidth]{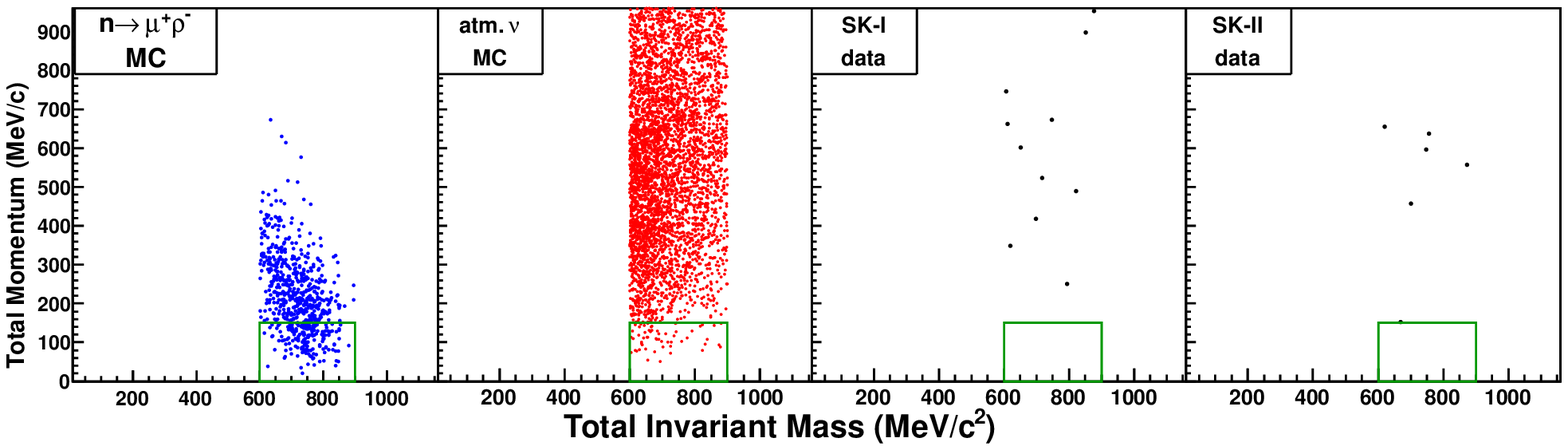}
     \caption{Total momentum versus total invariant mass distributions
     of proton decay MC (SK-I+II combined), the atmospheric neutrino MC
     (SK-I+II combined; 11.25 megaton$\cdot$years for each SK-I and
     SK-II) and SK-I (91.7 kiloton$\cdot$years) and SK-II
     (49.2 kiloton$\cdot$years) data, 
     from top to bottom: \eqerho, \eqmurho, \eqerhom\ and \eqmurhom.
     These events satisfy the event selection criteria for each mode
     except for the selection on total momentum and total mass.
     The boxes in figures indicate the total momentum and mass criteria.
     For the \eqmurhom\ mode, no total invariant mass cut is applied
     and the sharp cutoffs on the total invariant mass correspond to the
     $\rho^-$ invariant mass cut threshold.
     A candidate for each \eqmurho\ and \eqerhom\ was found in the SK-I data.
     }
     \label{fig:rho_mptot_fcdst}
    \end{center}
   \end{figure*}

   The total momentum and total invariant mass distributions for the
   $p\rightarrow l^+\rho^0$ modes are shown in
   Fig.~\ref{fig:rho_mptot_fcdst}.
   The \eqerho\ mode had a relatively higher background rate than that of
   \eqmurho\ mode.
   In order to reduce the background,
   a tighter total momentum cut of $P_\textrm{tot}<150$ MeV/$c$ was
   applied for the \eqerho\ mode.

   \subsubsection{$n\rightarrow l^+ \rho^-$ Mode Search}

   All of the neutrons in an H$_2$O molecule are bound in a nucleus.
   Because of this, all generated charged pions in neutron decay modes suffer from nuclear effects,
   and detection efficiencies for neutron decay searches tend to be lower 
   compared with their corresponding proton decay searches.

   In the neutron decay of $n\rightarrow l^+ \rho^-$,
   the $\rho^-$ meson decays into $\pi^-\pi^0$.
   The $\pi^0$ decays into two $\gamma$-rays.
   Accordingly, two shower-type rings and one non shower-type ring were
   required  from the $\rho^-$ meson decay.
   One more shower-type ring from the positron was required for the \eqerhom\
   mode search,
   while one more non shower-type ring from the muon was not required for
   the \eqmurhom\ mode search.
   In order to find the low momentum non shower-type ring,
   the PID with both a Cherenkov ring pattern and an
   opening angle was used for these modes.
   The fraction of 4-ring (3-ring) events was 9\% (23\%) for \eqerhom\ (\eqmurhom).

   The invariant mass of the $\rho^-$ meson was reconstructed using two
   shower-type rings and one non shower-type ring and required to be
   between 600 and 900~MeV/$c^2$.
   In addition to the $\rho$ meson mass,
   a $\pi^0$ invariant mass cut was also applied.
   This additional cut reduced
   the atmospheric neutrino background by a factor of 2,
   while the loss of detection efficiency was about 10\%.

   The total momentum and total invariant mass distributions for the
   $n\rightarrow l^+\rho^-$ modes are shown in
   Fig.~\ref{fig:rho_mptot_fcdst}.
   The tighter total momentum cut of $P_\textrm{tot}<150$~MeV/$c$ was
   applied to reduce the background for the \eqmurhom\ mode because the
   nucleon invariant mass cannot be reconstructed for this mode due to the
   invisible muon.

   \subsubsection{$n\rightarrow l^+ \pi^-$ Mode Search}

   The momentum of the charged lepton and charged pion in this mode is about
   460~MeV/$c$, almost the same as that of $p\rightarrow l^+ \pi^0$.
   In these modes,
   a Cherenkov ring from the charged pion generated from the nucleon
   decay can be observed.
   Approximately 50\% of the neutron decay events in SK-I were 2-ring
   events from the charged lepton and the charged pion.
   There was no event selection on the meson invariant mass for
   these modes.

\begin{figure*}[htbp]
    \begin{center}
     \includegraphics[width=\linewidth]{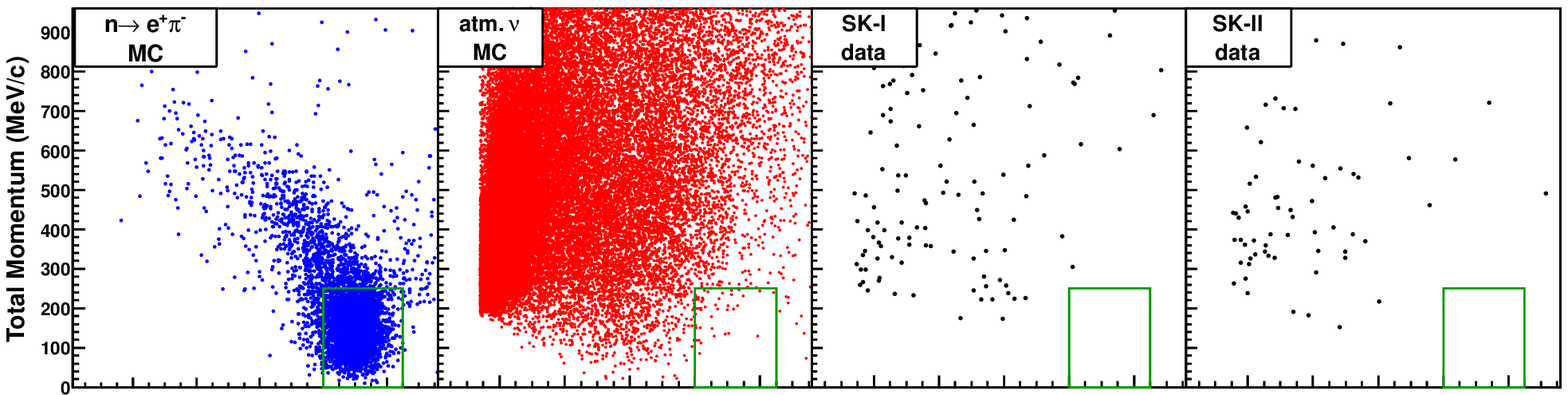}
     \includegraphics[width=\linewidth]{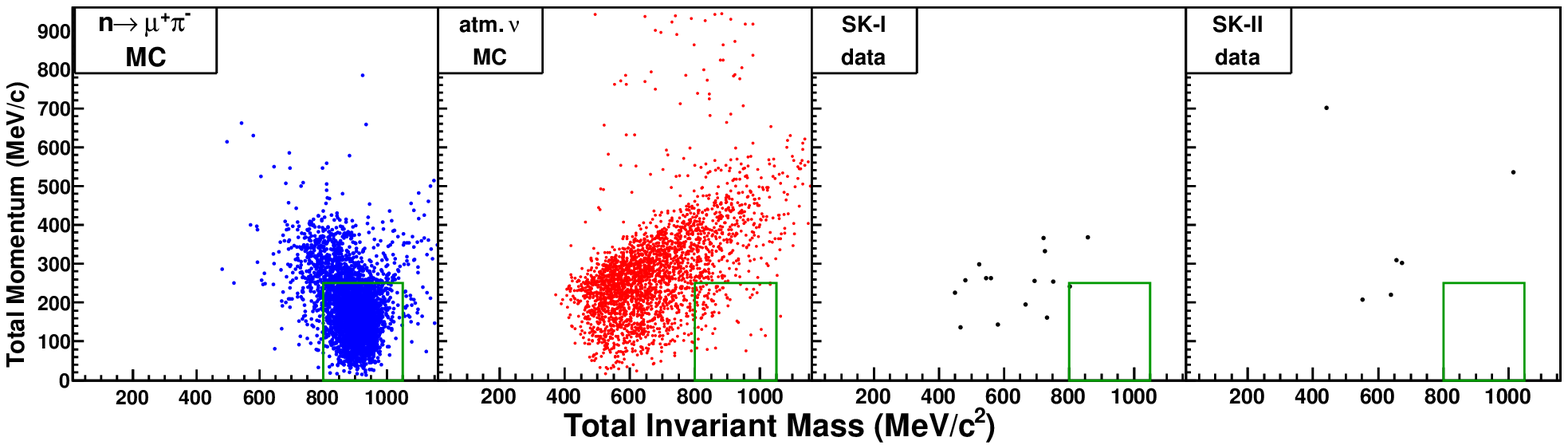}
     \caption{Total momentum versus total invariant mass distributions
     of proton decay MC (SK-I+II combined), the atmospheric neutrino MC
     (SK-I+II combined; 11.25 megaton$\cdot$years for each SK-I and
     SK-II) and SK-I (91.7 kiloton$\cdot$years) and SK-II
     (49.2 kiloton$\cdot$years) data, 
     for \eqepim\ (top) and \eqmupim\ (bottom).
     These events satisfy the event selection criteria for each mode
     except for the selection on the total momentum and total mass.
     The boxes in the figures indicate the total momentum and mass
     criteria.
     A candidate for \eqmupim\ was found in the SK-I data.
     }
     \label{fig:epim_mptot_fcdst}
    \end{center}
   \end{figure*}

   For the total invariant mass reconstruction in the \eqmupim\ mode,
   it is necessary to determine which ring was made by the muon or the charged pion.
   As described before, the ring combination which falls closest to the neutron mass was
   selected.
   This resulted in a narrower invariant mass distribution for \eqmupim\
   than that for the \eqepim\ MC,
   as shown in Fig.~\ref{fig:epim_mptot_fcdst}.

   \subsection{Search Results in the SK-I and SK-II data}

   The result of the nucleon decay searches are summarized in
   Table~\ref{tab:summaryall}.
   The detection efficiencies, expected backgrounds and number of
   candidate events are shown in the table.
   The number of events and detection efficiencies at each event
   selection step are shown in Fig.~\ref{fig:nevent_step} for the
   three typical nucleon decay mode searches.
   Figure~\ref{fig:nevent_step} shows the consistency between the
   observed data and the atmospheric neutrino background in terms of the
   number of events.

   Because of the difficulty in detecting charged pions in a
   large water Cherenkov detector, the efficiencies for modes with
   charged pions are relatively lower.
   For the modes with only Cherenkov rings from a charged
   lepton ($e$ or $\mu$) and $\gamma$, high efficiencies were achieved.
   The highest efficiency mode is \eqepi0. 
   Its efficiency is 87\% for free proton decay events.
   Also, the efficiencies for the SK-I free proton decay events in \eqeeta\
   ($2\gamma$), \eqeeta\ ($3\pi^0$) and \eqeomega\ ($\pi^0\gamma$)
   are 74\%, 67\% and 83\%, respectively.
   Nevertheless, the total efficiencies for these modes are much lower
   than \eqepi0\ due to the lower branching ratios for the meson decay
   modes.

   The detection efficiencies in SK-II are lower than in SK-I.
   However, 
   the difference is a only few percent for the modes with high detection
   efficiencies (\eqepi0, \eqeeta, etc.) and less than about 17\% even
   for the modes with low efficiencies as shown in Table~\ref{tab:summaryall}.
   Therefore, 
   nucleon decay searches using SK-II data are comparable to those
   using SK-I data.
   The estimated background rates of SK-I and SK-II are also comparable
   to each other.

   \begin{table*}[htbp]
    \begin{center}
     \begin{tabular}{l||rr|rr|c|cc||c}
      \hline \hline
      & \multicolumn{2}{c|}{Eff.(\%)} 
      & \multicolumn{3}{c|}{Background} 
      & \multicolumn{2}{c||}{Candidate}
      & \multicolumn{1}{c}{Lifetime Limit}
      \\
      & & & \multicolumn{2}{c}{NEUT} & NUANCE& & & ($\times 10^{33}$ years)\\
      Modes& SK-I & SK-II & \multicolumn{1}{c}{SK-I} &
		      \multicolumn{1}{c}{SK-II} & SK-I+II& SK-I & SK-II
      & at 90\% CL\\
      \hline
      \eqepi0  & 44.6 & 43.5 & 0.20 $\pm$ 0.04 (2.1) & 0.11 $\pm$ 0.02
		      (2.2) & 0.27 $\pm$ 0.10  & 0 & 0 & 8.2\\ 
      \eqmupi0 & 35.5 & 34.7 & 0.23 $\pm$ 0.04 (2.5) & 0.11 $\pm$ 0.02
		      (2.2) & 0.27 $\pm$ 0.09 & 0 & 0 & 6.6\\
      \eqeeta  &&&&&&& & 4.2\\
      \multicolumn{1}{r||}{($\eta\rightarrow 2\gamma$)} 
      & 18.8 & 18.2 &0.19 $\pm$ 0.04 (2.1) & 0.09 $\pm$ 0.02 (1.8) &
      0.29 $\pm$ 0.10 & 0 & 0 \\
      \multicolumn{1}{r||}{($\eta\rightarrow 3\pi^0$)}
      &  8.1 &  7.6 & 0.08 $\pm$ 0.03 (0.9) & 0.08 $\pm$ 0.02 (1.7) &
      0.32 $\pm$ 0.11 & 0 & 0\\
      \eqmueta &&&&&&& & 1.3\\
      \multicolumn{1}{r||}{($\eta\rightarrow 2\gamma$)} 
      & 12.4 & 11.7 & 0.03 $\pm$ 0.02 (0.3) & 0.01 $\pm$ 0.01 (0.2) &
      0.04 $\pm$ 0.04 & 0 & 0\\      
      \multicolumn{1}{r||}{($\eta\rightarrow 3\pi^0$)}
      &  6.1 &  5.4 & 0.30 $\pm$ 0.05 (3.3) & 0.15 $\pm$ 0.03 (2.9) &
      0.44 $\pm$ 0.13 & 0 & 2\\
      \eqerho  &  4.9 &  4.2 & 0.23 $\pm$ 0.05 (2.5) & 0.12 $\pm$ 0.02
		      (2.4) & 0.34 $\pm$ 0.11 & 0 & 0 & 0.71\\
      \eqmurho &  1.8 &  1.5 & 0.30 $\pm$ 0.05 (3.3) & 0.12 $\pm$ 0.02
		      (2.5) & 0.46 $\pm$ 0.12 & 1 & 0 & 0.16\\
      \eqeomega &&&&&&& & 0.32\\
      \multicolumn{1}{r||}{($\omega \rightarrow \pi^0\gamma$)} 
      &  2.4 &  2.2 & 0.10 $\pm$ 0.03 (1.1) & 0.04 $\pm$ 0.01 (0.9) &
      0.29 $\pm$ 0.10 & 0&0\\
      \multicolumn{1}{r||}{($\omega \rightarrow 3\pi$)} 
      &  2.5 &  2.3 & 0.26 $\pm$ 0.05 (2.9) & 0.13 $\pm$ 0.02 (2.6) &
      0.30 $\pm$ 0.11 & 1 & 0\\
      \eqmuomega &&&&&&& & 0.78\\
      \multicolumn{1}{r||}{($\omega \rightarrow \pi^0\gamma$)} 
      &  2.8 &  2.8 & 0.24 $\pm$ 0.05 (2.6) & 0.07 $\pm$ 0.02 (1.4) &
      0.37 $\pm$ 0.11 & 0&0\\
      \multicolumn{1}{r||}{($\omega \rightarrow 3\pi$)} 
      &  2.7 &  2.4 & 0.10 $\pm$ 0.03 (1.1) & 0.07 $\pm$ 0.02 (1.3) &
      0.05 $\pm$ 0.04 & 0 & 0\\
      \eqepim  & 19.4 & 19.3 & 0.16 $\pm$ 0.04 (1.7) & 0.11 $\pm$ 0.02
		      (2.2) & 0.37 $\pm$ 0.12 & 0 & 0 & 2.0\\
      \eqmupim & 16.7 & 15.6 & 0.30 $\pm$ 0.05 (3.3) & 0.13 $\pm$ 0.02
		      (2.6) & 0.44 $\pm$ 0.12 & 1 & 0 & 1.0\\
      \eqerhom &  1.8 &  1.6 & 0.25 $\pm$ 0.05 (2.7) & 0.13 $\pm$ 0.02
		      (2.7) & 0.44 $\pm$ 0.12 & 1 & 0 & 0.070\\
      \eqmurhom&  1.1 & 0.94 & 0.19 $\pm$ 0.04 (2.1) & 0.10 $\pm$ 0.02
		      (1.9) & 0.69 $\pm$ 0.14 & 0 & 0 & 0.036\\
      \hline \hline
     \end{tabular}
     \caption{Summary of the nucleon decay searches. Numbers in
     parentheses in backgrounds are exposure-normalized background rates in
     (megaton$\cdot$years)$^{-1}$. Background errors shown here are
     statistical errors of finite MC statistics.}
     \label{tab:summaryall}
    \end{center}
   \end{table*}
   
\begin{figure}[htbp]
    \begin{center}
     \includegraphics[width=\linewidth]{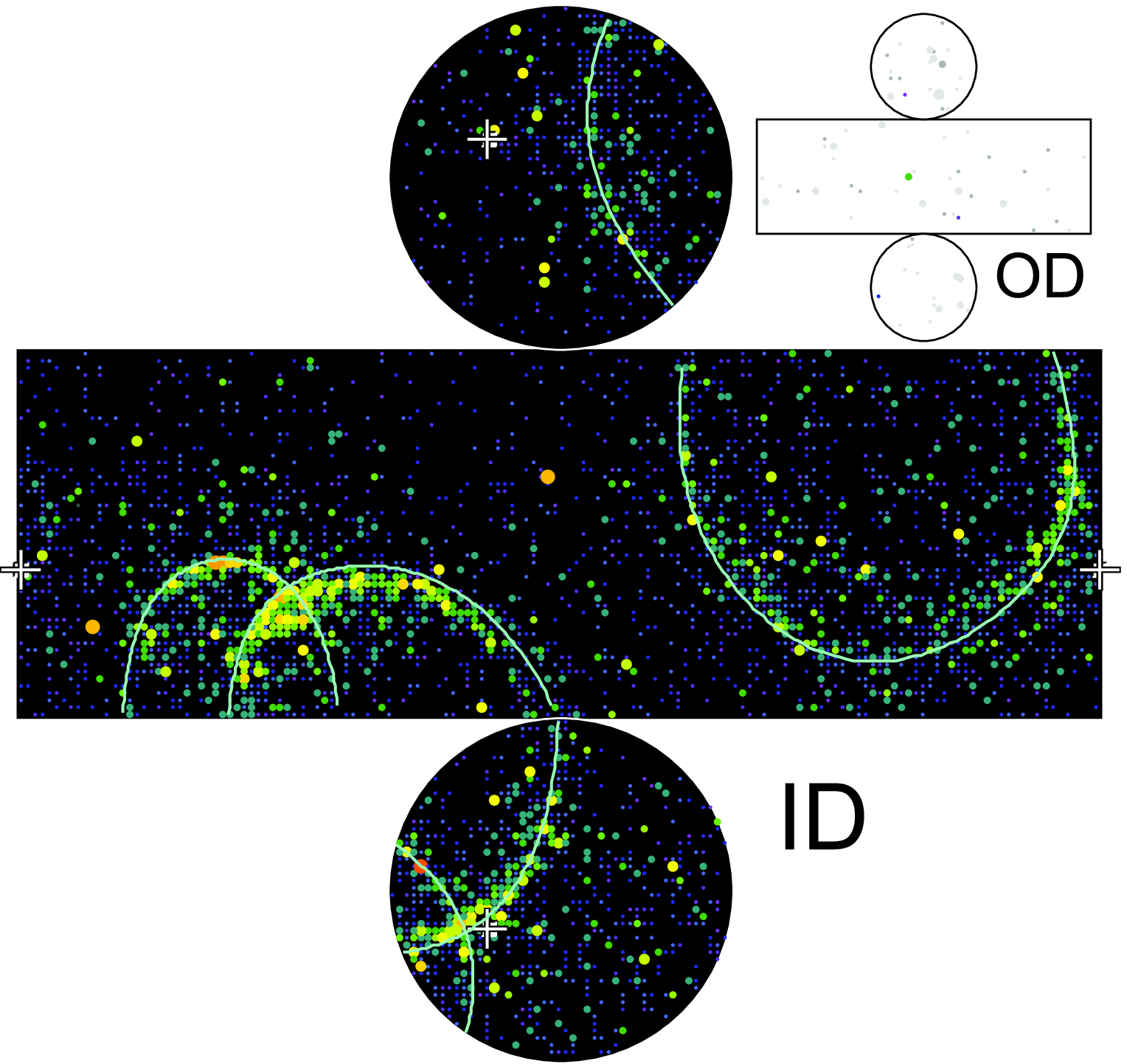}
     \caption{An example of a background event for the \eqepi0\ mode in the
     atmospheric neutrino MC. 
     An electron neutrino interacts with a neutron by CCQE and produces
     an electron and a proton.
     One ring from the electron
     and the
     other two rings from the decay of a $\pi^0$, 
     which is produced by a secondary interaction of the proton in water,
     mimic a proton decay signal from \eqepi0.
     Solid line circles show reconstructed rings.
     }
     \label{fig:display_epi0_BG}
    \end{center}
   \end{figure}

   Charged current pion production from neutrino interactions was the dominant
   background source
   for most of the studied nucleon decay modes. 
   The breakdown of background events is shown in
   Table~\ref{tab:breakdown}.
   Pions produced by neutrino interactions and/or hadronic interactions
   in nuclei and/or in water can mimic the event signatures of nucleon
   decay signals with a charged lepton.
   There was also a considerable contribution for the $p\rightarrow
   l^+\pi^0$ mode background from charged current
   quasi-elastic scattering (CCQE). 
   A typical background event from CCQE is shown
   in Fig.~\ref{fig:display_epi0_BG}.
   This is because a highly energetic
   proton ($>$ 1 GeV/$c$) produced by the interaction and scattered in water
   produces a secondary pion. 
   Backgrounds of high ring
   multiplicity modes (four-ring or five-ring events) such
   as $p\rightarrow l^+\eta (3\pi)$ or $p\rightarrow l^+\omega (3\pi)$
   have relatively higher fractions of multi-pion
   production events and has less dependency on whether the interaction
   is charged current or neutral current.
   
   \begin{table}[htbp]
    \begin{center}
     \begin{tabular}{lccccc}
      \hline \hline
      Mode & $p\rightarrow e^+\pi^0$  
	  & $N\rightarrow l^+\pi$ 
	      &	$p\rightarrow l^+\eta $
		  & $p\rightarrow l^+\omega $
		      & $N\rightarrow l^+\rho$\\
      \hline
      CCQE                        & 28\% & 21\% &  5\% &  4\% &  9\% \\ 
      CC 1-$\pi$             & 32\% & 51\% & 20\% & 25\% & 45\% \\
      CC multi-$\pi$              & 19\% & 14\% & 24\% & 29\% & 14\% \\
      CC others                   &  2\% &  6\% & 13\% &  7\% &  4\% \\
      NC                          & 19\% &  9\% & 37\% & 35\% & 28\% \\
      \hline \hline
    \end{tabular}
      \caption{The breakdown of the neutrino interaction modes of the
     background events. The breakdowns are calculated by simply adding
     background events of the modes decaying into each meson except for
     that of \eqepi0.}
     \label{tab:breakdown}
    \end{center}
   \end{table}

   The consistency of 
   the neutrino interactions and the nuclear effects
   was checked by comparing background estimations
   between our MC and NUANCE~\cite{Casper:2002sd}.
   NUANCE has different models of neutrino interactions and nuclear
   effects.
   The atmospheric neutrino MC generated with NUANCE used the same atmospheric
   neutrino flux equivalent to 100 years (2.25 megaton$\cdot$years exposure)
   and the same detector simulation.
   The results of these two estimates are also shown in Table~\ref{tab:summaryall}.
   Estimates obtained with NUANCE MC were consistent with results
   from our MC in most of the modes.
   
\begin{figure}[htbp]
    \begin{center}
     \includegraphics[width=\linewidth]{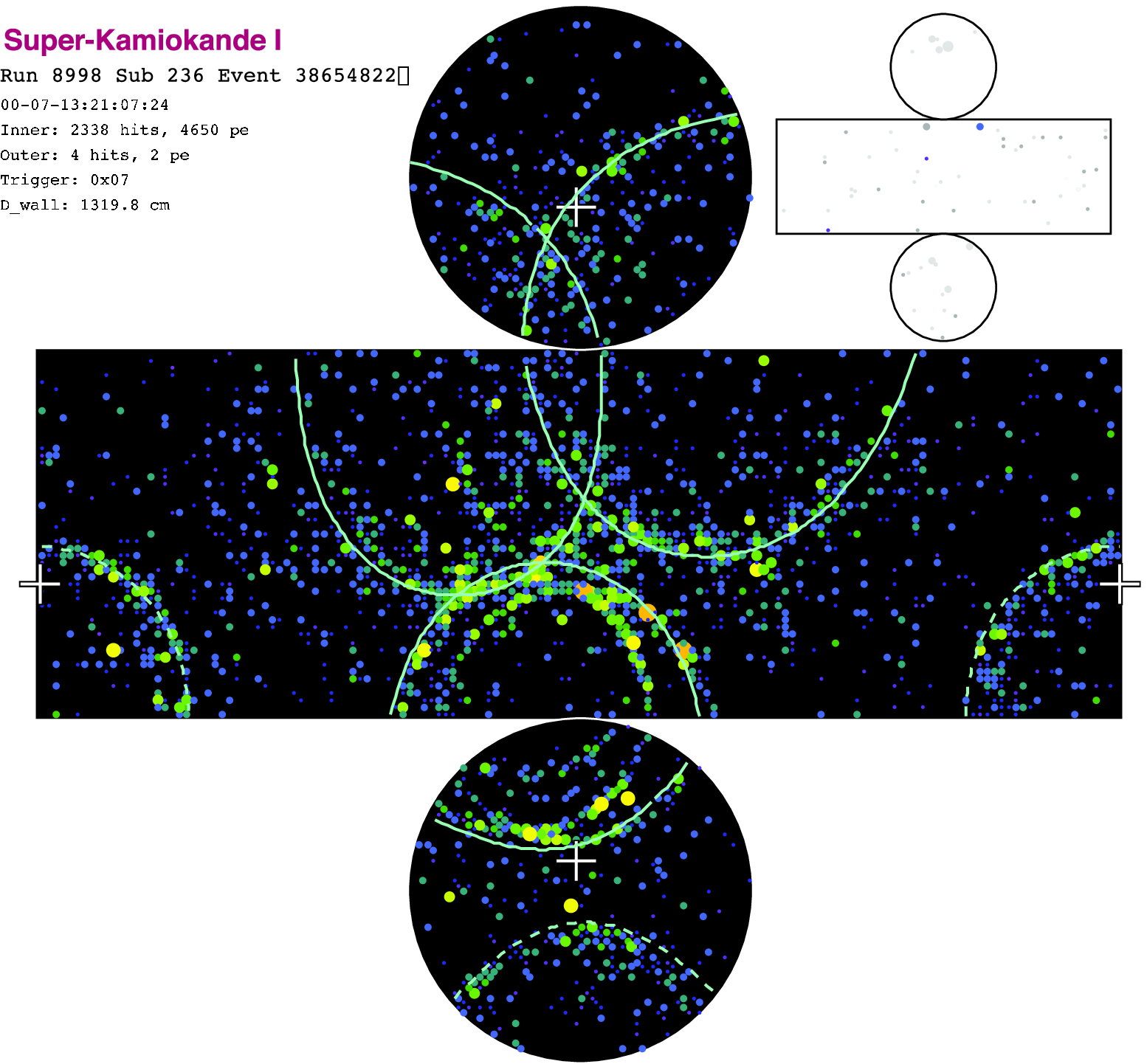}
     \caption{A candidate event for \eqeomega\ ($3\pi$) in the SK-I data.
     Reconstructed total momentum and invariant mass were 137~MeV/$c$ and
     796~MeV/$c^2$, respectively.
     Solid (dashed) lines correspond to reconstructed rings classified
     as shower (non shower) type.
     A Cherenkov ring at the center-bottom of the display was classified
     as a shower-type ring in the PID algorithm, but its Cherenkov ring
     edge looks as sharp as a non shower-type ring. 
     }
     \label{fig:display_eomega_candidate}
    \end{center}
   \end{figure}

   In total, six candidate events were found in the SK-I and SK-II data
   in the modes; \eqmueta\ ($3\pi^0$), \eqmurho, \eqeomega\ ($3\pi$),
   \eqmupim\ and \eqerhom.
   As can be seen from Figures~\ref{fig:eta_mptot_fcdst}, \ref{fig:omega_mptot_fcdst},
   \ref{fig:rho_mptot_fcdst}, \ref{fig:epim_mptot_fcdst}, every
   candidate event was around the threshold of the selection window of total momentum
   and total invariant mass cut.
   A candidate event for \eqeomega\ ($3\pi$)
   (Fig.~\ref{fig:display_eomega_candidate}) has a vertex around the
   center of the detector and the smallest total momentum of all
   candidates,
   although there might be a PID mis-identification for
   this event, which is described in the caption of the figure.

   These candidates were found in the five modes.
   For the \eqmueta\ ($3\pi^0$) mode, there were two candidate events.
   The probability of observing more than two events from the expected
   background was calculated without systematic errors to be 7.5\%.
   For the other modes with a candidate event, probabilities to observe
   one candidate from the expectations are about 30\% for each.
   Because the probabilities for observing one (two) or more background
   events in our sample are not so small for the 16 decay modes, we
   cannot take the candidates as serious evidence of nucleon decay.
   Moreover, the number of background events in total was 4.7 events.
   Consequently, the number of candidates is consistent with the
   estimation by the atmospheric neutrino MC.
   Therefore, nucleon partial lifetime limits were calculated in
   Section~\ref{sec:lifetime}.

   \subsection{Systematic Errors}
   Systematic errors for detection efficiencies and background estimations
   are described in this section.
   As for the exposure, the systematic errors of detector size and
   livetime are
   less than 1\% and negligible.

   \subsubsection{Systematic Errors of Detection Efficiency}

    \paragraph{Nuclear Effect}
    In most of the modes, meson ($\pi$, $\eta$, $\omega$) nuclear
    effects (meson-nucleon interactions in a nucleus) 
    have large effects on detection
    efficiencies and can be a dominant error source.

    \paragraph{$\pi$ nuclear effect}
    The systematic uncertainties of $\pi$ nuclear effects were estimated by
    comparing the nuclear effect simulation with another simulation
    result based on the model used by the IMB
    experiment~\cite{Bertini:1972vz}
    because there are no suitable experimental data which can be used
    for the systematic error estimation with enough precision at the
    moment.
    The comparison of the fraction of final states for the proton
    decay of \eqepi0\ in $^{16}$O nuclei is shown in
    Table~\ref{tab:epi0_imb_comparison}.
    The detection efficiency directly depends on the probability of the $\pi^0$
    escaping without any scattering.
    There is a 10\% difference in that probability between the
    two models.
    This difference corresponds to a 15\% difference in the total
    detection efficiencies of \eqepi0\ and \eqmupi0.

    \begin{table}[htbp]
     \begin{center}
      \begin{tabular}{lrr}
       \hline \hline
       nuclear effect & Our MC & IMB\\
       \hline
       no interaction & 44\% & 54\%\\
       absorption  & 22\% & 22\%\\
       charge exchange & 15\% & 10\%\\
       scattered      &  19\% & 14\%\\
       \hline \hline
      \end{tabular}
      \caption{Fraction of the final states of $\pi^0$ from the proton
      decay of \eqepi0 in $^{16}$O compared with the simulation used
      in the IMB experiment.}
      \label{tab:epi0_imb_comparison}
     \end{center}
    \end{table}

    The escape probability difference described above for the \eqepi0\ mode
    is equivalent to $\sim$~40\% difference in the total cross-section 
    of the $\pi$ nuclear effect.
    Thus, this difference in the total cross-section was used for the
    systematic error estimates from $\pi$ nuclear effects for the other
    nucleon decay modes,
    since detailed results of the IMB simulation were not available for
    those modes.
    As for the \eqepim\ mode, 
    the effect on the $\pi^-$ escape probability from a 40\% uncertainty
    on the $\pi$ nuclear effect was estimated by the nucleon decay
    simulation.
    This effect corresponds to a 20\% difference in the total detection
    efficiency for \eqepim.
    It is greater than that for \eqepi0\ because there are no free (unbound)
    nucleon decay events for \eqepim.

    \paragraph{$\eta$ nuclear effect}
    The systematic uncertainty of $\eta$ nuclear effects
    was estimated 
    in Section~\ref{subsec:etanuc}
    by comparing the experimental $\eta$ photoproduction cross-section 
    with the simulated cross-section.
    The estimated error for the $\eta$-nucleon cross-section in $^{16}$O
    nuclei was a factor of 2.

    \begin{table}[htbp]
     \begin{center}
      \begin{tabular}{lrrr}
       \hline \hline
       $\eta$ nuclear effect & $p\rightarrow l^+\eta$
       & $\sigma\times\frac{1}{2}$ & $\sigma\times 2$ \\
       \hline
       no interaction & 56\% & 73\% & 43\%\\
       scattered      &  6\% &   4\% & 5\%\\
       no $\eta$ survived & 38\% & 23\% & 53\%\\
       \hline \hline
      \end{tabular}
      \caption{Fraction of the final states of $\eta$ meson from the proton
      decay of $p\rightarrow l^+\eta$ in $^{16}$O.
      }
      \label{tab:sys_eta_nuc}
     \end{center}
    \end{table}

    Table~\ref{tab:sys_eta_nuc} shows the effects
    of the systematic uncertainties of the cross-section
    on the fraction of $\eta$ meson final states from a proton decay in
    $^{16}$O.
    The events in which no $\eta$ meson escapes from a nucleus hardly
    passed
    the selection criteria.
    The efficiency for the events in which the $\eta$ meson was
    scattered in the nucleus
    was less than $\sim$ 1/4 of the efficiency for the events without
    any interactions in a nucleus.
    Therefore, effects on the detection efficiency could be
    estimated almost entirely by
    the change of the escape probability with no interaction in a nucleus.
    The estimated errors 
    corresponded to a $\sim$~20\% error in total detection
    efficiencies for \eqeeta\ and \eqmueta.
    
    \paragraph{$\omega$ nuclear effect}
    The systematic uncertainty of $\omega$ nuclear effects was a factor of 3,
    which was estimated by the comparison of the $\omega$-nucleon cross-section 
    for the theoretical calculation and the extracted data
    from the $\omega$ photoproduction experiment
    as
    described in Section~\ref{subsec:omeganuc}.

    \begin{table}[htbp]
     \begin{center}
      \begin{tabular}{l||rr|rr}
       \hline \hline
       $\omega$ nuclear effect in $^{16}$O 
       & \eqeomega & $\sigma\times 3$ & \eqmuomega & $\sigma\times 3$\\
       \hline
       no interaction & 19\% & 11\% & 17\% & 10\%\\
       scattered      &  2\% & 2\% & 1\% & 1\%\\
       decay in a nucleus&  53\% & 35\% & 56\% & 37\%\\
       $\omega N \rightarrow N + meson(\neq \omega)$ 
       & 26\% & 52\% & 25\% & 51\%\\
       \hline \hline
      \end{tabular}
      \caption{Fraction of the final states of the $\omega$ meson from
      the
      proton
      decay of \eqeomega\ or \eqmuomega\ in $^{16}$O.
      }
      \label{tab:sys_omega_nuc}
     \end{center}
    \end{table}

    The effect of the factor of 3 uncertainty of the cross-section has
    a
    large effect on the fraction of $\omega$ meson final states from
    \eqeomega\ and \eqmuomega\ in an $^{16}$O nucleus,
    as shown in Table~\ref{tab:sys_omega_nuc}.
    However, 
    these effects correspond to only a $\sim$~20\% error on the total
    detection
    efficiency,
    since the fraction of free proton decay events, which do not
    suffer from nuclear effects, 
    in the total
    surviving events 
    was 40 $\sim$ 60 \% for the \eqeomega\ and \eqmuomega\ search.

   \paragraph{Hadron Propagation in Water}
   In some of the nucleon decay mode searches,
   Cherenkov rings from charged pions were required to be found.
   Charged pions strongly interact with nucleons in water.
   Thus, whether charged pion rings can be observed or not depends
   on their hadronic interactions in water.
   
   The uncertainty of the charged pion hadronic interaction cross-section
   in water was
   considered to be 10\% by comparing the detector simulation with
   experimental data~\cite{Ingram:1982bn,Albanese:1980np}.

   Only for the systematic errors on the background, 
   the uncertainty of the pion production probability by high momentum
   ($>$~$\sim$1~GeV/$c$) hadrons was also considered.
   This error of the production probability was conservatively set to 100\%.

   \paragraph{Fraction of $N$-$N$ Correlated Decay} 
   As described in Section~\ref{sec:sim_ndk},
   10\% of nucleons in an $^{16}$O nucleus are assumed to 
   correlate with another nucleon.
   Such a decay is calculated as a three-body decay.
   The detection efficiency for this decay can be very low.
   The uncertainty for this fraction was conservatively set to
   100\%.   

   \paragraph{Fermi Motion}
   The total momentum of a bound nucleon decay event corresponds to the
   Fermi motion of the source nucleon.
   The systematic error from the uncertainty of the Fermi motion was
   estimated by comparing the distributions used in the simulation with
   the Fermi gas model
   or by changing the momentum by $\pm$20\%.
   If a tighter total momentum cut is used, 
   the systematic error from the Fermi motion can be large.   

    \paragraph{Fiducial Volume}
    The systematic errors from the fiducial volume were estimated by the
    difference in the number of events for reconstructed and true
    vertices of multi-ring events.
    The estimated errors were 3\% and 2\% for SK-I and SK-II, respectively.
    This error can directly affect the detection efficiencies,
    but its magnitude is negligible compared with other systematic errors.
    
    \paragraph{Momentum Scale}
    The uncertainty of the 
    momentum 
    scale was estimated to be 1.1\%
    (1.7\%) in SK-I (SK-II) by the quadratic sum of the uncertainties of
    the absolute
    momentum 
    scale and the time variation
    described in Section~\ref{sec:calibration}.
    The
    momentum
    scale non-uniformity in the detector was within
    $\pm$~0.6\%.
    The non-uniformity can cause momentum imbalance of an event,
    leading to a 1.2\% error for the total momentum.
    
    The systematic errors on the efficiencies were estimated by the
    changing the threshold of the momentum and mass in the selection criteria.
    These two errors have negligible effects.
    
    \paragraph{PID, Ring Counting and Cherenkov Opening Angle}
    The systematic error of the particle identification was estimated by
    comparing the likelihood difference (non shower-type likelihood -
    shower-type likelihood) distributions 
    of observed data and atmospheric neutrino MC.
    The ring counting systematic error was also estimated by the same
    method.
    Those likelihood difference distributions of the data and MC agree
    well with each other, and
    systematic errors from the PID and the ring counting were
    estimated to be negligible.

    As for the Cherenkov opening angle,
    the systematic error of the Cherenkov opening angle was estimated
    to be
    0.7 (0.5) degrees in SK-I (SK-II)
    by
    comparing the opening angle distributions of the observed data and
    atmospheric neutrino MC
    for non-primary (not most energetic) non shower-type rings in fully contained
     sub-GeV events (in which visible energy is below 1.33 GeV).
    Charged pion momentum reconstruction largely depends on the opening
    angle.
    Thus, the systematic errors for the modes which need to reconstruct
    the charged pion momenta can be relatively large.

    \paragraph{Vertex Shift}
    The vertex position was reconstructed at the beginning of the event
    reconstruction.
    The reconstructed vertex can affect all of the following reconstruction
    algorithms like the ring counting, the PID, the momentum
    determination, etc.

    The systematic error from the uncertainty of the vertex position was
    estimated by shifting the reconstructed vertex by 30~cm
    along the direction of the most energetic particle momentum.
    This was a conservative estimate since the systematic shift on that
    direction could cause largest error on the other event reconstructions.
    The shift length 30~cm corresponds to the typical resolution scale for
    the nucleon decay events as well
    as the scale of the uncertainty
    of the vertex position estimated by comparing different vertex
    fitting algorithms.

    \paragraph{Summary of the Systematic Errors of Detection Efficiencies}
    Table~\ref{tab:syserr_eff} summarizes the
   results of
   the systematic error estimation for detection efficiencies of all
   modes.   
   The systematic errors for detection efficiencies were about
   20~$\sim$~30\% except for the much large error of the $n\rightarrow
   l^+\rho^-$ mode.
   
   Systematic uncertainties of nuclear effects are the dominant error
   sources for most of the modes.
   For the \eqerhom\ and \eqmurhom\ modes,
   the systematic errors from the pion nuclear effects are very large
   because both of the pions from the $\rho$ meson decay were
   required to escape from the nucleus.
   Positive and negative errors were individually estimated especially
   for these two modes.
   On the other hand, \eqerho\ and \eqmurho\ has a comparably smaller
   error
   since most of the surviving events in these modes are free proton
   decay events.
   The uncertainty of Fermi motion can be a dominant error source 
   especially in modes using a tight total momentum cut like
   \eqeeta\ ($3\pi^0$).
   Another important error is the reconstruction biases from the
   Cherenkov opening angle reconstruction and the vertex shift.
   These biases mostly contribute errors for the modes which
   require charged pion momentum reconstruction.

   These dominant errors are mostly common in SK-I and SK-II.
   There are no significant differences between the
   systematic errors of SK-I and SK-II.

   \begin{table*}[htbp]
    \begin{center}
     \begin{tabular}{l||ccccc|c}
      \hline \hline
          & meson       & hadron      & $N$-$N$      & Fermi    & Detector
      & Total\\
          & nuclear     & propagation & correlated & momentum & Performances\\
      Mode& effect      & in water    & decay      &          & \\
      \hline
      \eqepi0   & 15\% & - & 7\% & 8\% & 4\% & 19\%\\
      \eqmupi0  & 15\% & - & 7\% & 8\% & 4\% & 19\%\\
      \eqeeta\ ($2\gamma$) & 20\% & - & 7\% & 13\% & 5\% & 25\%\\
      \eqmueta\ ($2\gamma$) & 18\%& - & 7\% & 14\% & 4\% & 24\%\\
      \eqeeta\ ($3\pi^0$) & 15\%  & - & 5\% & 26\% & 9\% & 32\%\\
      \eqmueta\ ($3\pi^0$) & 20\% & - & 7\% & 14\% & 10\% & 28\%\\
      \eqerho  & 8\% & 17\% & 2\% & 10\% & 18\% & 28\%\\
      \eqmurho & 9\% & 24\% & 2\% & 6\% & 11\% & 29\%\\
      \eqeomega\ ($\pi^0\gamma$)  & 21\% & - & 5\% & 24\% & 9\% & 33\%\\
      \eqmuomega\ ($\pi^0\gamma$) & 23\% & - & 6\% & 13\% & 7\% & 28\%\\
      \eqeomega\ ($\pi^+\pi^-\pi^0$)  & 19\% & 13\% & 5\% & 12\% & 20\% & 34\%\\
      \eqmuomega\ ($\pi^+\pi^-\pi^0$) & 19\% & 15\% & 5\% &  2\% & 16\% & 29\%\\
      \eqepim & 20\% & 9\% & 11\% &  12\% & 12\% & 30\%\\
      \eqmupim & 24\% & 6\% & 11\% &  7\% & 17\% & 33\%\\
      \eqerhom  & +51\% -23\%& 9\% & 11\% &  15\% & 19\% & +59\% -37\%\\
      \eqmurhom & +51\% -25\%& 14\% & 10\% &  27\% & 23\% & +65\% -47\%\\
      \hline \hline
     \end{tabular}
     \caption{Systematic errors for detection efficiencies.
     The errors of SK-I and SK-II were separately estimated and averaged
     by the livetime.
     }
     \label{tab:syserr_eff}
    \end{center}
   \end{table*}

   \subsubsection{Systematic Errors of Background Estimation}\label{sec:sys_bg}
   
   For the background estimations from atmospheric neutrinos,
   the uncertainties of atmospheric neutrino flux and neutrino cross-sections 
   were considered.
   The systematic uncertainties of pion nuclear effects, hadron
   propagation in
   water and the detector performance were also considered as well as
   the detection efficiencies.
   
   Even with the large statistics of the atmospheric neutrino MC,
   only a few tens of events can survive the nucleon decay event selection
   criteria.
   In order to reduce statistical errors for the systematic error
   estimation of the background,
   systematic errors were estimated by averaging the estimations of SK-I
   and SK-II
   because they were basically common as described in the error estimation
   for the detection efficiencies.
   Moreover, for the modes with the same meson decay modes,
   systematic errors from the uncertainties of pion nuclear effects and
   hadron interactions in water were estimated by averaging the estimations
   of different charged lepton modes.  
   
   Almost the same systematic uncertainties 
   as the atmospheric neutrino oscillation analysis in the
   Super-Kamiokande experiment  
   were considered for the errors
   from the neutrino flux and the neutrino interaction.
   The details of the source of the uncertainties are
   given in~\cite{Ashie:2005ik,Wendell:2010md}.
   The systematic errors from the neutrino flux were estimated to be
   6~$\sim$~8\%, and mostly due to the uncertainty of the energy spectrum.
   These errors are negligible compared with the other much larger errors.
   The systematic errors from the neutrino interactions were also
   estimated to be negligible, 8~$\sim$16~\%.

   The estimated systematic errors for the backgrounds are shown in 
   Table~\ref{tab:syserr_bg}.
   The systematic errors for the backgrounds ranged from about
   40 to 70\%.
   One of the dominant errors comes from the uncertainty of the 
   pion-nucleon cross-section and the pion production probability in water.
   The errors from the detector and event reconstruction performances
   also have non-negligible contributions.
   The number of background events is very sensitive to the error of
   energy and momentum, 
   since surviving background events are usually distributed around
   the threshold of the selection window of momentum and invariant mass.
   Therefore,
   the systematic error from the energy scale stability,
   which was negligible for the detection efficiencies,
   was estimated to
   be about 10~$\sim$~20\%.
   The systematic shift of the reconstructed vertex 
   can cause errors in the Cherenkov opening angle and opening angles
   between two particles, which are important for momentum and mass
   reconstruction.
   For the same reason as energy scale,
   the systematic error from the vertex shift was larger than that for
   the detection efficiency, and estimated to be about 10~$\sim$~50\%.

   \begin{table*}[htbp]
    \begin{center}
     \begin{tabular}{l||ccccc|c}
      \hline \hline
          & neutrino & neutrino & pion    & hadron      & Detector      & Total\\
          & flux     & cross    & nuclear & propagation & Performances\\
      Mode&          & section  & effect  & in water    & \\
      \hline
      \eqepi0   & 8\% & 8\% & \multirow{2}{*}{8\%} & \multirow{2}{*}{36\%} & 22\% & 44\%\\
      \eqmupi0  & 8\% & 8\% & &  & 43\% & 58\%\\
      \eqeeta\ ($2\gamma$) & 8\% & 11\% & \multirow{2}{*}{5\%} & \multirow{2}{*}{36\%} & 26\% & 47\%\\
      \eqmueta\ ($2\gamma$) & 8\%& 14\% & & & 28\% & 49\%\\
      \eqeeta\ ($3\pi^0$)  & 8\%  & 15\% & \multirow{2}{*}{18\%} & \multirow{2}{*}{67\%} & 13\% & 76\%\\
      \eqmueta\ ($3\pi^0$) & 8\%  & 11\% & & & 20\% & 73\%\\
      \eqerho  & 6\% & 13\% & \multirow{2}{*}{14\%} & \multirow{2}{*}{33\%} & 33\% & 51\%\\
      \eqmurho & 8\% & 15\% & &  & 23\% & 46\%\\
      \eqeomega\ ($\pi^0\gamma$)  & 8\% & 14\% & \multirow{2}{*}{13\%} & \multirow{2}{*}{41\%} & 37\% & 59\%\\
      \eqmuomega\ ($\pi^0\gamma$) & 8\% & 10\% & &  & 28\% & 53\%\\
      \eqeomega\ ($\pi^+\pi^-\pi^0$)  & 7\% & 14\% & \multirow{2}{*}{8\%} & \multirow{2}{*}{53\%} & 28\% & 63\%\\
      \eqmuomega\ ($\pi^+\pi^-\pi^0$) & 7\% & 11\% &  &  & 29\% & 63\%\\
      \eqepim   & 8\% & 15\% & \multirow{2}{*}{8\%} &  \multirow{2}{*}{36\%} & 46\% & 61\%\\
      \eqmupim  & 8\% & 16\% &  &  & 36\% & 55\%\\
      \eqerhom  & 8\%& 14\% & \multirow{2}{*}{12\%} &  \multirow{2}{*}{18\%} & 54\% & 60\%\\
      \eqmurhom & 6\%& 16\% & & & 27\% & 39\%\\
      \hline \hline
     \end{tabular}
     \caption{Systematic errors for background estimates.
     }
     \label{tab:syserr_bg}
    \end{center}
   \end{table*}
  
  \subsection{Lifetime Limit} \label{sec:lifetime}
  The observed data in SK-I and SK-II are consistent with the
  atmospheric neutrino MC.
  Consequently, lower limits on the nucleon partial lifetime were calculated.
  
  The partial lifetime limit for each mode is derived 
  from Bayes' theorem
  which incorporates systematic errors.
  Because the nucleon decay search is a counting experiment,
  the probability to detect $n$ events is given by Poisson statistics as
  follows.
  \begin{equation}
   {\bf P}(n|\Gamma \lambda \epsilon b) 
    = \frac{e^{-\left(\Gamma \lambda \epsilon+b\right)}
    \left(\Gamma \lambda \epsilon+b\right)^n}{n!}
  \end{equation}
  where $\Gamma$ is the true decay rate, $\lambda$ is the true exposure,
  $\epsilon$ is the true detection efficiency, $b$ is the true
  number of background events, and ${\bf P}(A|B)$ is the conditional
  probability of A, given that proposition B is true.
  
  Applying Bayes' theorem allows us to write:
  \begin{equation}
   {\bf P}(\Gamma \lambda \epsilon b|n) 
    = \frac{1}{A}
    {\bf P}(n|\Gamma \lambda \epsilon b){\bf P}(\Gamma \lambda \epsilon b)
  \end{equation}
  where $A$ is the constant to normalize ${\bf P}(\Gamma \lambda \epsilon
  b|n)$.
  Because the decay rate, the detection efficiency, the exposure and the
  background are independent, ${\bf P}(\Gamma \lambda \epsilon b)$ can be
  separated into constituents.
  \begin{equation}
   {\bf P}(\Gamma \lambda \epsilon b|n) 
    = \frac{1}{A} {\bf P}(n|\Gamma \lambda \epsilon b)
    {\bf P}(\Gamma){\bf P}(\epsilon){\bf P}(\lambda){\bf P}(b)
  \end{equation}
  
  The probability density function of $\Gamma$ can be defined as;
  \begin{eqnarray}
   {\bf P}(\Gamma |n) &=& \int\int\int{\bf P}(\Gamma \epsilon \lambda b|n)d\epsilon d\lambda db\nonumber\\
   &=& \frac{1}{A}
    \int\int\int \frac{e^{-\left(\Gamma \lambda \epsilon+b\right)}
    \left(\Gamma \lambda \epsilon+b\right)^n}{n!}\nonumber \\
   && \times {\bf P}(\Gamma){\bf P}(\epsilon){\bf P}(\lambda){\bf P}(b)
    d\epsilon d\lambda db \label{eq:pdf}
  \end{eqnarray}
  where 
  ${\bf P}(\Gamma)$, ${\bf P}(\lambda)$, ${\bf P}(\epsilon)$ and ${\bf
  P}(\Gamma)$ are the prior probability distributions,
  in which systematic uncertainties can be incorporated.

  The priors for the exposure, the detection efficiency and the
  background are assumed
  to be truncated Gaussian distributions defined as;
  \begin{eqnarray}
   &{\bf P}(x) \propto
    \left\{
     \begin{array}{ll}
      \exp\left(-\frac{(x-x_0)^2}{2\sigma_x^2}\right) & \left(x>0\right)\\
      0 &\left(x \leq 0\right)
     \end{array}
    \right.\\
   &(x = \lambda, \epsilon, b)\nonumber
  \end{eqnarray}
  where $\lambda_0$ ($\sigma_\lambda$), $\epsilon_0$ ($\sigma_\epsilon$) 
  and $b$ ($\sigma_b$)
  are the estimates (systematic errors)
  of the detection efficiency, the exposure and the background, respectively.
  If the systematic error for the detection efficiency is assumed to be asymmetric,
  the prior ${\bf P}(\epsilon)$ is an asymmetric Gaussian.

  The prior for the decay rate is assumed to be uniform.
  This is implicitly assumed when calculating limits by simple Poisson
  statistics without systematic errors.
  \begin{equation}
   {\bf P}(\Gamma) = 
    \left\{
     \begin{array}{ll}
      1 & (0 < \Gamma < \Gamma_\textrm{cut})\\
      0 & (\Gamma \leq 0 \textrm{ or } \Gamma\geq \Gamma_\textrm{cut})
     \end{array}
    \right.
  \end{equation}
  where $\Gamma_\textrm{cut}$ is the upper limit of the decay rate for
  the calculation of
  the normalization constant $A$ in order to avoid divergence.
  The upper limit
  $\Gamma_\textrm{cut}$ 
  is set to be $10^{-31}$
  years$^{-1}$, which is sufficiently larger than the limits by the
  previous experiments.

  By integrating Equation~(\ref{eq:pdf}) using the priors,
  the confidence level can be calculated as;
  \begin{equation}
   CL = \int_0^{\Gamma_\textrm{limit}}{\bf P}(\Gamma |n)d\Gamma. \label{eq:cl}
  \end{equation}

  Lifetime limits are obtained by:
  \begin{equation}
   \tau_\textrm{limit} = 1/\Gamma_\textrm{limit}.\label{eq:limit}
  \end{equation}
  
  The combined result of SK-I and SK-II is also derived by the method
  described above.
  The probability to detect $n_1$ events in SK-I and $n_2$ events in
  SK-II is the product of the two Poisson probabilities.
  
  \begin{equation}
   {\bf P}(n_1, n_2|\Gamma \lambda_1 \epsilon_1 b_1 \lambda_2 \epsilon_2
    b_2) = {\bf P}(n_1|\Gamma \lambda_1 \epsilon_1 b_1){\bf P}(n_2|\Gamma \lambda_2 \epsilon_2 b_2)
  \end{equation}

  We apply Bayes' theorem assuming that the decay rate, the exposure,
  the detection
  efficiency and the background are independent:
  
  \begin{eqnarray}
   &{\bf P}(\Gamma \lambda_1 \epsilon_1 b_1\lambda_2 \epsilon_2 b_2|n_1, n_2)
    = \frac{1}{A} 
    {\bf P}(n_1,n_2|\Gamma \lambda_1 \epsilon_1 b_1
    \lambda_2 \epsilon_2 b_2 )\nonumber\\
   &\times{\bf P}(\Gamma){\bf P}(\epsilon_1,\epsilon_2)
    {\bf P}(\lambda_1,\lambda_2){\bf P}(b_1,b_2).
  \end{eqnarray}
  
  Most of the dominant systematic uncertainties like the nuclear effects
  and
  hadron propagation in water are common between SK-I and SK-II.
  Therefore, the systematic errors of SK-I and SK-II are  assumed to be
  fully correlated with each other,
  which gives conservative lifetime limits for this method.
  Then, the priors for the exposure, the detection efficiency and the
  background of SK-I and SK-II are expressed as:

  \begin{eqnarray}
   &{\bf P}(x_1,x_2) = {\bf P}(\delta_x) 
    \propto \exp\left(-\frac{\delta_x^2}{2}  \right)\\
   &(x = \lambda, \epsilon, b)\nonumber
  \end{eqnarray}
  where $\delta_x$ is a correlated error factor for SK-I and SK-II
  defined as follows:
  \begin{eqnarray}
   &\delta_{x} = \frac{\left(x_1 - x_{01} \right)}{\sigma_{x 1}} = \frac{\left(x_2 - x_{02} \right)}{\sigma_{x 2}} \\
   &(x = \lambda, \epsilon, b)\nonumber
  \end{eqnarray}
  $\lambda_{0i}$, $\epsilon_{0i}$ and $b_{0i}$ are the estimated
  exposure, detection efficiency and background for SK-I and SK-II,
  respectively.
  The statistical error of the background MC is ignored
  because of the much larger systematic errors.

 Finally the probability density function can be expressed as:
 
  \begin{eqnarray}
   {\bf P}(\Gamma |n_1, n_2) 
    &=&
    \frac{1}{A}
    \int\int\int 
    \frac{e^{-\left(\Gamma \lambda_1 \epsilon_1+b_1\right)}
    \left(\Gamma \lambda_1 \epsilon_1+b_1\right)^{n_1}}{n_1!}\nonumber\\
   &&\ \times\frac{e^{-\left(\Gamma \lambda_2 \epsilon_2+b_2\right)}
    \left(\Gamma \lambda_2 \epsilon_2+b_2\right)^{n_2}}{n_2!}\nonumber\\
   &&\ \times
    {\bf P}(\Gamma){\bf P}(\delta_\epsilon)
    {\bf P}(\delta_\lambda){\bf P}(\delta_b)
    d\delta_\epsilon d\delta_\lambda d\delta_b.
  \end{eqnarray}
  
  The confidence level and lifetime limit are calculated by integrating this probability density
  function as in Equations \ref{eq:cl} and \ref{eq:limit}.

  For a mode with two different meson decay modes like
  $\eta\rightarrow 2\gamma$ and $\eta\rightarrow 3\pi^0$,
  the efficiencies and the backgrounds were simply added
  because such different meson decay mode searches are independent.
  The systematic errors on the detection efficiencies and the
  backgrounds were also simply added 
  assuming 100\% correlation.

\begin{figure}[htbp]
    \begin{center}
     \includegraphics[width=\linewidth]{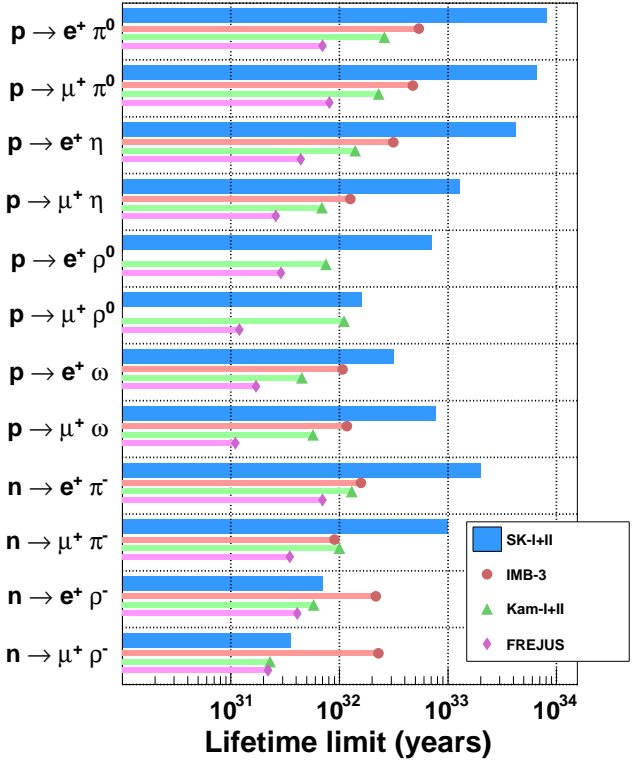}
     \caption{Explored ranges and lower limits (at 90\% confidence level)
     of nucleon partial lifetime      
     with the results of the previous experiments;
     IMB-3~\cite{McGrew:1999nd}, KAMIOKANDE-I+II~\cite{Hirata:1989kn} and
     FREJUS~\cite{Berger:1990kg}.}
     \label{fig:result} 
    \end{center}
   \end{figure}

  The nucleon partial lifetime limits at 90\% confidence level were
  obtained and 
  summarized in Table~\ref{tab:summaryall}
  and also shown in Fig.~\ref{fig:result}.
  
\section{Conclusion}
   Nucleon decays into a charged anti-lepton ($e^+$ and $\mu^+$) plus a
   light meson ($\pi^0$, $\pi^-$, $\eta$, $\rho^0$, $\rho^-$ and $\omega$) were searched for in 91.7
   and 49.2 kiloton$\cdot$year exposures of the SK-I and SK-II
   data, respectively.

   Performances for nucleon decay searches were compared between SK-I and
   SK-II.
   The observation in the SK-II period had similar performance
   to that in the SK-I period even though the photocathode coverage was
   half of SK-I.

   No evidence for proton decays via the \eqepi0\ mode was found,
   though this mode has the highest detection efficiency and is
   the dominant proton decay mode in various GUT models.
   Six candidate events were found in the SK-I and SK-II data for the
   five largest-background modes.
    The total expected background from atmospheric
    neutrinos was 4.7 events.
    The number and features of candidate events
    are consistent with the background estimate
    by the atmospheric neutrino MC.
    
   Nucleon partial lifetime limits were calculated based on Bayes'
   theorem.
   The lower limit on the partial lifetime of the proton via the \eqepi0\ mode
   was 
   calculated to be
   $8.2\times10^{33}$ years at 90\% confidence level.

   As for the \eqepi0\ and \eqmupi0\ modes, we have
   applied the same set of cuts to an increased exposure of the
   Super-Kamiokande detector and found no candidates. The new exposure
   includes running periods of SK-III and SK-IV. SK-III has restored
   photocoverage of 11100 PMTs, but with the acrylic shields
   introduced for SK-II. The SK-IV running includes new
   electronics. The same proton decay signal and atmospheric neutrino
   Monte Carlo methods were used to estimate the signal efficiency and
   background. We found for SK-III the efficiency for \eqepi0\ to be
   45.2\% and the efficiency for \eqmupi0\ to be 36.3\%, with
   background rates of 1.9 events/megaton$\cdot$year and 2.5
   events/megaton$\cdot$year respectively.  For SK-IV we found the
   efficiency for \eqepi0\ to be 45.0\% and the efficiency for
   \eqmupi0\ to be 43.9\%, with background rates of 1.7
   events/megaton$\cdot$year and 3.6 events/megaton$\cdot$year
   respectively. These numbers are consistent with the efficiencies
   and background rates presented in this paper, with the exception
   that the increased efficiency for \eqmupi0\ in SK-IV is attributed
   to an increased efficiency for muon-decay electron finding due to
   the improved electronics of SK-IV. The distributions of total
   invariant mass and total momentum are substantially similar to
   those for SK-I and SK-II presented in Fig.~\ref{fig:epi0_mptot}.
   As a result of finding no candidates, we also report an updated
   lifetime limit for 219.7 kt-years of SK-I, II, III, and IV
   exposure, finding $\tau/B > 1.29\times 10^{34}$ years for \eqepi0\ and
   $\tau/B > 1.08\times 10^{34}$ years for \eqmupi0\ at 90\% CL. 
   
   The obtained lower partial lifetime limits via the other modes 
   except for \eqerhom\ and \eqmurhom\
   are  also more stringent than the previous limits by IMB-3 or
   KAMIOKANDE-I+II.
   They range from 1.6$\times 10^{32}$ to $6.6\times 10^{33}$ years.
   The obtained  lifetime limits for 
   the \eqerhom\ and \eqmurhom\ modes
   are 7.0$\times10^{31}$ and 3.6 $\times 10^{31}$ years, respectively.
   They are less stringent than the IMB-3 result. For the case of the
   \eqerhom\ mode, the signal efficiency and estimated background in
   IMB-3 are 49\% (without uncertainty) and $6.3\times 10^2$
   events/megaton$\cdot$year,
   respectively, with the cuts optimized to obtain the best lifetime
   limit expectation. On the other hand, we applied tighter selection
   criteria to reduce the huge backgrounds to 2.7 events/megaton$\cdot$year,
   resulting in a smaller signal efficiency with an assigned uncertainty
   of $1.7^{+1.0} _{-0.7}$\%. The smaller efficiency and determined
   uncertainty
   are the main reasons why the obtained limit is less stringent than
   IMB-3. The same applies to the \eqmurhom\ mode.

   This systematic study does not rule out specific GUT models, 
   such as SUSY SU(5), SO(10) and so on,
   but can constrain parameters
   relevant to nucleon decay mediated by a super-heavy gauge boson.
 
\begin{acknowledgments}

We gratefully acknowledge the cooperation of the Kamioka Mining and
Smelting Company. The Super-Kamiokande experiment has been built and
operated from funding by the Japanese Ministry of Education, Culture,
Sports, Science and Technology, the United States Department of Energy,
and the U.S. National Science Foundation. 
Some of us have been supported by funds from 
the Korean Research Foundation (BK21), 
the National Research Foundation of Korea (NRF-20110024009),
the State Committee for Scientific Research in Poland
(grant1757/B/H03/2008/35),
the Japan Society for the Promotion of Science,
and the National Natural Science Foundation of China under Grants
No. 10575056.

\end{acknowledgments}

\bibliography{bibliography}

\begin{thebibliography}{43}%
\makeatletter
\providecommand \@ifxundefined [1]{%
 \@ifx{#1\undefined}
}%
\providecommand \@ifnum [1]{%
 \ifnum #1\expandafter \@firstoftwo
 \else \expandafter \@secondoftwo
 \fi
}%
\providecommand \@ifx [1]{%
 \ifx #1\expandafter \@firstoftwo
 \else \expandafter \@secondoftwo
 \fi
}%
\providecommand \natexlab [1]{#1}%
\providecommand \enquote  [1]{``#1''}%
\providecommand \bibnamefont  [1]{#1}%
\providecommand \bibfnamefont [1]{#1}%
\providecommand \citenamefont [1]{#1}%
\providecommand \href@noop [0]{\@secondoftwo}%
\providecommand \href [0]{\begingroup \@sanitize@url \@href}%
\providecommand \@href[1]{\@@startlink{#1}\@@href}%
\providecommand \@@href[1]{\endgroup#1\@@endlink}%
\providecommand \@sanitize@url [0]{\catcode `\\12\catcode `\$12\catcode
  `\&12\catcode `\#12\catcode `\^12\catcode `\_12\catcode `\%12\relax}%
\providecommand \@@startlink[1]{}%
\providecommand \@@endlink[0]{}%
\providecommand \url  [0]{\begingroup\@sanitize@url \@url }%
\providecommand \@url [1]{\endgroup\@href {#1}{\urlprefix }}%
\providecommand \urlprefix  [0]{URL }%
\providecommand \Eprint [0]{\href }%
\providecommand \doibase [0]{http://dx.doi.org/}%
\providecommand \selectlanguage [0]{\@gobble}%
\providecommand \bibinfo  [0]{\@secondoftwo}%
\providecommand \bibfield  [0]{\@secondoftwo}%
\providecommand \translation [1]{[#1]}%
\providecommand \BibitemOpen [0]{}%
\providecommand \bibitemStop [0]{}%
\providecommand \bibitemNoStop [0]{.\EOS\space}%
\providecommand \EOS [0]{\spacefactor3000\relax}%
\providecommand \BibitemShut  [1]{\csname bibitem#1\endcsname}%
\let\auto@bib@innerbib\@empty
\bibitem [{\citenamefont {Georgi}\ and\ \citenamefont
  {Glashow}(1974)}]{Georgi:1974sy}%
  \BibitemOpen
  \bibfield  {author} {\bibinfo {author} {\bibfnamefont {H.}~\bibnamefont
  {Georgi}}\ and\ \bibinfo {author} {\bibfnamefont {S.~L.}\ \bibnamefont
  {Glashow}},\ }\href {\doibase 10.1103/PhysRevLett.32.438} {\bibfield
  {journal} {\bibinfo  {journal} {Phys. Rev. Lett.}\ }\textbf {\bibinfo
  {volume} {32}},\ \bibinfo {pages} {438} (\bibinfo {year} {1974})}\BibitemShut
  {NoStop}%
\bibitem [{\citenamefont {Langacker}(1981)}]{Langacker:1980js}%
  \BibitemOpen
  \bibfield  {author} {\bibinfo {author} {\bibfnamefont {P.}~\bibnamefont
  {Langacker}},\ }\href {\doibase 10.1016/0370-1573(81)90059-4} {\bibfield
  {journal} {\bibinfo  {journal} {Phys. Rept.}\ }\textbf {\bibinfo {volume}
  {72}},\ \bibinfo {pages} {185} (\bibinfo {year} {1981})}\BibitemShut
  {NoStop}%
\bibitem [{\citenamefont {Langacker}(1994)}]{Langacker:1994vf}%
  \BibitemOpen
  \bibfield  {author} {\bibinfo {author} {\bibfnamefont {P.}~\bibnamefont
  {Langacker}},\ }\href@noop {} {\  (\bibinfo {year} {1994})},\ \Eprint
  {http://arxiv.org/abs/hep-ph/9411247} {arXiv:hep-ph/9411247} \BibitemShut
  {NoStop}%
\bibitem [{\citenamefont {McGrew}\ \emph {et~al.}(1999)\citenamefont {McGrew}
  \emph {et~al.}}]{McGrew:1999nd}%
  \BibitemOpen
  \bibfield  {author} {\bibinfo {author} {\bibfnamefont {C.}~\bibnamefont
  {McGrew}} \emph {et~al.},\ }\href {\doibase 10.1103/PhysRevD.59.052004}
  {\bibfield  {journal} {\bibinfo  {journal} {Phys. Rev.}\ }\textbf {\bibinfo
  {volume} {D59}},\ \bibinfo {pages} {052004} (\bibinfo {year}
  {1999})}\BibitemShut {NoStop}%
\bibitem [{\citenamefont {Hirata}\ \emph {et~al.}(1989)\citenamefont {Hirata}
  \emph {et~al.}}]{Hirata:1989kn}%
  \BibitemOpen
  \bibfield  {author} {\bibinfo {author} {\bibfnamefont {K.~S.}\ \bibnamefont
  {Hirata}} \emph {et~al.} (\bibinfo {collaboration} {KAMIOKANDE-II}),\ }\href
  {\doibase 10.1016/0370-2693(89)90058-0} {\bibfield  {journal} {\bibinfo
  {journal} {Phys. Lett.}\ }\textbf {\bibinfo {volume} {B220}},\ \bibinfo
  {pages} {308} (\bibinfo {year} {1989})}\BibitemShut {NoStop}%
\bibitem [{\citenamefont {Pati}\ and\ \citenamefont
  {Salam}(1974)}]{Pati:1974yy}%
  \BibitemOpen
  \bibfield  {author} {\bibinfo {author} {\bibfnamefont {J.~C.}\ \bibnamefont
  {Pati}}\ and\ \bibinfo {author} {\bibfnamefont {A.}~\bibnamefont {Salam}},\
  }\href {\doibase 10.1103/PhysRevD.10.275} {\bibfield  {journal} {\bibinfo
  {journal} {Phys. Rev.}\ }\textbf {\bibinfo {volume} {D10}},\ \bibinfo {pages}
  {275} (\bibinfo {year} {1974})}\BibitemShut {NoStop}%
\bibitem [{\citenamefont {Lee}\ \emph {et~al.}(1995)\citenamefont {Lee},
  \citenamefont {Mohapatra}, \citenamefont {Parida},\ and\ \citenamefont
  {Rani}}]{Lee:1994vp}%
  \BibitemOpen
  \bibfield  {author} {\bibinfo {author} {\bibfnamefont {D.~G.}\ \bibnamefont
  {Lee}}, \bibinfo {author} {\bibfnamefont {R.~N.}\ \bibnamefont {Mohapatra}},
  \bibinfo {author} {\bibfnamefont {M.~K.}\ \bibnamefont {Parida}}, \ and\
  \bibinfo {author} {\bibfnamefont {M.}~\bibnamefont {Rani}},\ }\href {\doibase
  10.1103/PhysRevD.51.229} {\bibfield  {journal} {\bibinfo  {journal} {Phys.
  Rev.}\ }\textbf {\bibinfo {volume} {D51}},\ \bibinfo {pages} {229} (\bibinfo
  {year} {1995})},\ \Eprint {http://arxiv.org/abs/hep-ph/9404238}
  {arXiv:hep-ph/9404238} \BibitemShut {NoStop}%
\bibitem [{\citenamefont {Shaban}\ and\ \citenamefont
  {Stirling}(1992)}]{Shaban:1992vv}%
  \BibitemOpen
  \bibfield  {author} {\bibinfo {author} {\bibfnamefont {N.~T.}\ \bibnamefont
  {Shaban}}\ and\ \bibinfo {author} {\bibfnamefont {W.~J.}\ \bibnamefont
  {Stirling}},\ }\href {\doibase 10.1016/0370-2693(92)91046-C} {\bibfield
  {journal} {\bibinfo  {journal} {Phys. Lett.}\ }\textbf {\bibinfo {volume}
  {B291}},\ \bibinfo {pages} {281} (\bibinfo {year} {1992})}\BibitemShut
  {NoStop}%
\bibitem [{\citenamefont {Pati}(2003)}]{Pati:2003qia}%
  \BibitemOpen
  \bibfield  {author} {\bibinfo {author} {\bibfnamefont {J.~C.}\ \bibnamefont
  {Pati}},\ }\href {\doibase 10.1142/S0217751X03017427} {\bibfield  {journal}
  {\bibinfo  {journal} {Int. J. Mod. Phys.}\ }\textbf {\bibinfo {volume}
  {A18}},\ \bibinfo {pages} {4135} (\bibinfo {year} {2003})},\ \Eprint
  {http://arxiv.org/abs/hep-ph/0305221} {arXiv:hep-ph/0305221} \BibitemShut
  {NoStop}%
\bibitem [{\citenamefont {Kim}\ and\ \citenamefont {Raby}(2003)}]{Kim:2002im}%
  \BibitemOpen
  \bibfield  {author} {\bibinfo {author} {\bibfnamefont {H.~D.}\ \bibnamefont
  {Kim}}\ and\ \bibinfo {author} {\bibfnamefont {S.}~\bibnamefont {Raby}},\
  }\href@noop {} {\bibfield  {journal} {\bibinfo  {journal} {JHEP}\ }\textbf
  {\bibinfo {volume} {01}},\ \bibinfo {pages} {056} (\bibinfo {year} {2003})},\
  \Eprint {http://arxiv.org/abs/hep-ph/0212348} {arXiv:hep-ph/0212348}
  \BibitemShut {NoStop}%
\bibitem [{\citenamefont {Buchmuller}\ \emph {et~al.}(2004)\citenamefont
  {Buchmuller}, \citenamefont {Covi}, \citenamefont {Emmanuel-Costa},\ and\
  \citenamefont {Wiesenfeldt}}]{Buchmuller:2004eg}%
  \BibitemOpen
  \bibfield  {author} {\bibinfo {author} {\bibfnamefont {W.}~\bibnamefont
  {Buchmuller}}, \bibinfo {author} {\bibfnamefont {L.}~\bibnamefont {Covi}},
  \bibinfo {author} {\bibfnamefont {D.}~\bibnamefont {Emmanuel-Costa}}, \ and\
  \bibinfo {author} {\bibfnamefont {S.}~\bibnamefont {Wiesenfeldt}},\ }\href
  {\doibase 10.1088/1126-6708/2004/09/004} {\bibfield  {journal} {\bibinfo
  {journal} {JHEP}\ }\textbf {\bibinfo {volume} {09}},\ \bibinfo {pages} {004}
  (\bibinfo {year} {2004})},\ \Eprint {http://arxiv.org/abs/hep-ph/0407070}
  {arXiv:hep-ph/0407070} \BibitemShut {NoStop}%
\bibitem [{\citenamefont {Ellis}\ \emph {et~al.}(2002)\citenamefont {Ellis},
  \citenamefont {Nanopoulos},\ and\ \citenamefont {Walker}}]{Ellis:2002vk}%
  \BibitemOpen
  \bibfield  {author} {\bibinfo {author} {\bibfnamefont {J.~R.}\ \bibnamefont
  {Ellis}}, \bibinfo {author} {\bibfnamefont {D.~V.}\ \bibnamefont
  {Nanopoulos}}, \ and\ \bibinfo {author} {\bibfnamefont {J.}~\bibnamefont
  {Walker}},\ }\href {\doibase 10.1016/S0370-2693(02)02956-8} {\bibfield
  {journal} {\bibinfo  {journal} {Phys. Lett.}\ }\textbf {\bibinfo {volume}
  {B550}},\ \bibinfo {pages} {99} (\bibinfo {year} {2002})},\ \Eprint
  {http://arxiv.org/abs/hep-ph/0205336} {arXiv:hep-ph/0205336} \BibitemShut
  {NoStop}%
\bibitem [{\citenamefont {Nishino}\ \emph {et~al.}(2009)\citenamefont {Nishino}
  \emph {et~al.}}]{Nishino:2009gd}%
  \BibitemOpen
  \bibfield  {author} {\bibinfo {author} {\bibfnamefont {H.}~\bibnamefont
  {Nishino}} \emph {et~al.} (\bibinfo {collaboration} {Super-Kamiokande}),\
  }\href {\doibase 10.1103/PhysRevLett.102.141801} {\bibfield  {journal}
  {\bibinfo  {journal} {Phys. Rev. Lett.}\ }\textbf {\bibinfo {volume} {102}},\
  \bibinfo {pages} {141801} (\bibinfo {year} {2009})},\ \Eprint
  {http://arxiv.org/abs/0903.0676} {arXiv:0903.0676 [hep-ex]} \BibitemShut
  {NoStop}%
\bibitem [{\citenamefont {Raaf}\ \emph {et~al.}()\citenamefont {Raaf} \emph
  {et~al.}}]{epi0_sk1234}%
  \BibitemOpen
  \bibfield  {author} {\bibinfo {author} {\bibfnamefont {J.~L.}\ \bibnamefont
  {Raaf}} \emph {et~al.},\ }\href@noop {} {}\bibinfo {note} {Talk given at
  "Fundamental Physics at the Intensity Frontier" (Rockville, Maryland, USA,
  2011)}\BibitemShut {NoStop}%
\bibitem [{\citenamefont {Machacek}(1979)}]{Machacek:1979tx}%
  \BibitemOpen
  \bibfield  {author} {\bibinfo {author} {\bibfnamefont {M.}~\bibnamefont
  {Machacek}},\ }\href {\doibase 10.1016/0550-3213(79)90325-0} {\bibfield
  {journal} {\bibinfo  {journal} {Nucl. Phys.}\ }\textbf {\bibinfo {volume}
  {B159}},\ \bibinfo {pages} {37} (\bibinfo {year} {1979})}\BibitemShut
  {NoStop}%
\bibitem [{\citenamefont {Gavela}\ \emph {et~al.}(1981)\citenamefont {Gavela},
  \citenamefont {Le~Yaouanc}, \citenamefont {Oliver}, \citenamefont {Pene},\
  and\ \citenamefont {Raynal}}]{Gavela:1980at}%
  \BibitemOpen
  \bibfield  {author} {\bibinfo {author} {\bibfnamefont {M.~B.}\ \bibnamefont
  {Gavela}}, \bibinfo {author} {\bibfnamefont {A.}~\bibnamefont {Le~Yaouanc}},
  \bibinfo {author} {\bibfnamefont {L.}~\bibnamefont {Oliver}}, \bibinfo
  {author} {\bibfnamefont {O.}~\bibnamefont {Pene}}, \ and\ \bibinfo {author}
  {\bibfnamefont {J.~C.}\ \bibnamefont {Raynal}},\ }\href {\doibase
  10.1016/0370-2693(81)90366-X} {\bibfield  {journal} {\bibinfo  {journal}
  {Phys. Lett.}\ }\textbf {\bibinfo {volume} {B98}},\ \bibinfo {pages} {51}
  (\bibinfo {year} {1981})}\BibitemShut {NoStop}%
\bibitem [{\citenamefont {Donoghue}(1980)}]{Donoghue:1979pr}%
  \BibitemOpen
  \bibfield  {author} {\bibinfo {author} {\bibfnamefont {J.~F.}\ \bibnamefont
  {Donoghue}},\ }\href {\doibase 10.1016/0370-2693(80)90313-5} {\bibfield
  {journal} {\bibinfo  {journal} {Phys. Lett.}\ }\textbf {\bibinfo {volume}
  {B92}},\ \bibinfo {pages} {99} (\bibinfo {year} {1980})}\BibitemShut
  {NoStop}%
\bibitem [{\citenamefont {Buccella}\ \emph {et~al.}(1989)\citenamefont
  {Buccella}, \citenamefont {Miele}, \citenamefont {Rosa}, \citenamefont
  {Santorelli},\ and\ \citenamefont {Tuzi}}]{Buccella:1989za}%
  \BibitemOpen
  \bibfield  {author} {\bibinfo {author} {\bibfnamefont {F.}~\bibnamefont
  {Buccella}}, \bibinfo {author} {\bibfnamefont {G.}~\bibnamefont {Miele}},
  \bibinfo {author} {\bibfnamefont {L.}~\bibnamefont {Rosa}}, \bibinfo {author}
  {\bibfnamefont {P.}~\bibnamefont {Santorelli}}, \ and\ \bibinfo {author}
  {\bibfnamefont {T.}~\bibnamefont {Tuzi}},\ }\href {\doibase
  10.1016/0370-2693(89)90637-0} {\bibfield  {journal} {\bibinfo  {journal}
  {Phys. Lett.}\ }\textbf {\bibinfo {volume} {B233}},\ \bibinfo {pages} {178}
  (\bibinfo {year} {1989})}\BibitemShut {NoStop}%
\bibitem [{\citenamefont {Berezinsky}\ \emph {et~al.}(1981)\citenamefont
  {Berezinsky}, \citenamefont {Ioffe},\ and\ \citenamefont
  {Kogan}}]{Berezinsky:1981qb}%
  \BibitemOpen
  \bibfield  {author} {\bibinfo {author} {\bibfnamefont {V.~S.}\ \bibnamefont
  {Berezinsky}}, \bibinfo {author} {\bibfnamefont {B.~L.}\ \bibnamefont
  {Ioffe}}, \ and\ \bibinfo {author} {\bibfnamefont {Y.~I.}\ \bibnamefont
  {Kogan}},\ }\href {\doibase 10.1016/0370-2693(81)90034-4} {\bibfield
  {journal} {\bibinfo  {journal} {Phys. Lett.}\ }\textbf {\bibinfo {volume}
  {B105}},\ \bibinfo {pages} {33} (\bibinfo {year} {1981})}\BibitemShut
  {NoStop}%
\bibitem [{\citenamefont {Kobayashi}\ \emph {et~al.}(2005)\citenamefont
  {Kobayashi} \emph {et~al.}}]{Kobayashi:2005pe}%
  \BibitemOpen
  \bibfield  {author} {\bibinfo {author} {\bibfnamefont {K.}~\bibnamefont
  {Kobayashi}} \emph {et~al.} (\bibinfo {collaboration} {Super-Kamiokande}),\
  }\href {\doibase 10.1103/PhysRevD.72.052007} {\bibfield  {journal} {\bibinfo
  {journal} {Phys. Rev.}\ }\textbf {\bibinfo {volume} {D72}},\ \bibinfo {pages}
  {052007} (\bibinfo {year} {2005})},\ \Eprint
  {http://arxiv.org/abs/hep-ex/0502026} {arXiv:hep-ex/0502026} \BibitemShut
  {NoStop}%
\bibitem [{\citenamefont {Fukuda}\ \emph {et~al.}(2003)\citenamefont {Fukuda}
  \emph {et~al.}}]{Fukuda:2002uc}%
  \BibitemOpen
  \bibfield  {author} {\bibinfo {author} {\bibfnamefont {Y.}~\bibnamefont
  {Fukuda}} \emph {et~al.},\ }\href {\doibase 10.1016/S0168-9002(03)00425-X}
  {\bibfield  {journal} {\bibinfo  {journal} {Nucl. Instrum. Meth.}\ }\textbf
  {\bibinfo {volume} {A501}},\ \bibinfo {pages} {418} (\bibinfo {year}
  {2003})}\BibitemShut {NoStop}%
\bibitem [{\citenamefont {Nakamura}\ \emph {et~al.}(1976)\citenamefont
  {Nakamura} \emph {et~al.}}]{Nakamura:1976mb}%
  \BibitemOpen
  \bibfield  {author} {\bibinfo {author} {\bibfnamefont {K.}~\bibnamefont
  {Nakamura}} \emph {et~al.},\ }\href {\doibase 10.1016/0375-9474(76)90539-X}
  {\bibfield  {journal} {\bibinfo  {journal} {Nucl. Phys.}\ }\textbf {\bibinfo
  {volume} {A268}},\ \bibinfo {pages} {381} (\bibinfo {year}
  {1976})}\BibitemShut {NoStop}%
\bibitem [{\citenamefont {Yamazaki}\ and\ \citenamefont
  {Akaishi}(1999)}]{Yamazaki:1999gz}%
  \BibitemOpen
  \bibfield  {author} {\bibinfo {author} {\bibfnamefont {T.}~\bibnamefont
  {Yamazaki}}\ and\ \bibinfo {author} {\bibfnamefont {Y.}~\bibnamefont
  {Akaishi}},\ }\href {\doibase 10.1016/S0370-2693(99)00163-X} {\bibfield
  {journal} {\bibinfo  {journal} {Phys. Lett.}\ }\textbf {\bibinfo {volume}
  {B453}},\ \bibinfo {pages} {1} (\bibinfo {year} {1999})}\BibitemShut
  {NoStop}%
\bibitem [{gea(1993)}]{geant}%
  \BibitemOpen
  \href@noop {} {\enquote {\bibinfo {title} {Geant detector description and
  simulation tool},}\ } (\bibinfo {year} {1993}),\ \bibinfo {note} {{CERN
  Program Library Long Writeup W5013}}\BibitemShut {NoStop}%
\bibitem [{\citenamefont {Gabriel}\ \emph {et~al.}(1989)\citenamefont
  {Gabriel}, \citenamefont {Brau},\ and\ \citenamefont
  {Bishop}}]{Gabriel:1989ri}%
  \BibitemOpen
  \bibfield  {author} {\bibinfo {author} {\bibfnamefont {T.~A.}\ \bibnamefont
  {Gabriel}}, \bibinfo {author} {\bibfnamefont {J.~E.}\ \bibnamefont {Brau}}, \
  and\ \bibinfo {author} {\bibfnamefont {B.~L.}\ \bibnamefont {Bishop}},\
  }\href {\doibase 10.1109/23.34394} {\bibfield  {journal} {\bibinfo  {journal}
  {IEEE Trans. Nucl. Sci.}\ }\textbf {\bibinfo {volume} {36}},\ \bibinfo
  {pages} {14} (\bibinfo {year} {1989})}\BibitemShut {NoStop}%
\bibitem [{\citenamefont {Nakahata}\ \emph {et~al.}(1986)\citenamefont
  {Nakahata} \emph {et~al.}}]{Nakahata:1986zp}%
  \BibitemOpen
  \bibfield  {author} {\bibinfo {author} {\bibfnamefont {M.}~\bibnamefont
  {Nakahata}} \emph {et~al.} (\bibinfo {collaboration} {KAMIOKANDE}),\ }\href
  {\doibase 10.1143/JPSJ.55.3786} {\bibfield  {journal} {\bibinfo  {journal}
  {J. Phys. Soc. Jap.}\ }\textbf {\bibinfo {volume} {55}},\ \bibinfo {pages}
  {3786} (\bibinfo {year} {1986})}\BibitemShut {NoStop}%
\bibitem [{\citenamefont {Salcedo}\ \emph {et~al.}(1988)\citenamefont
  {Salcedo}, \citenamefont {Oset}, \citenamefont {Vicente-Vacas},\ and\
  \citenamefont {Garcia-Recio}}]{Salcedo:1987md}%
  \BibitemOpen
  \bibfield  {author} {\bibinfo {author} {\bibfnamefont {L.~L.}\ \bibnamefont
  {Salcedo}}, \bibinfo {author} {\bibfnamefont {E.}~\bibnamefont {Oset}},
  \bibinfo {author} {\bibfnamefont {M.~J.}\ \bibnamefont {Vicente-Vacas}}, \
  and\ \bibinfo {author} {\bibfnamefont {C.}~\bibnamefont {Garcia-Recio}},\
  }\href {\doibase 10.1016/0375-9474(88)90310-7} {\bibfield  {journal}
  {\bibinfo  {journal} {Nucl. Phys.}\ }\textbf {\bibinfo {volume} {A484}},\
  \bibinfo {pages} {557} (\bibinfo {year} {1988})}\BibitemShut {NoStop}%
\bibitem [{\citenamefont {Rowe}\ \emph {et~al.}(1978)\citenamefont {Rowe},
  \citenamefont {Salomon},\ and\ \citenamefont {Landau}}]{Rowe:1978fb}%
  \BibitemOpen
  \bibfield  {author} {\bibinfo {author} {\bibfnamefont {G.}~\bibnamefont
  {Rowe}}, \bibinfo {author} {\bibfnamefont {M.}~\bibnamefont {Salomon}}, \
  and\ \bibinfo {author} {\bibfnamefont {R.~H.}\ \bibnamefont {Landau}},\
  }\href {\doibase 10.1103/PhysRevC.18.584} {\bibfield  {journal} {\bibinfo
  {journal} {Phys. Rev.}\ }\textbf {\bibinfo {volume} {C18}},\ \bibinfo {pages}
  {584} (\bibinfo {year} {1978})}\BibitemShut {NoStop}%
\bibitem [{\citenamefont {Roebig-Landau}\ \emph {et~al.}(1996)\citenamefont
  {Roebig-Landau} \emph {et~al.}}]{RoebigLandau:1996xa}%
  \BibitemOpen
  \bibfield  {author} {\bibinfo {author} {\bibfnamefont {M.}~\bibnamefont
  {Roebig-Landau}} \emph {et~al.},\ }\href {\doibase
  10.1016/0370-2693(96)00125-6} {\bibfield  {journal} {\bibinfo  {journal}
  {Phys. Lett.}\ }\textbf {\bibinfo {volume} {B373}},\ \bibinfo {pages} {45}
  (\bibinfo {year} {1996})}\BibitemShut {NoStop}%
\bibitem [{sim()}]{sim:etaphotopro-nucleon}%
  \BibitemOpen
  \href@noop {} {}\bibinfo {howpublished}
  {\url{http://gwdac.phys.gwu.edu}}\BibitemShut {NoStop}%
\bibitem [{\citenamefont {Krusche}\ \emph {et~al.}(1995)\citenamefont {Krusche}
  \emph {et~al.}}]{Krusche:1995zx}%
  \BibitemOpen
  \bibfield  {author} {\bibinfo {author} {\bibfnamefont {B.}~\bibnamefont
  {Krusche}} \emph {et~al.},\ }\href {\doibase 10.1016/0370-2693(95)00985-T}
  {\bibfield  {journal} {\bibinfo  {journal} {Phys. Lett.}\ }\textbf {\bibinfo
  {volume} {B358}},\ \bibinfo {pages} {40} (\bibinfo {year}
  {1995})}\BibitemShut {NoStop}%
\bibitem [{\citenamefont {Lykasov}\ \emph {et~al.}(1999)\citenamefont
  {Lykasov}, \citenamefont {Cassing}, \citenamefont {Sibirtsev},\ and\
  \citenamefont {Rzyanin}}]{Lykasov:1998ma}%
  \BibitemOpen
  \bibfield  {author} {\bibinfo {author} {\bibfnamefont {G.~I.}\ \bibnamefont
  {Lykasov}}, \bibinfo {author} {\bibfnamefont {W.}~\bibnamefont {Cassing}},
  \bibinfo {author} {\bibfnamefont {A.}~\bibnamefont {Sibirtsev}}, \ and\
  \bibinfo {author} {\bibfnamefont {M.~V.}\ \bibnamefont {Rzyanin}},\ }\href
  {\doibase 10.1007/s100500050319} {\bibfield  {journal} {\bibinfo  {journal}
  {Eur. Phys. J.}\ }\textbf {\bibinfo {volume} {A6}},\ \bibinfo {pages} {71}
  (\bibinfo {year} {1999})},\ \Eprint {http://arxiv.org/abs/nucl-th/9811019}
  {arXiv:nucl-th/9811019} \BibitemShut {NoStop}%
\bibitem [{\citenamefont {Kotulla}\ \emph {et~al.}(2008)\citenamefont {Kotulla}
  \emph {et~al.}}]{Kotulla:2008xy}%
  \BibitemOpen
  \bibfield  {author} {\bibinfo {author} {\bibfnamefont {M.}~\bibnamefont
  {Kotulla}} \emph {et~al.} (\bibinfo {collaboration} {CBELSA/TAPS}),\ }\href
  {\doibase 10.1103/PhysRevLett.100.192302} {\bibfield  {journal} {\bibinfo
  {journal} {Phys. Rev. Lett.}\ }\textbf {\bibinfo {volume} {100}},\ \bibinfo
  {pages} {192302} (\bibinfo {year} {2008})},\ \Eprint
  {http://arxiv.org/abs/0802.0989} {arXiv:0802.0989 [nucl-ex]} \BibitemShut
  {NoStop}%
\bibitem [{\citenamefont {Hayato}(2002)}]{Hayato:2002sd}%
  \BibitemOpen
  \bibfield  {author} {\bibinfo {author} {\bibfnamefont {Y.}~\bibnamefont
  {Hayato}},\ }\href {\doibase 10.1016/S0920-5632(02)01759-0} {\bibfield
  {journal} {\bibinfo  {journal} {Nucl. Phys. B Proc. Suppl.}\ }\textbf
  {\bibinfo {volume} {112}},\ \bibinfo {pages} {171} (\bibinfo {year}
  {2002})}\BibitemShut {NoStop}%
\bibitem [{\citenamefont {Honda}\ \emph {et~al.}(2007)\citenamefont {Honda},
  \citenamefont {Kajita}, \citenamefont {Kasahara}, \citenamefont
  {Midorikawa},\ and\ \citenamefont {Sanuki}}]{Honda:2006qj}%
  \BibitemOpen
  \bibfield  {author} {\bibinfo {author} {\bibfnamefont {M.}~\bibnamefont
  {Honda}}, \bibinfo {author} {\bibfnamefont {T.}~\bibnamefont {Kajita}},
  \bibinfo {author} {\bibfnamefont {K.}~\bibnamefont {Kasahara}}, \bibinfo
  {author} {\bibfnamefont {S.}~\bibnamefont {Midorikawa}}, \ and\ \bibinfo
  {author} {\bibfnamefont {T.}~\bibnamefont {Sanuki}},\ }\href {\doibase
  10.1103/PhysRevD.75.043006} {\bibfield  {journal} {\bibinfo  {journal} {Phys.
  Rev.}\ }\textbf {\bibinfo {volume} {D75}},\ \bibinfo {pages} {043006}
  (\bibinfo {year} {2007})},\ \Eprint {http://arxiv.org/abs/astro-ph/0611418}
  {arXiv:astro-ph/0611418} \BibitemShut {NoStop}%
\bibitem [{\citenamefont {Ashie}\ \emph {et~al.}(2005)\citenamefont {Ashie}
  \emph {et~al.}}]{Ashie:2005ik}%
  \BibitemOpen
  \bibfield  {author} {\bibinfo {author} {\bibfnamefont {Y.}~\bibnamefont
  {Ashie}} \emph {et~al.} (\bibinfo {collaboration} {Super-Kamiokande}),\
  }\href {\doibase 10.1103/PhysRevD.71.112005} {\bibfield  {journal} {\bibinfo
  {journal} {Phys. Rev.}\ }\textbf {\bibinfo {volume} {D71}},\ \bibinfo {pages}
  {112005} (\bibinfo {year} {2005})},\ \Eprint
  {http://arxiv.org/abs/hep-ex/0501064} {arXiv:hep-ex/0501064} \BibitemShut
  {NoStop}%
\bibitem [{\citenamefont {Mine}\ \emph {et~al.}(2008)\citenamefont {Mine} \emph
  {et~al.}}]{Mine:2008rt}%
  \BibitemOpen
  \bibfield  {author} {\bibinfo {author} {\bibfnamefont {S.}~\bibnamefont
  {Mine}} \emph {et~al.} (\bibinfo {collaboration} {K2K}),\ }\href {\doibase
  10.1103/PhysRevD.77.032003} {\bibfield  {journal} {\bibinfo  {journal} {Phys.
  Rev.}\ }\textbf {\bibinfo {volume} {D77}},\ \bibinfo {pages} {032003}
  (\bibinfo {year} {2008})}\BibitemShut {NoStop}%
\bibitem [{\citenamefont {Casper}(2002)}]{Casper:2002sd}%
  \BibitemOpen
  \bibfield  {author} {\bibinfo {author} {\bibfnamefont {D.}~\bibnamefont
  {Casper}},\ }\href {\doibase 10.1016/S0920-5632(02)01756-5} {\bibfield
  {journal} {\bibinfo  {journal} {Nucl. Phys. Proc. Suppl.}\ }\textbf {\bibinfo
  {volume} {112}},\ \bibinfo {pages} {161} (\bibinfo {year} {2002})},\ \Eprint
  {http://arxiv.org/abs/hep-ph/0208030} {arXiv:hep-ph/0208030} \BibitemShut
  {NoStop}%
\bibitem [{\citenamefont {Bertini}(1972)}]{Bertini:1972vz}%
  \BibitemOpen
  \bibfield  {author} {\bibinfo {author} {\bibfnamefont {H.~W.}\ \bibnamefont
  {Bertini}},\ }\href {\doibase 10.1103/PhysRevC.6.631} {\bibfield  {journal}
  {\bibinfo  {journal} {Phys. Rev.}\ }\textbf {\bibinfo {volume} {C6}},\
  \bibinfo {pages} {631} (\bibinfo {year} {1972})}\BibitemShut {NoStop}%
\bibitem [{\citenamefont {Ingram}\ \emph {et~al.}(1983)\citenamefont {Ingram}
  \emph {et~al.}}]{Ingram:1982bn}%
  \BibitemOpen
  \bibfield  {author} {\bibinfo {author} {\bibfnamefont {C.~H.~Q.}\
  \bibnamefont {Ingram}} \emph {et~al.},\ }\href {\doibase
  10.1103/PhysRevC.27.1578} {\bibfield  {journal} {\bibinfo  {journal} {Phys.
  Rev.}\ }\textbf {\bibinfo {volume} {C27}},\ \bibinfo {pages} {1578} (\bibinfo
  {year} {1983})}\BibitemShut {NoStop}%
\bibitem [{\citenamefont {Albanese}\ \emph {et~al.}(1980)\citenamefont
  {Albanese} \emph {et~al.}}]{Albanese:1980np}%
  \BibitemOpen
  \bibfield  {author} {\bibinfo {author} {\bibfnamefont {J.~P.}\ \bibnamefont
  {Albanese}} \emph {et~al.},\ }\href {\doibase 10.1016/0375-9474(80)90461-3}
  {\bibfield  {journal} {\bibinfo  {journal} {Nucl. Phys.}\ }\textbf {\bibinfo
  {volume} {A350}},\ \bibinfo {pages} {301} (\bibinfo {year}
  {1980})}\BibitemShut {NoStop}%
\bibitem [{\citenamefont {Wendell}\ \emph {et~al.}(2010)\citenamefont {Wendell}
  \emph {et~al.}}]{Wendell:2010md}%
  \BibitemOpen
  \bibfield  {author} {\bibinfo {author} {\bibfnamefont {R.}~\bibnamefont
  {Wendell}} \emph {et~al.} (\bibinfo {collaboration} {Super-Kamiokande}),\
  }\href {\doibase 10.1103/PhysRevD.81.092004} {\bibfield  {journal} {\bibinfo
  {journal} {Phys. Rev.}\ }\textbf {\bibinfo {volume} {D81}},\ \bibinfo {pages}
  {092004} (\bibinfo {year} {2010})},\ \Eprint {http://arxiv.org/abs/1002.3471}
  {arXiv:1002.3471 [hep-ex]} \BibitemShut {NoStop}%
\bibitem [{\citenamefont {Berger}\ \emph {et~al.}(1991)\citenamefont {Berger}
  \emph {et~al.}}]{Berger:1990kg}%
  \BibitemOpen
  \bibfield  {author} {\bibinfo {author} {\bibfnamefont {C.}~\bibnamefont
  {Berger}} \emph {et~al.} (\bibinfo {collaboration} {Frejus}),\ }\href
  {\doibase 10.1007/BF01551450} {\bibfield  {journal} {\bibinfo  {journal} {Z.
  Phys.}\ }\textbf {\bibinfo {volume} {C50}},\ \bibinfo {pages} {385} (\bibinfo
  {year} {1991})}\BibitemShut {NoStop}%
\end{thebibliography}%

\end{document}